\documentclass[pra,aps,preprint,amsmath,amssymb,showpacs]{revtex4}
\usepackage{graphicx}
\usepackage{epsfig}
\usepackage{epstopdf}
\include{graphics}
\begin{document}

\title{Stable scalable control of soliton propagation 
in broadband nonlinear optical waveguides}

\author{Avner Peleg$^{1}$, Quan M. Nguyen$^{2}$, and Toan T. Huynh$^{3}$}

\affiliation{$^{1}$Department of Exact Sciences, Afeka College of Engineering, 
Tel Aviv 69988, Israel}

\affiliation{$^{2}$Department of Mathematics, International University, 
Vietnam National University-HCMC, Ho Chi Minh City, Vietnam}

\affiliation{$^{3}$Department of Mathematics, 
University of Medicine and Pharmacy-HCMC, Ho Chi Minh City, Vietnam}

%\date{\today}
% Paper on soliton propagation in the presence of narrowband nonlinear gain-loss 
% and broadband delayed Raman response.

\begin{abstract}
We develop a method for achieving scalable 
transmission stabilization and switching of $N$ colliding soliton sequences 
in optical waveguides with broadband delayed 
Raman response and narrowband nonlinear gain-loss. 
We show that dynamics of soliton amplitudes in $N$-sequence 
transmission is described by a generalized $N$-dimensional predator-prey model. 
Stability and bifurcation analysis for the predator-prey model are used to obtain 
simple conditions on the physical parameters for robust transmission 
stabilization as well as on-off and off-on switching of $M$ out of $N$ 
soliton sequences. Numerical simulations for single-waveguide 
transmission  with a system of $N$ coupled nonlinear 
Schr\"odinger equations with $2 \le N \le 4$ show excellent agreement with 
the predator-prey model's predictions and stable propagation 
over significantly larger distances compared with other 
broadband nonlinear single-waveguide systems. 
Moreover, stable on-off and off-on switching of multiple soliton sequences 
and stable multiple transmission switching events are demonstrated 
by the simulations. We discuss the reasons for the robustness and scalability of 
transmission stabilization and switching in waveguides 
with broadband delayed Raman response and narrowband nonlinear gain-loss, 
and explain their advantages compared with other 
broadband nonlinear waveguides.  
\end{abstract}
\maketitle
%\newpage

%\maketitle
\section{Introduction}
\label{Introduction}
The rates of information transmission through broadband optical waveguide links 
can be significantly increased by transmitting many pulse sequences 
through the same waveguide \cite{Agrawal2001,Tkach97,Mollenauer2006,Gnauck2008,Essiambre2010}.  
This is achieved by the wavelength-division-multiplexed (WDM) method, 
where each pulse sequence is characterized by the central frequency of its pulses, 
and is therefore called a frequency channel \cite{multisequence}.    
Applications of these WDM or multichannel systems 
include fiber optics transmission lines \cite{Tkach97,Mollenauer2006,Gnauck2008,Essiambre2010}, 
data transfer between computer processors through silicon waveguides \cite{Agrawal2007a,Dekker2007,Gaeta2008}, 
and multiwavelength lasers \cite{Chow96,Shi97,Zhang2009,Liu2013}. 
Since pulses from different frequency channels 
propagate with different group velocities, interchannel pulse collisions are very 
frequent, and can therefore lead to error generation and severe transmission 
degradation \cite{Agrawal2001,Tkach97,Mollenauer2006,Gnauck2008,Essiambre2010,Iannone98,MM98}. 
On the other hand, the significant collision-induced effects can be used for controlling 
the propagation, for tuning of optical pulse parameters, such as amplitude, frequency, and phase, 
and for transmission switching, i.e., the turning on or off of transmission of one or 
more of the pulse sequences \cite{NP2010,PNC2010,PC2012,CPJ2013,NPT2015,PNT2015}.   
A major advantage of multichannel waveguide systems compared with single-channel systems 
is that the former can simultaneously handle a large number of pulses using relatively low pulse energies.  
One of the most important challenges in multichannel transmission concerns the realization 
of stable scalable control of the transmission, which holds for an arbitrary number of frequency channels. 
In the current study we address this challenge, by showing that 
stable scalable transmission control can be achieved in multichannel optical waveguide systems 
with frequency dependent linear gain-loss, broadband delayed Raman response, 
and narrowband nonlinear gain-loss.

Interchannel crosstalk, which is the commonly used name for the energy exchange in 
interchannel collisions, is one of the main processes affecting pulse propagation 
in broadband waveguide systems. Two important crosstalk-inducing mechanisms 
are due to broadband delayed Raman response and broadband nonlinear gain-loss. 
Raman-induced interchannel crosstalk is an important impairment 
in WDM transmission lines employing silica glass fibers \cite{Chraplyvy84,Tkach95,Ho2000,Yamamoto2003,P2004,P2007,CP2008,Golovchenko2009,PC2012b}, but is also beneficially employed for amplification \cite{Islam2004,Agrawal2005}. 
Interchannel crosstalk due to cubic loss was shown to be a major factor in 
error generation in multichannel silicon nanowaveguide transmission \cite{Gaeta2010}. 
Additionally, crosstalk induced by quintic loss can lead to transmission degradation 
and loss of transmission scalability in multichannel optical waveguides due to the impact 
of three-pulse interaction on the crosstalk \cite{PC2012,PNG2014}.      
On the other hand, nonlinear gain-loss crosstalk can be used for achieving energy 
equalization, transmission stabilization, and transmission switching 
\cite{PNC2010,PC2012,CPJ2013,NPT2015}.

In several earlier studies \cite{NP2010,PNC2010,PC2012,CPJ2013,NPT2015,PNT2015}, 
we provided a partial solution to the key problem 
of achieving stable transmission control in multichannel nonlinear waveguide systems, 
considering solitons as an example for the optical pulses. 
Our approach was based on showing that the dynamics of soliton 
amplitudes in $N$-sequence transmission can be described by Lotka-Volterra (LV)  
models for $N$ species, where the specific form of the LV model depends on the nature 
of the dissipative processes in the waveguide.   
Stability and bifurcation analysis for the steady states of the LV models  
was used to guide a clever choice of the physical parameters, 
which in turn leads to transmission stabilization, i.e., 
the amplitudes of all propagating solitons approach 
desired predetermined values \cite{NP2010,PNC2010,PC2012,CPJ2013,NPT2015,PNT2015}. 
Furthermore, on-off and off-on transmission switching were demonstrated in two-channel 
waveguide systems with broadband nonlinear gain-loss \cite{CPJ2013,NPT2015}. 
The design of waveguide setups for transmission switching was also guided by 
stability and bifurcation analysis for the steady states of the LV models \cite{CPJ2013,NPT2015}.

The results of Refs. \cite{NP2010,PNC2010,PC2012,CPJ2013,NPT2015,PNT2015} 
provide the first steps toward employing crosstalk induced by delayed Raman response 
or by nonlinear gain-loss for transmission control, stabilization, and switching. 
However, these results are still quite limited, since they do not enable 
scalable transmission stabilization and switching for $N$ pulse sequences 
with a general $N$ value in a single optical waveguide.  
To explain this, we first note that 
in waveguides with broadband delayed Raman response, such as optical fibers, 
and in waveguides with broadband cubic loss, such as silicon waveguides, some 
or all of the soliton sequences propagate in the presence of net linear gain \cite{NP2010,PNC2010,PNT2015}. 
This leads to transmission destabilization at intermediate distances 
due to radiative instability and growth of small amplitude waves.  
As a result, the distances along which stable propagation is 
observed in these single-waveguide multichannel systems 
are relatively small even for small values of the Raman and cubic loss coefficients \cite{PNC2010,PNT2015}.  
The radiative instability observed in optical fibers and silicon waveguides  
can be mitigated by employing waveguides with linear loss, cubic gain, and quintic loss,  
i.e., waveguides with a Ginzburg-Landau (GL) gain-loss profile \cite{PC2012,CPJ2013,NPT2015}. 
However, the latter waveguides suffer from another serious limitation 
because of the broadband nature of the waveguides nonlinear gain-loss. 
More specifically, due to the presence of broadband quintic loss, 
three-pulse interaction gives an important contribution to   
collision-induced amplitude shifts \cite{PC2012,PNG2014}. 
The complex nature of three-pulse interaction in generic 
three-soliton collisions in this case (see Ref. \cite{PNG2014}) 
leads to a major difficulty in extending the LV model for amplitude dynamics from $N=2$ 
to a general $N$ value in waveguides with broadband nonlinear gain-loss. 
In the absence of a general $N$-dimensional LV model, 
it is unclear how to design setups for stable transmission stabilization and switching in $N$-sequence 
systems with $N>2$. For this reason, transmission stabilization and switching 
in waveguides with broadband nonlinear gain-loss 
were so far limited to two-sequence systems \cite{PC2012,CPJ2013,NPT2015}.

%=========================
%  Description of the results   
%=========================

In view of the limitations of the waveguides studied in Refs. 
\cite{NP2010,PNC2010,PC2012,CPJ2013,NPT2015,PNT2015}, 
it is important to look for new routes for realizing scalable transmission 
stabilization and switching, which work for $N$-sequence transmission 
with a general $N$ value. In the current paper we take on this task, 
by studying propagation of $N$ soliton sequences in nonlinear waveguides 
with frequency dependent linear gain-loss, broadband delayed Raman response, 
and narrowband nonlinear gain-loss. 
Due to the narrowband nature of the nonlinear gain-loss, 
it affects only single-pulse propagation and intrasequence interaction, 
but does not affect intersequence soliton collisions.  
We show that the combination of 
Raman-induced amplitude shifts in intersequence soliton collisions and single-pulse 
amplitude shifts due to gain-loss with properly chosen physical parameter values 
can be used to realize robust scalable trasmission stabilization and switching. 
For this purpose, we first obtain the generalized $N$-dimensional predator-prey 
model for amplitude dynamics in an $N$-sequence system. 
We then use stability and bifurcation analysis for the predator-prey model to obtain 
simple conditions on the values of the physical parameters, which lead to robust  
transmission stabilization as well as on-off and off-on switching of $M$ out of $N$ 
soliton sequences. The validity of the predator-prey model's predictions
is checked by carrying out numerical simulations with the full propagation model, 
which consists of a system of $N$ perturbed coupled nonlinear Schr\"odinger 
(NLS) equations. Our numerical simulations with $2 \le N \le 4$ 
soliton sequences show excellent agreement with 
the predator-prey model's predictions and stable propagation 
over significantly larger distances compared with other 
broadband nonlinear single-waveguide systems. 
Moreover, stable on-off and off-on switching of multiple soliton sequences 
and stable multiple transmission switching events are demonstrated 
by the simulations. We discuss the reasons for the robustness 
and scalability of transmission stabilization and switching in waveguides 
with broadband delayed Raman response and narrowband nonlinear gain-loss, 
and explain their advantages compared with other 
broadband nonlinear waveguides.

The rest of the paper is organized as follows. 
In Section \ref{models}, we present the coupled-NLS model 
for propagation of $N$ pulse sequences through waveguides 
with frequency dependent linear gain-loss, broadband delayed Raman response, 
and narrowband nonlinear gain-loss. 
In addition, we present the corresponding generalized $N$-dimensional 
predator-prey model for amplitude dynamics. 
In Section \ref{stability}, we carry out stability and bifurcation  
analysis for the steady states of the predator-prey model, 
and use the results to derive conditions on the values of the physical parameters 
for achieving scalable transmission stabilization and switching. 
In Section \ref{simu}, we present the results of numerical simulations 
with the coupled-NLS model for transmission stabilization, 
single switching events, and multiple transmission switching.  
We also analyze these results in comparison   
with the predictions of the predator-prey model. 
In Section \ref{Discussion}, we discuss the underlying reasons for 
the robustness and scalability of transmission stabilization and switching 
in waveguides with broadband delayed Raman response and narrowband nonlinear gain-loss. 
Section \ref{conclusions} is reserved for conclusions.

\section{Coupled-NLS and predator-prey models}
\label{models} 
\subsection{A coupled-NLS model for pulse propagation}
We consider $N$ sequences of optical pulses, each characterized by pulse frequency, 
propagating in an optical waveguide in the presence of second-order dispersion, 
Kerr nonlinearity, frequency dependent linear gain-loss, broadband delayed Raman response, 
and narrowband nonlinear gain-loss. We assume that the net linear gain-loss is the difference 
between amplifier gain and waveguide loss and that the frequency differences between 
all sequences are much larger than the spectral width of the pulses. 
Under these assumptions, the propagation is described by 
the following system of $N$ perturbed coupled-NLS equations: 
\begin{eqnarray} &&
i\partial_z\psi_{j}+\partial_{t}^2\psi_{j}+2|\psi_{j}|^2\psi_{j}
+4\sum_{k=1}^{N}(1-\delta_{jk})|\psi_{k}|^2\psi_{j}=
\nonumber \\&&
ig_{j}\psi_{j}/2+iL(|\psi_j|^{2})\psi_{j}
-\epsilon_{R}\psi_{j}\partial_{t}|\psi_{j}|^2 
%\nonumber \\&&
-\epsilon_{R}\sum_{k=1}^{N}(1-\delta_{jk})
\left[\psi_{j}\partial_{t}|\psi_{k}|^2
+\psi_{k}\partial_{t}(\psi_{j}\psi_{k}^{\ast})\right],  
\label{global1}
\end{eqnarray}             
where $\psi_{j}$ is proportional to the envelope of the electric field of the $j$th sequence, 
$1 \le j \le N$, $z$ is propagation distance, and $t$ is time.  
In Eq. (\ref{global1}), $g_{j}$ is the linear gain-loss coefficient for the $j$th sequence,  
$\epsilon_{R}$ is the Raman coefficient, and $L(|\psi_j|^{2})$ is a polynomial of $|\psi_j|^{2}$, 
describing the waveguide's nonlinear gain-loss profile. 
The values of the $g_{j}$ coefficients are determined 
by the $N$-dimensional predator-prey model for amplitude dynamics, 
by looking for steady-state transmission with equal amplitudes for all sequences. 
The second term on the left-hand side of Eq. (\ref{global1}) is due to second-order dispersion, 
while the third and fourth terms represent intrasequence and intersequence interaction 
due to Kerr nonlinearity. The first term on the right-hand side of Eq. (\ref{global1}) 
is due to linear gain-loss,  the second corresponds to intrasequence interaction due to nonlinear gain-loss, 
the third describes Raman-induced intrasequence interaction, 
while the fourth and fifth describe Raman-induced intersequence interaction.   
Since we consider waveguides with broadband delayed Raman response 
and narrowband nonlinear gain-loss, Raman-induced intersequence interaction 
is taken into account, while intersequence interaction due to nonlinear gain-loss 
is neglected. The polynomial $L$ in Eq. (\ref{global1}) 
can be quite general. In the current paper, we consider two central examples 
for waveguide systems with nonlinear gain-loss: 
(1) waveguides with a GL gain-loss profile, (2) waveguides with linear gain-loss and cubic loss.   
The expression for $L(|\psi_{j}|^2)$ for waveguides with a GL gain-loss profile is  
\begin{eqnarray} &&      
L_{1}(|\psi_{j}|^2)=\epsilon _{3}^{(1)}|\psi_{j}|^2-\epsilon _{5}|{\psi_{j}}|^4, 
\label{global2}
\end{eqnarray} 
where $\epsilon _{3}^{(1)}$ and $\epsilon _{5}$ are the cubic gain and quintic loss coefficients.
The expression for $L(|\psi_{j}|^2)$ for waveguides with linear gain-loss and cubic loss is
\begin{eqnarray} &&      
L_{2}(|\psi_{j}|^2)=-\epsilon _{3}^{(2)}\left|{\psi_{j}}\right|^2,  
\label{global3}
\end{eqnarray} 
where $\epsilon _{3}^{(2)}$ is the cubic loss coefficient. We emphasize, however, 
that our approach can be employed to treat a general form of the polynomial $L$.    
Note that since some of the perturbation terms 
in the propagation model (\ref{global1}) are nonlinear gain or loss terms, 
the model can also be regarded as a coupled system of GL equations.

%=====================
% Relation between dimensional and dimensionless quantities 
%=====================
The dimensional and dimensionless physical quantities are related by the 
standard scaling laws for NLS solitons \cite{Agrawal2001}. 
Exactly the same scaling relations were used in our previous works 
on soliton propagation in broadband nonlinear waveguide systems 
\cite{PNC2010,PC2012,CPJ2013,NPT2015,PNT2015}.
In these scaling relations, the dimensionless distance $z$ in Eq. (\ref{global1})  is 
$z=X/(2L_{D})$, where $X$ is the dimensional distance, 
$L_{D}=\tau_{0}^{2}/|\tilde\beta_{2}|$ is the dimensional dispersion length,    
$\tau_{0}$ is the soliton width, and $\tilde\beta_{2}$ is the second-order
dispersion coefficient. The dimensionless retarded time is
$t=\tau/\tau_{0}$, where $\tau$ is the retarded time.  
The solitons spectral width is $\nu_{0}=1/(\pi^{2}\tau_{0})$ and the 
frequency difference between adjacent channels is $\Delta\nu=(\pi\Delta\beta\nu_{0})/2$. 
$\psi_{j}=E_{j}/\sqrt{P_{0}}$, where $E_{j}$ is proportional to the 
electric field of the $j$th pulse sequence and $P_{0}$ is the peak power. 
The dimensionless second order dispersion coefficient is 
$d=-1=\tilde\beta_{2}/(\gamma P_{0}\tau_{0}^{2})$, 
where $\gamma$ is the Kerr nonlinearity coefficient.
The dimensional linear gain-loss coefficient for the $j$th sequence 
$\rho_{1j}^{(l)}$ is related to the dimensionless coefficient via      
$g_{j}^{(l)}=2\rho_{1j}^{(l)}/(\gamma P_{0})$. The coefficients 
$\epsilon_3^{(1)}$, $\epsilon_3^{(2)}$, and $\epsilon_{5}$ are 
related to the dimensional cubic gain $\rho_{3}^{(1)}$, 
cubic loss $\rho_{3}^{(2)}$, and quintic loss $\rho_{5}$, 
by $\epsilon_3^{(1)}=2\rho_{3}^{(1)}/\gamma$, $\epsilon_3^{(2)}=2\rho_{3}^{(2)}/\gamma$, 
and $\epsilon_5=2\rho_{5}P_{0}/\gamma$, respectively \cite{NPT2015}.   
The dimensionless Raman coefficient is $\epsilon_{R}=2\tau_{R}/\tau_{0}$, 
where $\tau_{R}$ is a dimensional time constant, 
characterizing the waveguide's delayed Raman response \cite{Agrawal2001,Chi89}. 
The time constant $\tau_{R}$ can be determined 
from the slope of the Raman gain curve of the waveguide \cite{Agrawal2001,Chi89}.

\begin{figure}[ptb]
\begin{tabular}{cc}
\epsfxsize=12cm  \epsffile{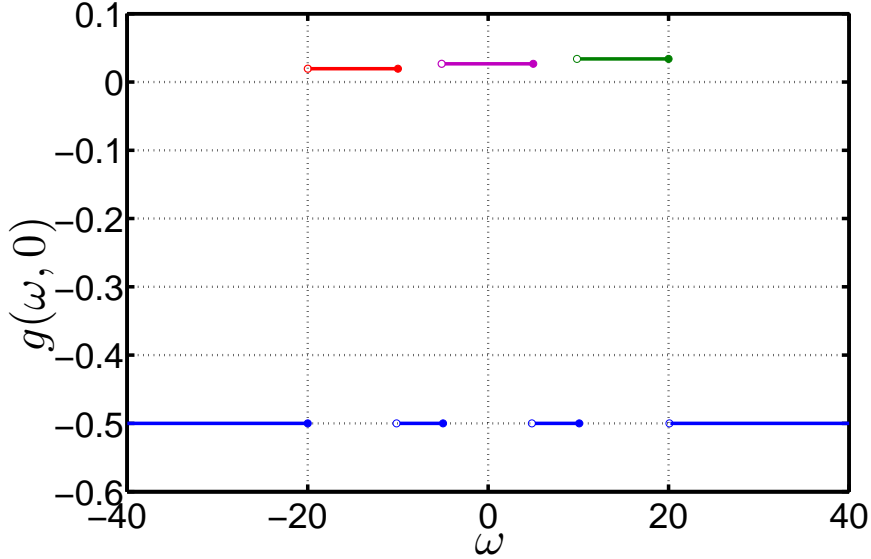} 
\end{tabular}
\caption{An example for the frequency dependent linear gain-loss 
function $g(\omega,z)$ of Eq. (\ref{global3a}) at $z=0$ 
in a three-channel system.}
 \label{fig_add1}
\end{figure}

We note that for waveguides with linear gain-loss and cubic loss, 
some or all of the pulse sequences 
propagate in the presence of net linear gain. This leads to transmission destabilization due to 
radiation emission. The radiative instability can be partially mitigated by employing 
frequency dependent linear gain-loss $g(\omega,z)$. In this case, the first term 
on the right hand side of Eq. (\ref{global1}) is replaced by $i{\cal F}^{-1}(g(\omega,z) \hat\psi_{j})/2$, 
where $\hat\psi$ is the Fourier transform of $\psi$ with respect to time, 
and ${\cal F}^{-1}$ stands for the inverse Fourier transform. 
The form of $g(\omega,z)$ is chosen such that existence of 
steady-state transmission with equal amplitudes for all sequences is retained, 
while radiation emission effects are minimized. 
More specifically, $g(\omega,z)$ is equal to a value $g_{j}$, required to balance 
amplitude shifts due to nonlinear gain-loss and Raman crosstalk, 
inside a frequency interval of width $W$ centered about the 
frequency of the $j$th-channel solitons at distance $z$, $\beta_{j}(z)$, 
and is equal to a negative value $g_{L}$ elsewhere \cite{g_omega_z}. 
Thus, $g(\omega,z)$ is given by:      
\begin{eqnarray} &&
g(\omega,z) =
%\nonumber \\&&
\left\{ \begin{array}{l l}
g_{j} &  \mbox{if} \;\;\beta_{j}(z)-W/2 < \omega \le \beta_{j}(z)+W/2 \;\; 
\mbox{for} \;\; 1\le j \le N,\\
g_{L} &  \mbox{elsewhere,}\\
\end{array} \right. 
\nonumber \\&&
\label{global3a}
\end{eqnarray} 
where $g_{L}<0$. The width $W$ in Eq. (\ref{global3a}) satisfies  $1 < W \le \Delta\beta$, 
where $\Delta\beta=\beta_{j+1}(0)-\beta_{j}(0)$ for $1 \le j \le N-1$. 
The values of the $g_{j}$ coefficients are determined 
by the generalized predator-prey model for collision-induced amplitude dynamics, 
such that amplitude shifts due to Raman crosstalk and nonlinear gain-loss 
are compensated for by the linear gain-loss. 
The values of $g_{L}$ and $W$ are determined by carrying out numerical simulations 
with Eqs. (\ref{global1}), (\ref{global3}), and (\ref{global3a}), while looking for the set, 
which yields the longest stable propagation distance \cite{g_omega_z}. 
Figure \ref{fig_add1} shows a typical example for the 
frequency dependent linear gain-loss function $g(\omega,z)$ at $z=0$ for a three-channel system 
with $g_{1}=0.0195$, $g_{2}=0.0267$, $g_{3}=0.0339$, $g_{L}=-0.5$,  
$\beta_{1}(0)=-15$, $\beta_{2}(0)=0$, $\beta_{3}(0)=15$, and $W=10$. 
These parameter values are used in the numerical simulations, 
whose results are shown in Fig. \ref{fig3}.

The optical pulses in the $j$th sequence are fundamental solitons of the unperturbed NLS equation 
with central frequency $\beta_{j}$. The envelopes of these solitons are given by 
$\psi_{sj}(t,z)=\eta_{j}\exp(i\chi_{j})\mbox{sech}(x_{j})$,
where $x_{j}=\eta_{j}\left(t-y_{j}-2\beta_{j} z\right)$, 
$\chi_{j}=\alpha_{j}+\beta_{j}(t-y_{j})+
\left(\eta_{j}^2-\beta_{j}^{2}\right)z$, and $\eta_{j}$, $y_{j}$, and $\alpha_{j}$ 
are the soliton amplitude, position, and phase. 
Due to the large frequency differences between the pulse sequences, 
the solitons undergo a large number of fast intersequence collisions. 
The energy exchange in the collisions, induced by broadband delayed Raman response, 
can lead to significant amplitude shifts and to transmission degradation.  On the other hand, 
the combination of Raman-induced amplitude shifts in intersequence collisions   
and single-pulse amplitude shifts due to frequency dependent linear gain-loss  
and narrowband nonlinear gain-loss with properly chosen coefficients can be 
used to realize  robust scalable transmission stabilization and switching. 
In the current paper, we demonstrate that such stable scalable transmission control can 
indeed be achieved, even with the simple nonlinear gain-loss profiles (\ref{global2}) and (\ref{global3}).

\subsection{A generalized $N$-dimensional predator-prey model for amplitude dynamics}

The design of waveguide setups for transmission stabilization and switching 
is based on the derivation of LV models for dynamics of soliton amplitudes. 
For this purpose, we consider propagation of $N$ soliton sequences in a waveguide loop,  
and assume that the frequency spacing $\Delta\beta$ between the sequences is a large constant, 
i.e., $\Delta\beta=|\beta_{j+1}(z)-\beta_{j}(z)|\gg 1$ for $1 \le j \le N-1$.  
Similar to Refs. \cite{NP2010,PNC2010}, we can show that amplitude 
dynamics of the $N$ sequences is approximately described by a generalized $N$-dimensional predator-prey model. 
The derivation of the predator-prey model is based on the following assumptions.
(1) The temporal separation $T$ between adjacent solitons in each sequence satisfies: 
$T \gg 1$. In addition, the amplitudes are equal for all solitons 
from the same sequence, but are not necessarily equal for solitons from 
different sequences. This setup corresponds, for example, 
to phase-shift-keyed soliton transmission.  
(2) As $T\gg 1$, intrasequence interaction is exponentially small and is neglected. 
(3) Delayed Raman response and gain-loss are assumed to be weak perturbations. 
As a result, high-order effects due to radiation emission are neglected, 
in accordance with single-collision analysis.

Since the pulse sequences are periodic, 
the amplitudes of all solitons in a given sequence undergo the same dynamics. 
Taking into account collision-induced amplitude shifts due to broadband delayed Raman response 
and single-pulse amplitude changes induced by gain and loss, we obtain the following equation 
for amplitude dynamics of the $j$th-sequence solitons 
(see Refs. \cite{NP2010,PNC2010} for similar derivations): 
\begin{eqnarray} &&
\frac{d \eta_{j}}{dz}=
\eta_{j}\left[g_{j} + F(\eta_{j}) +C\sum_{k=1}^{N}(k-j)f(|j-k|)\eta_{k}\right],  
\label{global4}
\end{eqnarray}
where $1 \le j \le N$, and $C=4\epsilon_{R}\Delta\beta/T$. 
The function $F(\eta_{j})$ on the right hand side of Eq. (\ref{global4})
is a polynomial in $\eta_{j}$, 
whose form is determined by the form of $L(|\psi_j|^{2})$. 
For $L_{1}$ and $L_{2}$ given by Eqs. (\ref{global2}) and (\ref{global3}), 
we obtain $F_{1}(\eta_{j})=4\epsilon_{3}^{(1)}\eta_{j}^{2}/3-16\epsilon_{5}\eta_{j}^{4}/15$ 
and  $F_{2}(\eta_{j})=-4\epsilon_{3}^{(2)}\eta_{j}^{2}/3$, respectively.            
The coefficients $f(|j-k|)$ on the right hand side of Eq. (\ref{global4}), 
which describe the strength of Raman interaction between $j$th- and 
$k$th-sequence solitons, are determined by the frequency dependence 
of the Raman gain. For the widely used triangular approximation for the Raman gain 
curve \cite{Agrawal2001,Chraplyvy84}, in which the gain is a piecewise linear function of the frequency,  
$f(|j-k|)=1$ for $1\le j \le N$ and $1\le k \le N$ \cite{NP2010}.

In order to demonstrate stable scalable control of soliton propagation, 
we look for an equilibrium state of the system (\ref{global4}) in the form
$\eta^{(eq)}_{j}=\eta>0$ for $1 \le j \le N$. 
Such equilibrium state corresponds to steady-state transmission with equal 
amplitudes for all sequences. This requirement leads to: 
\begin{eqnarray} &&
g_{j}=-F(\eta)-C\eta\sum_{k=1}^{N}(k-j)f(|j-k|).
\label{global4b}
\end{eqnarray}   
Consequently, Eq. (\ref{global4}) takes the form 
\begin{eqnarray} &&
\frac{d \eta_{j}}{dz}=
\eta_{j}\left[F(\eta_{j})-F(\eta)
%\right. 
%\nonumber \\&&
%\left.
+C\sum_{k=1}^{N}(k-j)f(|j-k|)(\eta_{k}-\eta)\right],   
\label{global5}
\end{eqnarray}   
which is a generalized predator-prey model for $N$ species \cite{Lotka25,Volterra28}. 
Notice that  $(\eta, \dots, \eta)$ and $(0, \dots, 0)$ are equilibrium states of the model 
for any positive values of $\epsilon _{3}^{(1)}$, $\epsilon _{3}^{(2)}$, $\epsilon _{5}$, 
$\eta$, and $C$.

We point out that the derivation of an $N$-dimensional predator-prey model 
with a general $N$ value is enabled by the narrow bandwidth of the waveguide's 
nonlinear gain-loss. Indeed, due to this property, the gain-loss does not contribute 
to amplitude shifts in interchannel collisions, and therefore, three-pulse interaction 
can be ignored. This makes the extension of the predator-prey model from $N=2$ 
to a general $N$ value straightforward. As a result, extending waveguide setup 
design from $N=2$ to a general $N$ value for waveguides with broadband delayed 
Raman response and narrowband nonlinear gain-loss is also straightforward. 
This situation is very different from the one encountered for waveguides 
with broadband nonlinear gain-loss. In the latter case, interchannel collisions 
are strongly affected by the nonlinear gain-loss, and three-pulse interaction 
gives an important contribution to the collision-induced amplitude shift \cite{PC2012,PNG2014}. 
Due to the complex nature of three-pulse interaction in generic 
three-soliton collisions in waveguides with broadband nonlinear gain or loss (see Ref. \cite{PNG2014}),  
it is very difficult to extend the LV model for amplitude dynamics from $N=2$ 
to a generic $N$ value for these waveguides. In the absence of an $N$-dimensional LV model, 
it is unclear how to design setups for stable transmission stabilization and switching 
in $N$-sequence systems with $N>2$.     
As a result, transmission stabilization and switching in waveguides with broadband nonlinear gain-loss 
have been so far limited to two-sequence systems \cite{PC2012,CPJ2013,NPT2015}.

\section{Stability analysis for the predator-prey model (\ref{global5})}
\label{stability}
\subsection{Introduction}
%======================
% Discussion of transmission switching 
%======================
Transmission stabilization and switching are guided by stability analysis 
of the equilibrium states of the predator-prey model (\ref{global5}). 
In particular, in transmission stabilization, 
we require that the equilibrium state $(\eta, \dots, \eta)$ 
is asymptotically stable, so that soliton amplitude values tend to $\eta$ with 
increasing propagation distance for all sequences. 
Furthermore, transmission switching is based 
on bifurcations of the equilibrium state $(\eta, \dots, \eta)$. 
To explain this, we denote by $\eta_{th}$ the value of the decision level, 
distinguishing between on and off transmission states, 
and consider transmission switching of $M$ sequences, for example.  
In off-on switching of $M$ sequences, the values of one or more of the physical parameters 
are changed at the switching distance $z_{s}$, such that $(\eta, \dots, \eta)$ 
turns from unstable to asymptotically stable. As a result, before switching, 
soliton amplitudes tend to values smaller than $\eta_{th}$ in $M$ sequences 
and to values larger than $\eta_{th}$ in $N-M$ sequences, while after the switching, 
soliton amplitudes in all $N$ sequences tend to $\eta$, where $\eta>\eta_{th}$.  
This means that transmission of $M$ sequences is turned on at $z>z_{s}$. 
On-off switching of $M$ sequences is realized by changing the physical parameters 
at $z=z_{s}$, such that $(\eta, \dots, \eta)$ turns from asymptotically stable to unstable, 
while another equilibrium state with $M$ components smaller than 
$\eta_{th}$ is asymptotically stable. Therefore, before switching, soliton amplitudes 
in all $N$ sequences tend to $\eta$, where $\eta>\eta_{th}$, while after the switching, 
soliton amplitudes tend to values smaller than $\eta_{th}$ in $M$ sequences and to values 
larger than $\eta_{th}$ in $N-M$ sequences. Thus, transmission of $M$ sequences 
is turned off at $z>z_{s}$ in this case. In both transmission stabilization and switching 
we require that the equilibrium state at the origin is asymptotically stable. 
This requirement is necessary 
in order to suppress radiative instability due to growth 
of small amplitude waves \cite{PC2012,CPJ2013,NPT2015}.

The setups of transmission switching that 
we develop and study in the current paper are different from the  
single-pulse switching setups that are commonly considered in nonlinear optics 
(see, e.g., Ref. \cite{Agrawal2001} for a description of the latter setups). 
We therefore point out the main differences between the two approaches 
to switching. First, in the common approach, the amplitude value in the off state is close to zero. 
In contrast, in our approach, 
the amplitude value in the off state only needs to be 
smaller than $\eta_{th}$, although the possibility to extend the switching 
to very small amplitude values does increase switching robustness.  
Second, in the common approach, the switching is based 
on a single collision or on a small number of collisions, and as a result, it often requires 
high-energy pulses for its implementation. In contrast, in our approach, 
the switching occurs as a result of the cumulative amplitude shift in a large number 
of fast interchannel collisions. Therefore, in this case pulse energies need not be high. 
Third and most important, in the common approach, the switching is carried out 
on a single pulse or on a few pulses. In contrast, in our approach, the switching 
is carried out on all pulses in the waveguide loop (or within a given waveguide span). 
As a result, the switching can be implemented with an arbitrary number of pulses. 
Because of this property, we can refer to transmission switching in our approach 
as channel switching. Since channel switching is carried out for all pulses inside the 
waveguide loop (or inside a given waveguide span), it can be much faster than  
conventional single-pulse switching. More specifically, 
channel switching can be faster by a factor of $M \times K$ 
compared with single-pulse switching, where $M$ is the number of channels, 
whose transmission is switched, and $K$ is the number of pulses per channel   
in the waveguide loop. For example, in a 100-channel system with $10^{4}$ 
pulses per channel, channel switching can be faster by a factor of $10^{6}$ 
compared with single-pulse switching.

Our channel switching approach can be used in any application, 
in which the same ``processing'' of all pulses within the same channel is required, 
where here processing can mean amplification, filtering, routing, signal processing, etc. 
A simple and widely known example for channel switching is provided by transmission recovery, 
i.e., the amplification of a sequence of pulses from small amplitudes values 
below $\eta_{th}$ to a desired final value above it.   
However, our channel switching approach can actually be used 
in a much more general and sophisticated manner. 
More specifically, let $p_{j}$ represent the transmission state of the $j$th channel, 
i.e., $p_{j}=0$ if the $j$th channel is off and $p_{j}=1$ if the $j$th channel is on. 
Then, the $N$-component vector $(p_{1}, ..., p_{j}, ..., p_{N})$, where $1 \le j \le N$, 
represents the transmission state of the entire $N$-channel system. 
One can then use this $N$-component vector to encode information 
about the processing to be carried out on different channels 
in the next ``processing station'' in the transmission line. 
After this processing has been carried out, 
the transmission state of the system can be switched to a new state, 
$(q_{1}, ..., q_{j}, ..., q_{N})$, 
which represents the type of processing to be carried out in the next processing station. 
Note that the channel switching approach is especially suitable 
for phase-shift-keyed transmission. Indeed, in this case, 
the phase is used for encoding the information, and therefore, 
no information is lost by operating with amplitude values smaller 
than $\eta_{th}$ \cite{AK_transmission}.      
%However, channel switching can also be implemented in 
%amplitude-keyed transmission. In this case, one should define 
%a second threshold level $\eta_{th2}$, satisfying $0<\eta_{th2}<\eta_{th}$. 
%The larger decision level $\eta_{th}$ is then used to determine 
%the transmission state of each channel for channel switching, 
%while the smaller decision level $\eta_{th2}$ is used to determine 
%the state of each time-slot within a given channel.         
%Thus, in amplitude-keyed transmission, the on and off states 
%for the $j$th channel are determined by the conditions 
%$\eta_{j}>\eta_{th}$ and $\eta_{th2}<\eta_{j}<\eta_{th}$, respectively, 
%where $\eta_{j}$ is the common amplitude value for pulses in occupied time slots.    

\subsection{Stability analysis for transmission stabilization and off-on switching}
Let us obtain the conditions on the values of the physical parameters for 
transmission stabilization and off-on switching. As explained above, in this case 
we require that both  $(\eta, \dots, \eta)$ and the origin are asymptotically stable 
equilibrium states of the predator-prey model (\ref{global5}).

We first analyze stability of the equilibrium state $(\eta, \dots, \eta)$ 
in a waveguide with a narrowband GL gain-loss profile, 
where $F(\eta_{j})=F_{1}(\eta_{j})$. For this purpose, we show that 
\begin{eqnarray} &&
V_{L}(\pmb{\eta})=
\sum_{j=1}^{N}\left[\eta_{j}-\eta
+\eta\ln\left(\eta/\eta_{j}\right)\right],
\label{global6}
\end{eqnarray}      
where $\pmb{\eta}=(\eta_{1},\dots,\eta_{j}, \dots, \eta_{N})$, 
is a Lyapunov function for Eq. (\ref{global5})  \cite{Lyapunov}. 
Indeed, we observe that $V_{L}(\pmb{\eta})\ge 0$ 
for any $\pmb{\eta}$ with $\eta_{j}>0$ for $1 \le j \le N$,
where equality holds only at the equilibrium point. 
Furthermore, the derivative of $V_{L}$ along trajectories 
of Eq. (\ref{global5}) satisfies: 
\begin{eqnarray} &&
d V_{L}/dz=-(16\epsilon_{5}/15)
\sum_{j=1}^{N}(\eta_{j}+\eta)(\eta_{j}-\eta)^2
%\nonumber \\&&
\times
\left(\eta_{j}^{2}+\eta^{2}-5\kappa/4\right), 
\label{global7}
\end{eqnarray}     
where $\kappa=\epsilon_{3}^{(1)}/\epsilon_{5}$ and $\epsilon_{5}\ne 0$. 
For asymptotic stability, we require $d V_{L}/dz < 0$. 
This condition is satisfied in a domain containing $(\eta, \dots, \eta)$ if $0< \kappa < 8\eta^{2}/5$. 
Thus, $V_{L}(\pmb{\eta})$ is a Lyapunov function for Eq. (\ref{global5}), 
and the equilibrium point $(\eta, \dots, \eta)$ is asymptotically stable, 
if $0< \kappa < 8\eta^{2}/5$ \cite{linear_stab}.   
When $0< \kappa \le 4\eta^{2}/5$, $(\eta, \dots, \eta)$ is globally asymptotically stable, 
since in this case, $d V_{L}/dz < 0$ for any initial condition with nonzero amplitude values. 
When  $4\eta^{2}/5 < \kappa < 8\eta^{2}/5$, $d V_{L}/dz < 0$ for amplitude values 
satisfying $\eta_{j} > (5\kappa/4-\eta^{2})^{1/2}$ for $1 \le j \le N$. 
Thus, in this case  the basin of attraction of $(\eta, \dots, \eta)$  
can be estimated by $((5\kappa/4-\eta^{2})^{1/2},\infty)$ for $1 \le j \le N$. 
For instability, we require $d V_{L}/dz > 0$ along trajectories of (\ref{global5}). 
This inequality is satisfied in a domain containing $(\eta, \dots, \eta)$ if $\kappa > 8\eta^{2}/5$. 
Therefore, $(\eta, \dots, \eta)$ is unstable for $\kappa > 8\eta^{2}/5$ \cite{linear_stab}.

Consider now the stability properties of the origin for $F(\eta_{j})=F_{1}(\eta_{j})$. 
Linear stability analysis shows that $(0, \dots, 0)$ is asymptotically stable (a stable node)
when $g_j < 0$ for $1 \le j \le N$, i.e., when all pulse sequences propagate in the presence 
of net linear loss. To slightly simplify the discussion, we now employ the widely 
accepted triangular approximation for the Raman gain curve \cite{Agrawal2001,Chraplyvy84}. 
In this case, $f(|j-k|)=1$ for $1\le j \le N$ and $1\le k \le N$ \cite{NP2010}, 
and therefore the net linear gain-loss coefficients take the form     
\begin{eqnarray} &&
g_{j}=-F_{1}(\eta) - CN(N+1)\eta/2 + CN\eta j.
\label{global7b}
\end{eqnarray}   
Since $g_{j}$ is increasing with increasing $j$, it is sufficient to require $g_{N}<0$. 
Substituting Eq. (\ref{global7b}) into this inequality, 
we find that the origin is asymptotically stable, provided that    
\begin{eqnarray} &&
\kappa > 4\eta^{2}/5+3CN(N-1)/(8\epsilon_{5}\eta).
\label{global7c}
\end{eqnarray}          
The same simple condition is obtained by showing 
that $V_{L}(\pmb{\eta})=\sum_{j=1}^{N} \eta_{j}^{2}$ 
is a Lyapunov function for Eq. (\ref{global5}).

Let us discuss the implications of stability analysis for $(\eta, \dots, \eta)$  
and the origin for transmission stabilization and off-on switching. 
Combining the requirements for asymptotic stability of both $(\eta, \dots, \eta)$ 
and the origin, we expect to observe stable long-distance propagation, 
for which soliton amplitudes in all sequences tend to their steady-state value $\eta$, 
provided the physical parameters satisfy 
\begin{eqnarray} &&
4\eta^{2}/5 + 3CN(N-1)/(8\epsilon_{5}\eta) 
< \kappa < 8\eta^{2}/5.
\label{global8}
\end{eqnarray}     
The same condition is required for realizing stable off-on transmission switching.  
Using inequality (\ref{global8}), we find that the smallest value of $\epsilon_{5}$, 
required for transmission stabilization and off-on switching, satisfies the simple condition 
\begin{eqnarray} &&
\epsilon_{5} > 15CN(N-1)/(32\eta^3). 
\label{global8b}
\end{eqnarray}  
As a result, the ratio $\epsilon_{R}/\epsilon_{5}$ should be a small parameter
in $N$-sequence transmission with $N \gg 1$. 
The independence of the stability condition for $(\eta, \dots, \eta)$ on $N$ and $\epsilon_{R}$ 
and the simple scaling properties of the stability condition for the origin are essential 
ingredients in enabling robust scalable transmission stabilization and switching.

Similar stability analysis can be carried out for waveguides  
with other forms of the nonlinear gain-loss $F(\eta_{j})$ \cite{Lyapunov}. 
Consider the central example of a waveguide with  narrowband cubic loss, 
where $F(\eta_{j})=F_{2}(\eta_{j})$. 
One can show that in this case $V_{L}(\pmb{\eta})$, 
given by Eq. (\ref{global6}), is a Lyapunov function 
for the predator-prey model (\ref{global5}), and that 
\begin{eqnarray} &&
d V_{L}/dz=-(4\epsilon_{3}^{(2)}/3)
\sum_{j=1}^{N}(\eta_{j}+\eta)(\eta_{j}-\eta)^2<0, 
\label{global9}
\end{eqnarray}         
for any trajectory with $\eta_{j}>0$ for $1\le j \le N$. 
Thus, $(\eta, \dots, \eta)$ is globally asymptotically stable, 
regardless of the values of $\eta$, $\epsilon_{R}$, $\epsilon_{3}^{(2)}$, and $N$. 
However, linear stability analysis shows that the origin is a saddle in this case, 
i.e., it is unstable. This instability is related to the fact that in waveguides with cubic loss, 
soliton sequences with $j$ values satisfying $j > (N+1)/2 - 4\epsilon_{3}^{(2)}\eta/(3CN)$ 
propagate under net linear gain, and are thus subject to radiative instability. 
The instability of the origin for uniform waveguides with cubic loss 
makes these waveguides unsuitable for long-distance transmission stabilization. 
On the other hand, the global stability of $(\eta, \dots, \eta)$
and its independence on the physical parameters, 
make waveguide spans with narrowband cubic loss 
very suitable for realizing robust scalable off-on switching in hybrid waveguides.   
To demonstrate this, consider a hybrid waveguide consisting of spans 
with linear gain-loss and cubic loss [$F(\eta_{j})=F_{2}(\eta_{j})$] 
and spans with a GL gain-loss profile [$F(\eta_{j})=F_{1}(\eta_{j})$]. 
In this case, the global stability of $(\eta, \dots, \eta)$ for spans 
with linear gain-loss and cubic loss can be used to bring amplitude 
values close to $\eta$ from small initial amplitude values, 
while the local stability of the origin for spans with a GL gain-loss profile 
can be employed to stabilize the propagation against radiation emission.

\subsection{Stability analysis for on-off switching}
We now describe stability analysis for on-off switching 
in waveguides with a GL gain-loss profile, 
considering the general case of switching off of $M$ out of $N$ 
soliton sequences. As explained in Subsection 3.1, 
in switching off of $M$ sequences, we require that $(\eta, \dots, \eta)$ is 
unstable, the origin is asymptotically stable, and another equilibrium state 
with $M$ components smaller than $\eta_{th}$ is also asymptotically stable. 
The requirement for instability of $(\eta, \dots, \eta)$ and asymptotic 
stability of the origin leads to the following condition on the physical parameter values: 
\begin{eqnarray} &&
\kappa > \mbox{max} \{8\eta^{2}/5,
4\eta^{2}/5+3CN(N-1)/(8\epsilon_{5}\eta)\}.
\label{global10}
\end{eqnarray}

In order to obtain guiding rules for choosing the on-off transmission switching setups, 
it is useful to consider first the case of switching off of N-1 out of N sequences.  
Suppose that we switch off the sequences $1 \le k \le j-1$  and $j+1 \le k \le N$. 
To realize such switching, we require that $(0, \dots, 0, \eta_{sj}, 0, \dots, 0)$ 
is a stable equilibrium point of Eq. (\ref{global5}).  
The value of $\eta_{sj}$ is determined by the equation 
\begin{eqnarray} &&
\eta_{sj}^{4}-5\kappa\eta_{sj}^{2}/4-15g_j/(16\epsilon_{5}) = 0.
\label{global11}
\end{eqnarray}     
Since the origin is a stable equilibrium point, transmission switching of 
$N-1$ sequences can be realized by requiring that Eq. (\ref{global11}) 
has two distinct roots on the positive half of the $\eta_{j}$-axis 
(the largest of which corresponds to $\eta_{sj}$). 
This requirement is satisfied, provided \cite{Negative_g_j}:
\begin{eqnarray} &&
\epsilon_{5} > 12|g_{j}|/(5\kappa^2). 
\label{global12}
\end{eqnarray} 
Assuming that $g_{1}<g_{2}<\dots<g_{N}<0$, we see that 
the switching off of the $N-1$ low-frequency sequences $1 \le j \le N-1$ is the least restrictive, 
since it can be realized with smaller $\epsilon_{5}$ values. 
For this reason, we choose to adopt the switching setup, 
in which sequences $1 \le j \le N-1$ are switched off.     
Employing inequality (\ref{global12}) and the triangular-approximation-based expression 
(\ref{global7b}) for $j=N$, we find that Eq. (\ref{global11}) 
has two distinct roots on the positive half of the $\eta_{N}$-axis, provided that    
\begin{eqnarray} &&
\!\!\!\!\!\!\!\!\!\!\!\!\!
\kappa > (8\eta/5)\left[5\kappa/4-\eta^{2}-15CN(N-1)
/(32\epsilon_{5}\eta)\right]^{1/2}.   
%\nonumber \\&&
\label{global13}  
\end{eqnarray}      
Therefore, the switching off of sequences $1 \le j \le N-1$ can be realized when 
conditions (\ref{global10}) and (\ref{global13}) are satisfied \cite{Eta_s_N}.

We now turn to discuss the general case, where transmission 
of $M$ out of $N$ sequences is switched off. 
Based on the discussion in the previous paragraph, one might expect 
that switching off of $M$ sequences can be most conveniently realized 
by turning off transmission of the low-frequency sequences, $1 \le j \le M$. 
This expectation is confirmed by numerical solution of the predator-prey 
model (\ref{global5}) and the coupled-NLS model (\ref{global1}). 
For this reason, we choose to employ switching off of 
$M$ sequences, in which transmission in the $M$ lowest frequency channels is turned off. 
Thus, we require that $(0, \dots, 0, \eta_{s(M+1)}, \dots, \eta_{sN})$ 
is an asymptotically stable equilibrium point of Eq. (\ref{global5}). 
The values of $\eta_{s(M+1)}, \dots, \eta_{sN}$ are determined by 
the following system of equations    
\begin{eqnarray}&& 
\!\!\!\!\!\!\!\!\!\!\!\!\!\!\!\!
\eta_{sj}^{4}-5\kappa\eta_{sj}^{2}/4-15g_j/(16\epsilon_{5}) 
%\nonumber \\&&
-15C/(16\epsilon_{5})
\!\!\!\!\sum_{k=M+1}^{N}\!\!\!\!(k-j)f(|j-k|)\eta_{sk} = 0,
\label{global14}
\end{eqnarray}     
where $M+1 \le j \le N$. Employing the triangular approximation for the 
Raman gain curve and using Eq. (\ref{global7b}), 
we can rewrite the system as: 
\begin{eqnarray} && 
\eta_{sj}^{4}-5\kappa\eta_{sj}^{2}/4
%\nonumber \\&&
-15C/(16\epsilon_{5}) \sum_{k=M+1}^{N}(k-j)\eta_{sk} 
-\eta^{4}+5\kappa\eta^{2}/4
\nonumber \\&&
+15CN[(N+1)/2-j]\eta/(16\epsilon_{5}) = 0.
\label{global15}
\end{eqnarray}        
Stability of $(0, \dots, 0, \eta_{s(M+1)}, \dots, \eta_{sN})$ is determined 
by calculating the eigenvalues of the Jacobian matrix $\mathcal {J}$ at this point. 
The calculation yields $\mathcal {J}_{jk}=0$ for $1 \le j \le M$ and $j \ne k$, 
\begin{eqnarray} && 
\mathcal {J}_{jj} = -4\epsilon_{3}^{(1)}\eta^{2}/3 + 16\epsilon_{5}\eta^{4}/15
%\nonumber \\&&
-C[N(N+1)\eta/2-\sum_{k=M+1}^{N}k\eta_{sk}] 
\nonumber \\&&
+C[N\eta-\sum_{k=M+1}^{N}\eta_{sk}]j
\;\;\; \mbox{for} \;\;\; 1 \le j \le M,     
\label{global16}
\end{eqnarray}
\begin{eqnarray} && 
\mathcal {J}_{jk} =C(k-j)\eta_{sj}
\;\;\; \mbox{for} \;\;\; M+1 \le j \le N \;\;\; \mbox{and} \;\;  j\ne k,  
\nonumber \\&&   
\label{global17}
\end{eqnarray}            
and 
\begin{eqnarray} && 
\!\!\!\!\!\!\!\!\!\!\!\!\!\!\!\!\!\!
\mathcal {J}_{jj} =
g_{j}+4\epsilon_{3}^{(1)}\eta_{sj}^{2} - 16\epsilon_{5}\eta_{sj}^{4}/3
%\nonumber \\&&   
+C\!\!\!\!\sum_{k=M+1}^{N}\!\!\!\! (k-j)\eta_{sk} 
\;\; \mbox{for} \;\; M+1 \le j \le N.     
\label{global18}
\end{eqnarray}
Note that the Raman triangular approximation was used to slightly simplify 
the form of Eqs. (\ref{global16})-(\ref{global18}).
Since $\mathcal {J}_{jk}=0$ for $1 \le j \le M$ and $j \ne k$, 
the first $M$ eigenvalues of the Jacobian matrix are 
$\lambda_{j}=\mathcal {J}_{jj}$, where the $\mathcal {J}_{jj}$ coefficients are  
given by Eq. (\ref{global16}). Furthermore, since $\mathcal {J}_{jj}$ is either 
monotonically increasing or monotonically decreasing with increasing $j$, 
to establish stability, it is sufficient to check that either $\mathcal {J}_{MM}<0$ 
or $\mathcal {J}_{11}<0$. To find the other $N-M$ eigenvalues of the Jacobian matrix, 
one needs to calculate the determinant of the $(N-M)\times(N-M)$ 
matrix, whose elements are $\mathcal {J}_{jk}$, where $M+1 \le j, k \le N$. 
The latter calculation can also be significantly simplified by noting that  
for $M+1 \le j \le N$, the diagonal elements are of order $\epsilon_{5}$, 
while the off-diagonal elements are of order $N\epsilon_{R}$ at most. 
Thus, the leading term in the expression for the determinant 
is of order $\epsilon_{5}^{N-M}$. The next term in the expansion 
is the sum of $N-M$ terms, each of which is of order   
$N^{2}\epsilon_{R}^{2}\epsilon_{5}^{N-M-2}$ at most.    
Therefore, the next term in the expansion of the determinant 
is of order $(N-M)N^{2}\epsilon_{R}^{2}\epsilon_{5}^{N-M-2}$ at most. 
Comparing the first and second terms, we see that the correction term 
can be neglected, provided that $\epsilon_{5} \gg N^{3/2}\epsilon_{R}$. 
We observe that the last condition is automatically satisfied by our 
on-off transmission switching setup for $N \gg 1$, since stability of the origin requires 
$\epsilon_{5} > N^{2}\epsilon_{R} \gg N^{3/2}\epsilon_{R}$ [see inequality (\ref{global10})].  
It follows that the other $N-M$ eigenvalues of the Jacobian matrix are well approximated 
by the diagonal elements $\mathcal {J}_{jj}$ for $M+1 \le j \le N$.    
Therefore, for $N \gg 1$, stability analysis of $(0, \dots, 0, \eta_{s(M+1)}, \dots, \eta_{sN})$ 
only requires the calculation of $N-M+1$ diagonal elements of the Jacobian matrix.

We point out that the preference for the turning off of transmission of low-frequency 
sequences in on-off switching is a consequence of the nature of the Raman-induced 
energy exchange in soliton collisions. Indeed, Raman crosstalk leads to energy transfer 
from high-frequency solitons to low-frequency ones 
\cite{Chi89,Malomed91,Kumar98,Kaup99,P2004,CP2005,NP2010b}. 
To compensate for this energy loss or gain,  
high-frequency sequences should be overamplified while low-frequency 
sequences should be underamplified compared to mid-frequency sequences \cite{NP2010,PNT2015}. 
As a result, the magnitude of the net linear loss is largest for the low-frequency 
sequences, and therefore, on-off switching is easiest to realize for these 
sequences. It follows that the presence of broadband delayed Raman response introduces a preference 
for turning off the transmission of the low-frequency sequences, and by this, enables 
systematic scalable on-off switching in $N$-sequence systems.

\section{Numerical simulations with the coupled-NLS model}
\label{simu}
The predator-prey model (\ref{global5}) is based on several simplifying assumptions, 
which might break down with increasing number of channels or at large propagation 
distances. In particular, Eq. (\ref{global5}) neglects the effects of pulse distortion, radiation emission, 
and intrasequence interaction that are incorporated in the full coupled-NLS model (\ref{global1}). 
These effects can lead to transmission destabilization and to the breakdown of the 
predator-prey model description \cite{PNC2010,PC2012,CPJ2013,NPT2015,PNT2015}.
In addition, during transmission switching, soliton amplitudes can become small, 
and as a result, the magnitude of the linear gain-loss term in Eq. (\ref{global1}) 
might become comparable to the magnitude of the Kerr nonlinearity terms. 
This can in turn lead to the breakdown of the perturbation theory, which is 
the basis for the derivation of the predator-prey model. 
It is therefore essential to test the validity of the predator-prey model's predictions 
by carrying out numerical simulations with the full coupled-NLS model (\ref{global1}).

The coupled-NLS system (\ref{global1}) is numerically integrated using the split-step method 
with periodic boundary conditions \cite{Agrawal2001}. Due to the usage of periodic 
boundary conditions, the simulations describe pulse propagation in a closed waveguide loop. 
The initial condition for the simulations consists of $N$ periodic sequences of $2K$ solitons 
with amplitudes $\eta_{j}(0)$, frequencies $\beta_{j}(0)$, and zero phases:    
\begin{eqnarray} &&
\psi_{j}(t,0)\!=\!\sum_{k=-K}^{K-1}
\frac{\eta_{j}(0)\exp\{i\beta_{j}(0)[t-(k+1/2)T-\delta_{j}]\}}
{\cosh\{\eta_{j}(0)[t-(k+1/2)T-\delta_{j}]\}}, 
%\nonumber \\&&   
\label{global21}
\end{eqnarray}
where the frequency differences satisfy 
$\Delta\beta=\beta_{j+1}(0)-\beta_{j}(0) \gg 1$, for $1\le j \le N-1$. 
The coefficients $\delta_{j}$ represent the initial position shift of the  
$j$th sequence solitons relative to pulses located at $(k+1/2)T$ for $-K\le k \le K-1$.   
To maximize propagation distance in the presence of delayed Raman response, 
we use $\delta_{j}=(j-1)T/N$ for $1\le j \le N$.  
As a concrete example, we present the results of numerical simulations 
for the following set of physical parameters:  $T=15$, $\Delta\beta=15$, and $K=1$. 
In addition, we employ the triangular approximation for the Raman 
gain curve, so that the coefficients $f(|j-k|)$ satisfy $f(|j-k|)=1$ for $1 \le j, k \le N$ \cite{NP2010,PNT2015}.   
We emphasize, however, that similar results are obtained with other choices of the physical parameter values, 
satisfying the stability conditions discussed in Section \ref{stability}.

\begin{figure*}[ptb]
\begin{tabular}{cc}
\epsfxsize=8.2cm  \epsffile{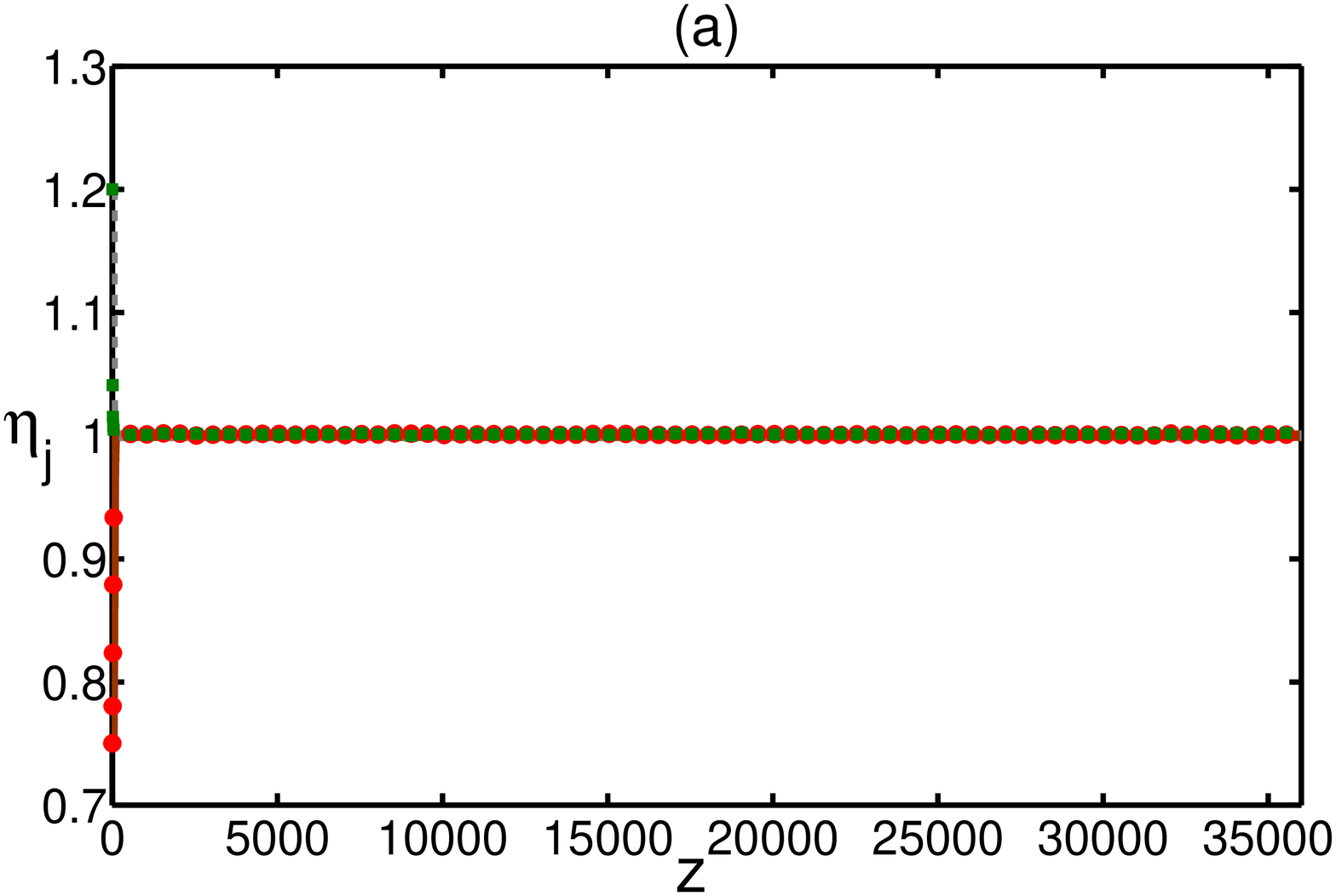} &
\epsfxsize=8.2cm  \epsffile{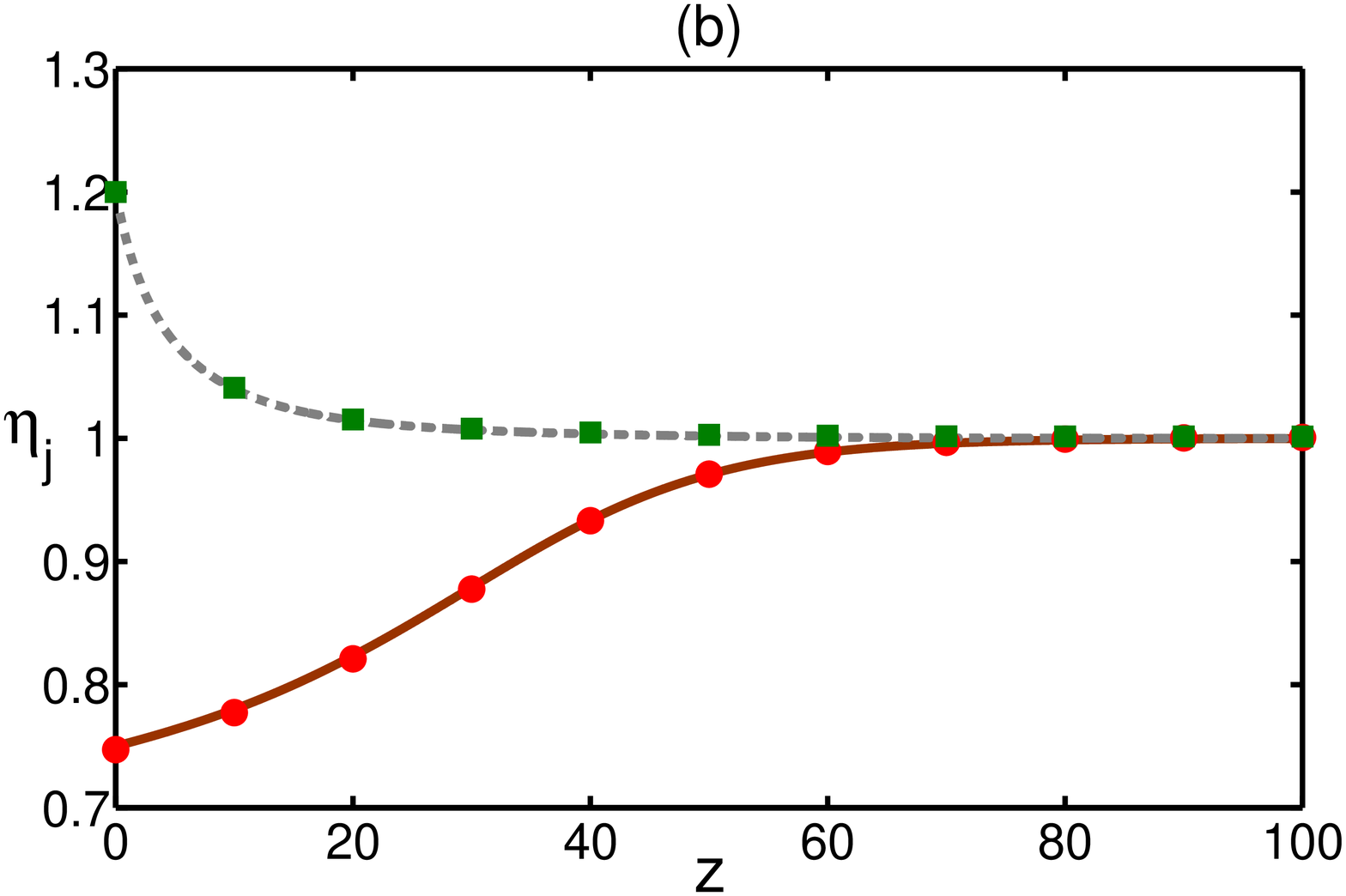} \\
\epsfxsize=8.2cm  \epsffile{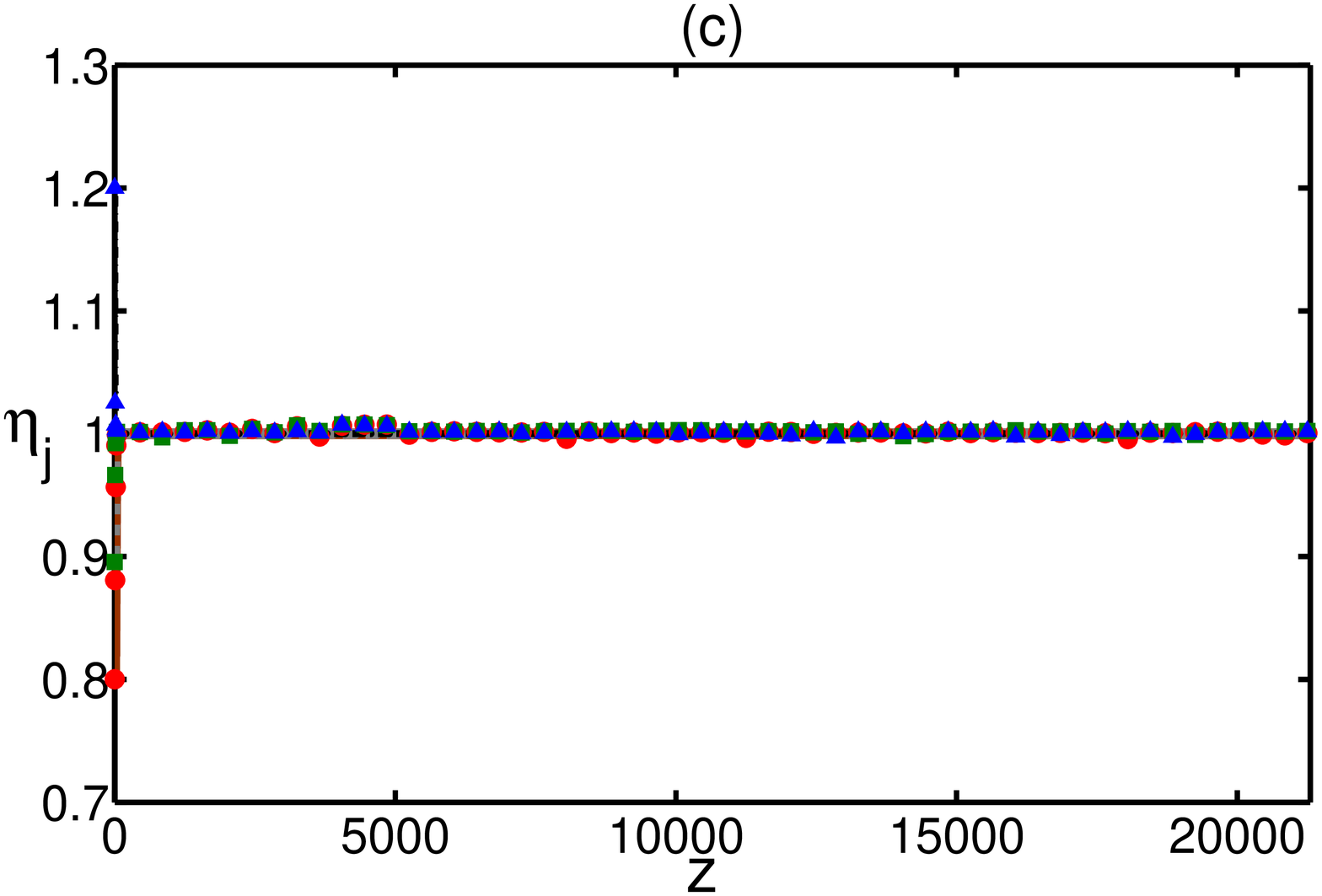} &
\epsfxsize=8.2cm  \epsffile{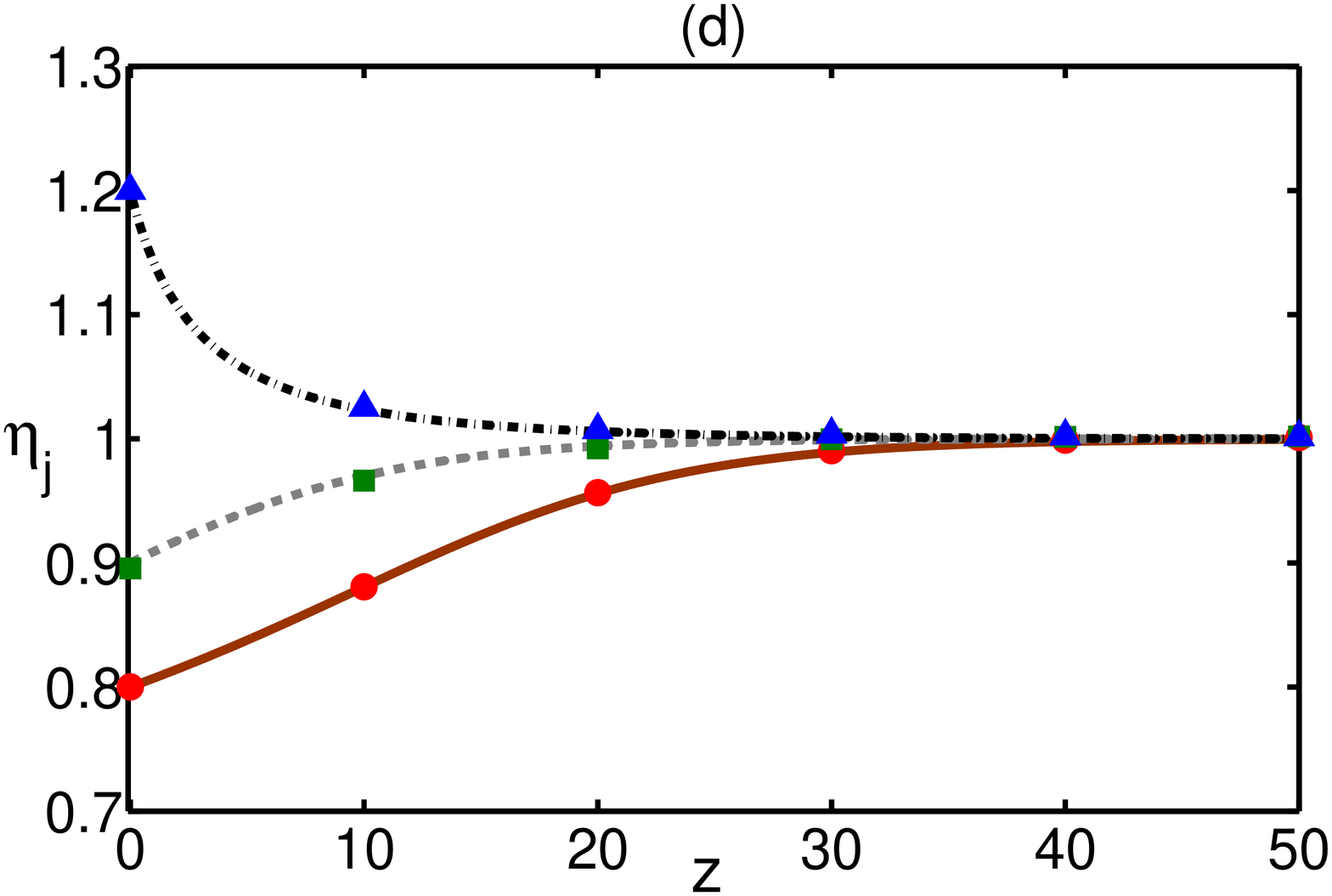}\\
\epsfxsize=8.2cm  \epsffile{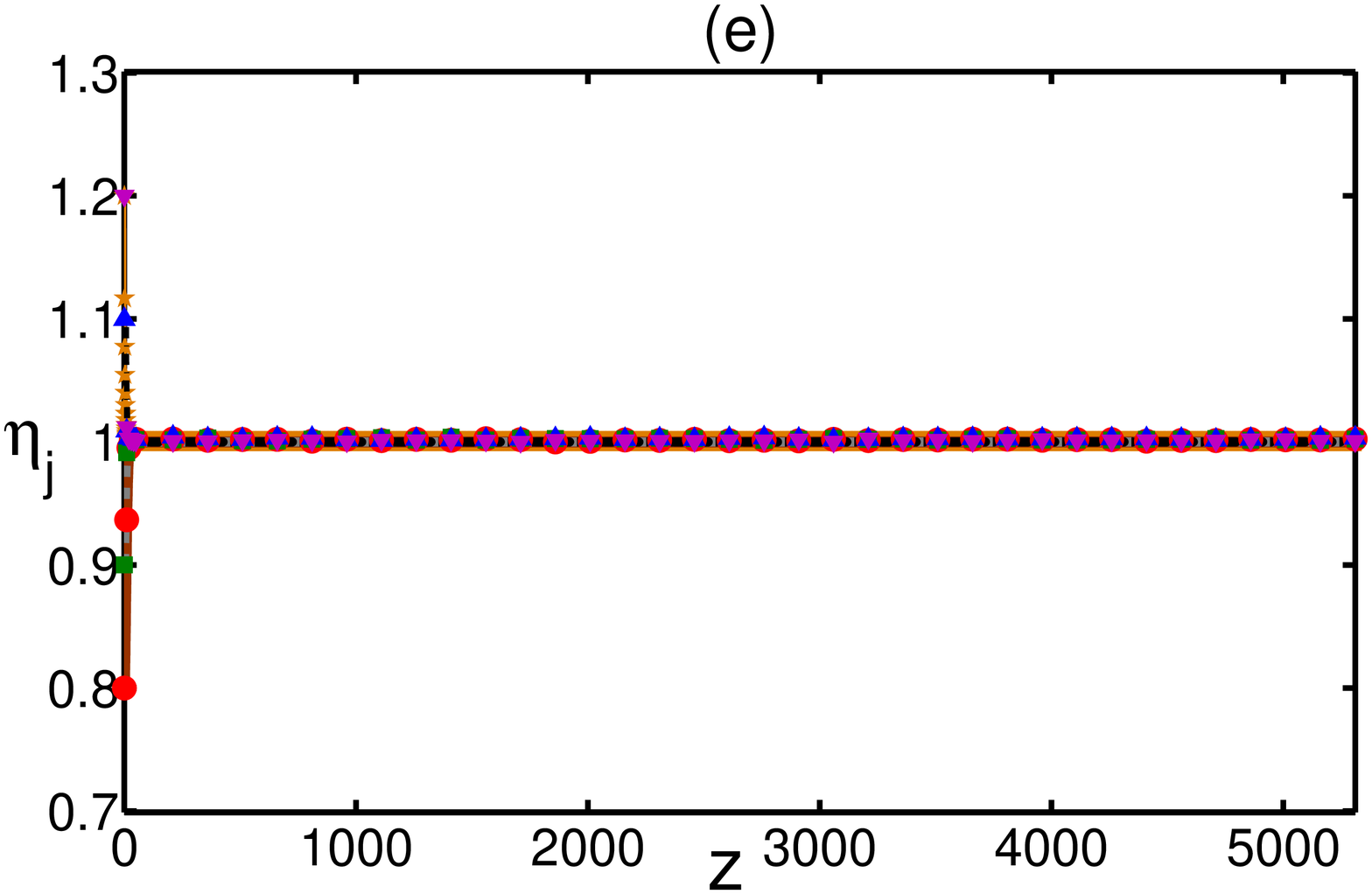} &
\epsfxsize=8.2cm  \epsffile{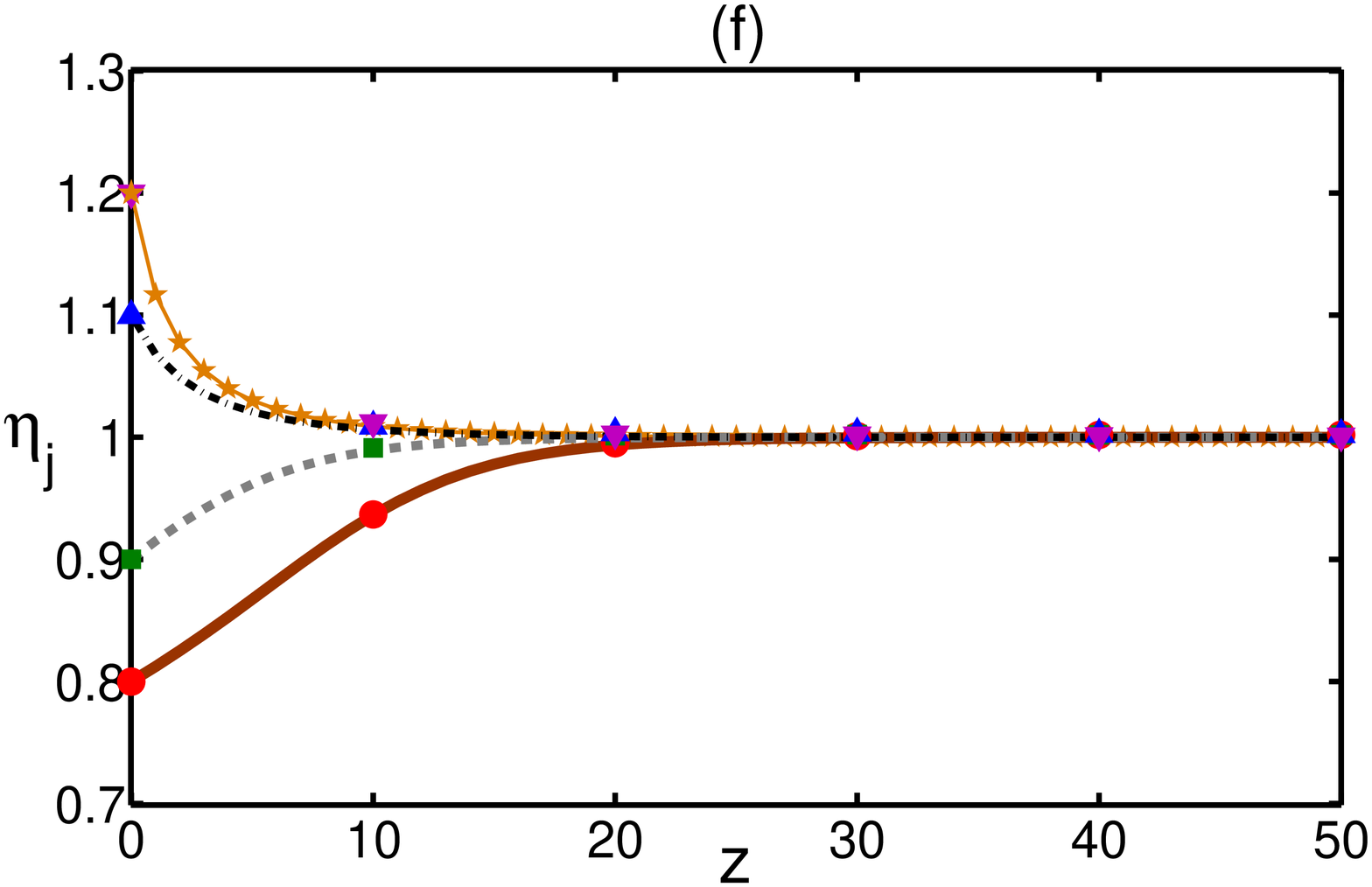}
\end{tabular}
\caption{The $z$ dependence of soliton amplitudes $\eta_{j}$ 
during transmission stabilization in waveguides with 
broadband delayed Raman response and narrowband GL gain-loss 
for two-sequence [(a) and (b)], three-sequence [(c) and (d)], 
and four-sequence [(e) and (f)] transmission. 
Graphs (b), (d), and (f) show magnified versions of 
the $\eta_{j}(z)$ curves in graphs (a), (c), and (e) at short distances.
The red circles, green squares, blue up-pointing triangles, 
and magenta down-pointing triangles 
represent $\eta_{1}(z)$, $\eta_{2}(z)$, $\eta_{3}(z)$, and $\eta_{4}(z)$, 
obtained by numerical simulations with Eqs. (\ref{global1}) and (\ref{global2}). 
The solid brown, dashed gray, dashed-dotted black, and solid-starred orange curves 
correspond to $\eta_{1}(z)$, $\eta_{2}(z)$, $\eta_{3}(z)$, and $\eta_{4}(z)$,  
obtained by the predator-prey model (\ref{global5}).} 
\label{fig1}
\end{figure*}

\begin{figure*}[ptb]
\begin{tabular}{cc}
\epsfxsize=8.2cm  \epsffile{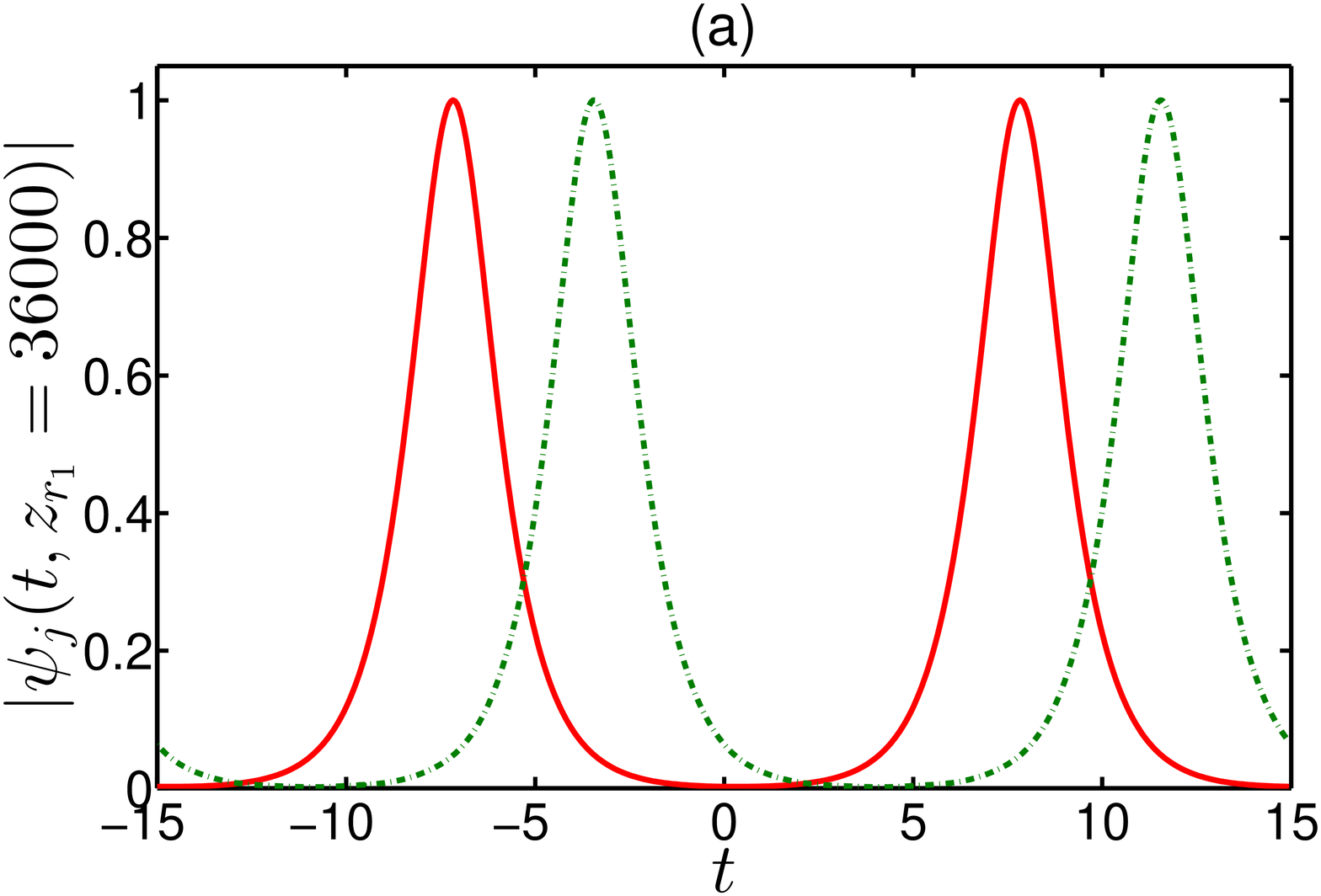} &
\epsfxsize=8.2cm  \epsffile{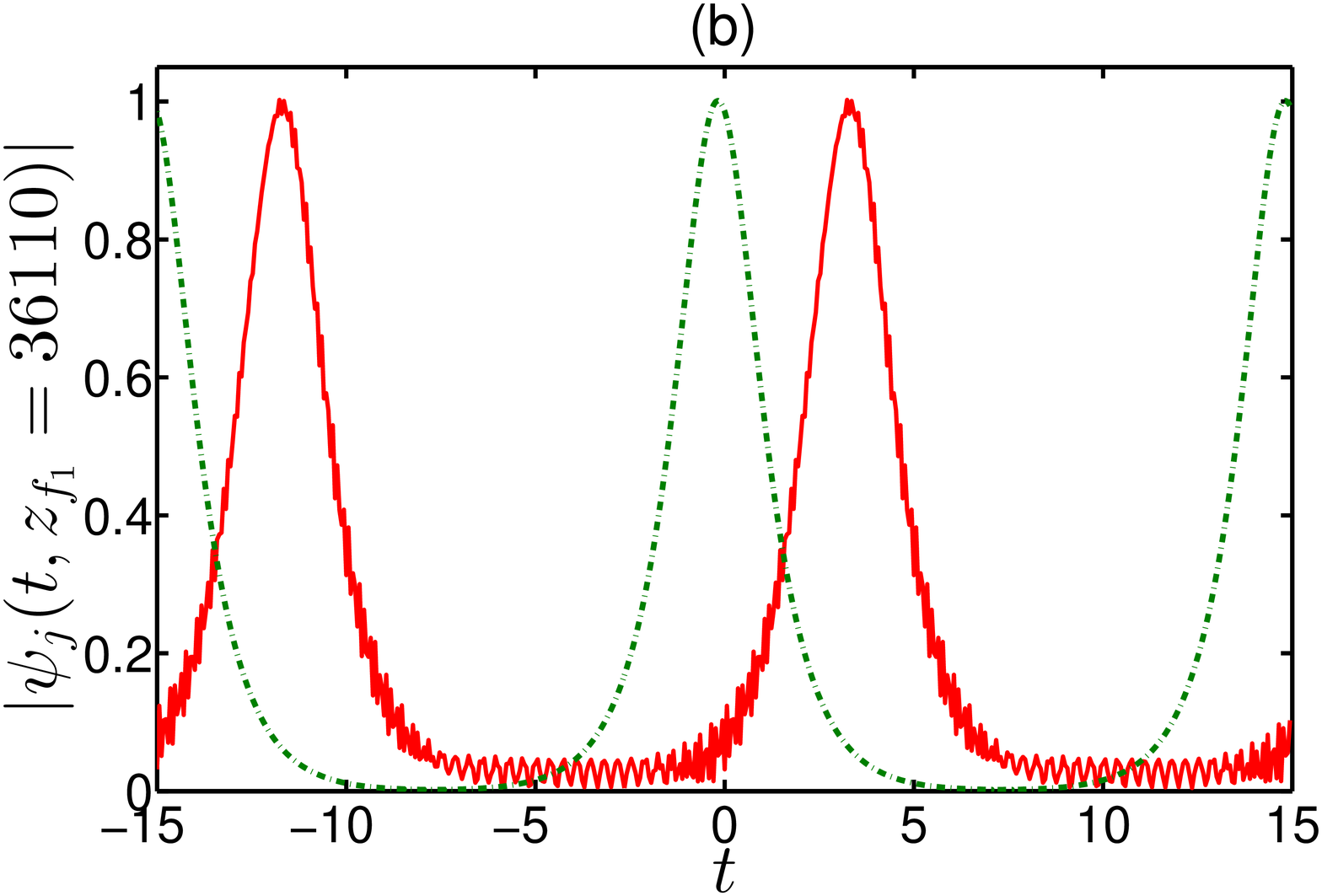} \\
\epsfxsize=8.2cm  \epsffile{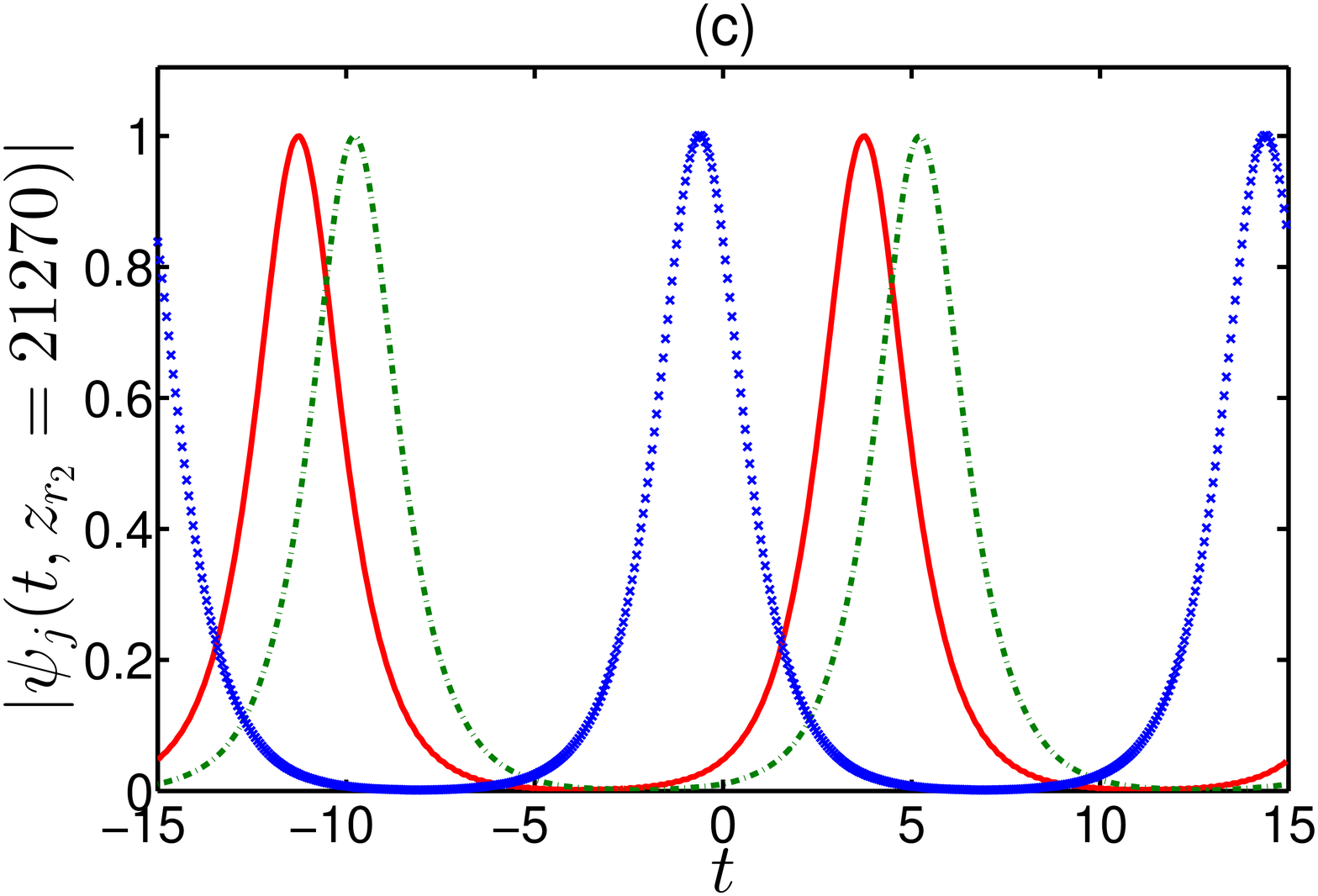} &
\epsfxsize=8.2cm  \epsffile{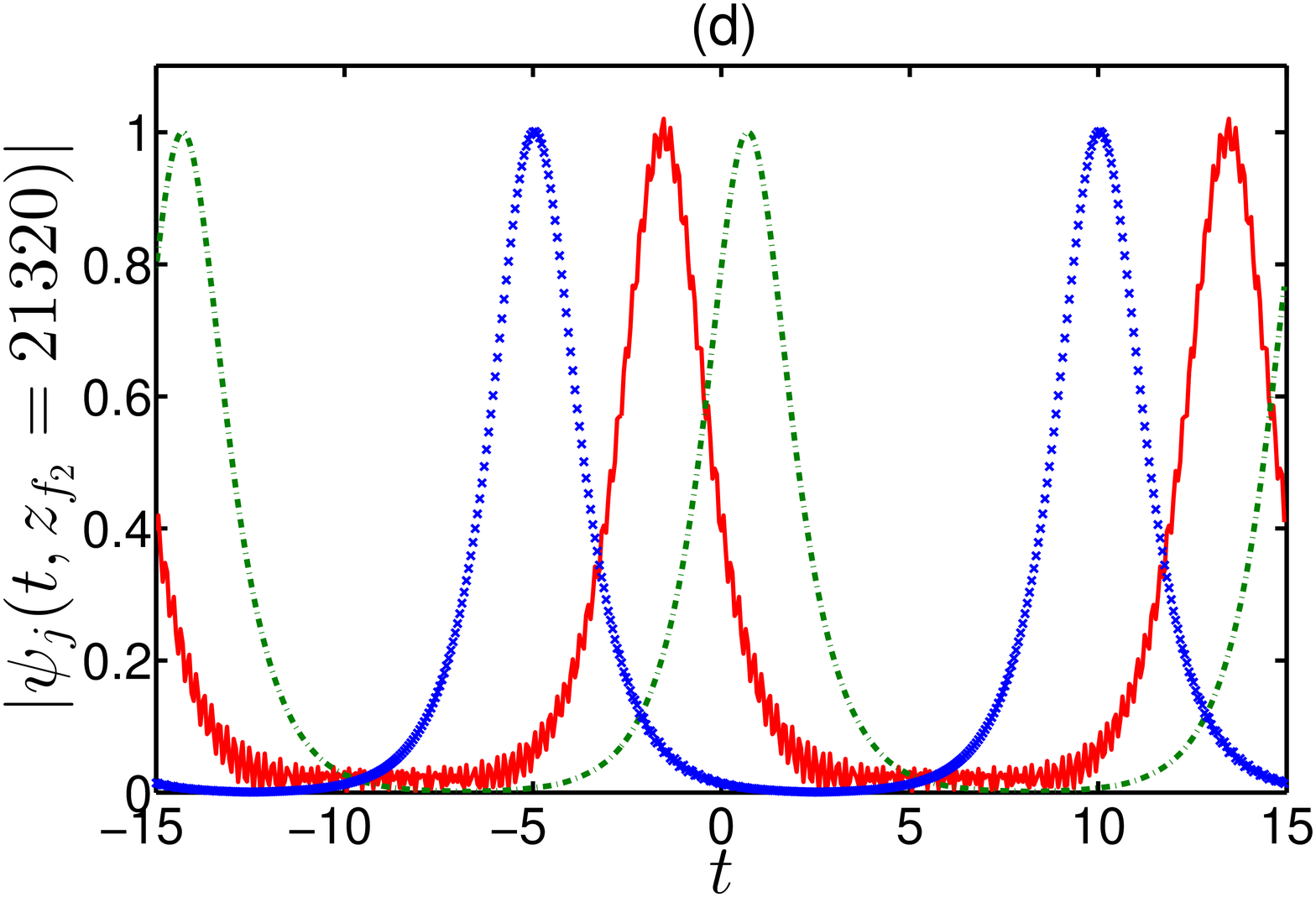} \\
\epsfxsize=8.2cm  \epsffile{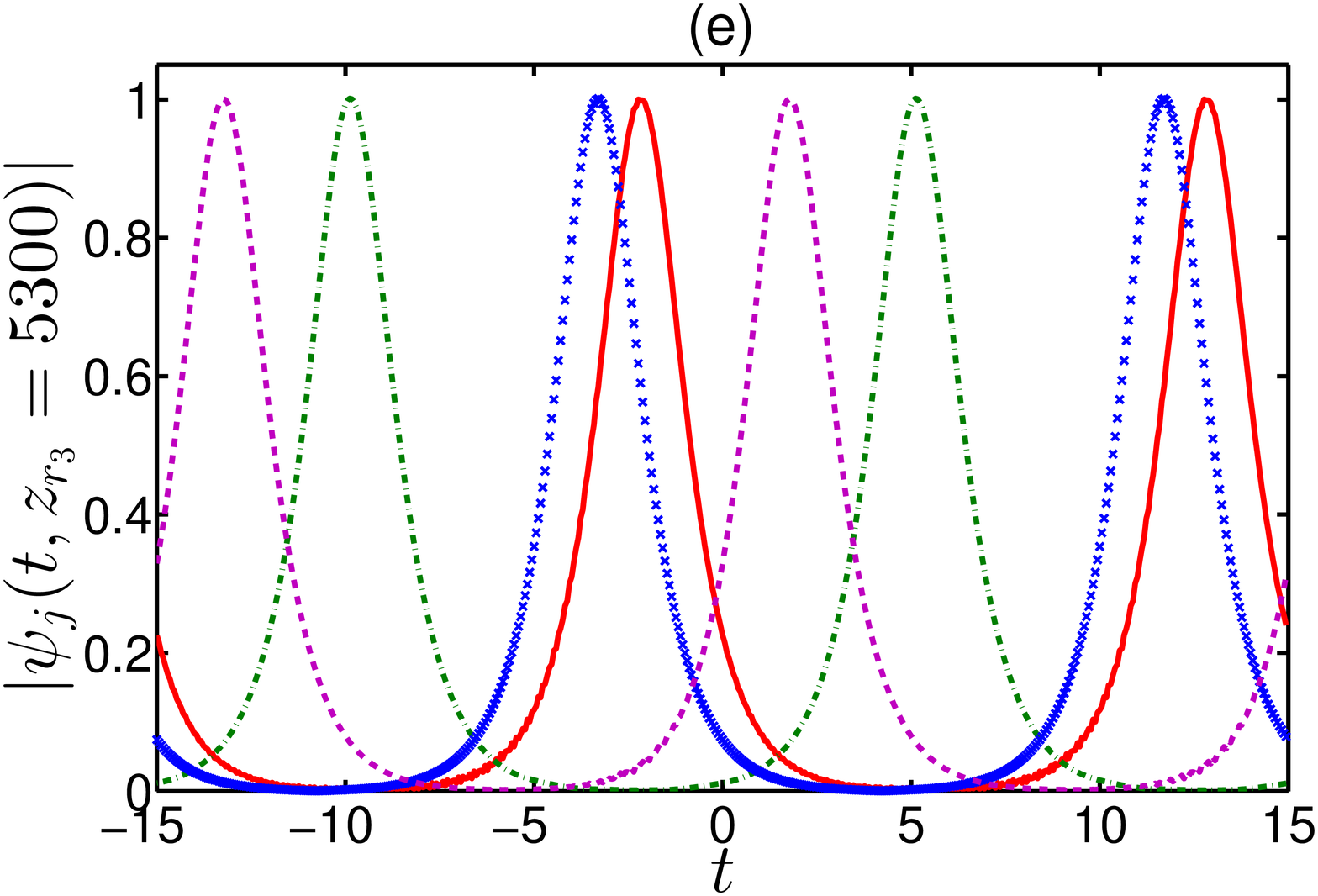} &
\epsfxsize=8.2cm  \epsffile{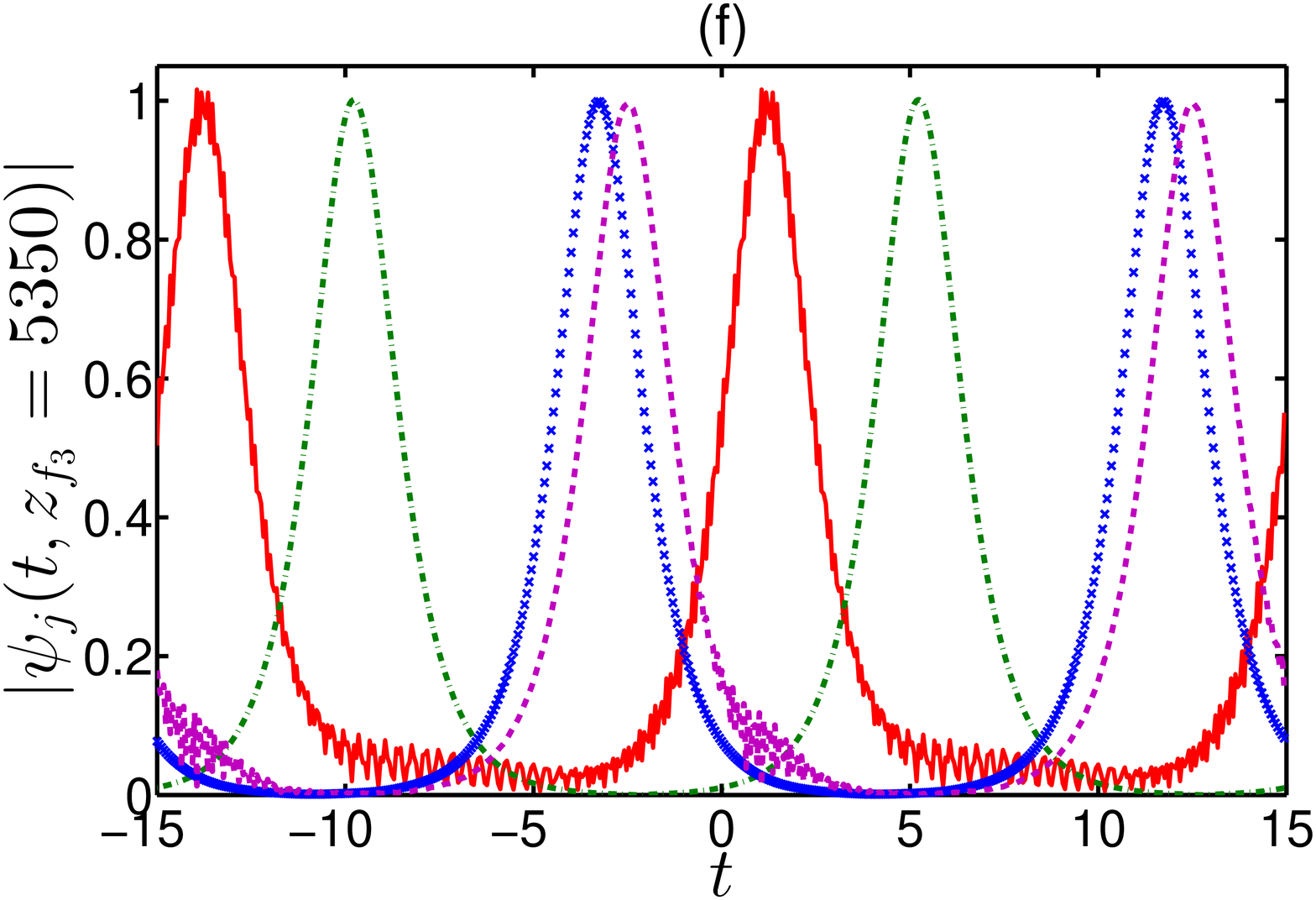}
\end{tabular}
\caption{The pulse patterns  $|\psi_{j}(t,z)|$ near the onset of transmission instability
for the two-sequence [(a) and (b)], three-sequence [(c) and (d)], 
and four-sequence [(e) and (f)] transmission setups considered in Fig. \ref{fig1}. 
Graphs (a), (c), and (e) show $|\psi_{j}(t,z)|$ before the onset of instability, 
while graphs (b), (d), and (f) show $|\psi_{j}(t,z)|$ at the onset of instability. 
The solid red curve, dashed-dotted green curve, blue crosses, and dashed magenta curve  
represent $|\psi_{j}(t,z)|$ with $j=1,2,3,4$, obtained by numerical solution of 
Eqs. (\ref{global1}) and (\ref{global2}). 
The propagation distances are $z=z_{r_1}=36000$ (a), $z=z_{f_1}=36110$ (b), 
$z=z_{r_2}=21270$ (c), $z=z_{f_2}=21320$ (d),        
$z=z_{r_3}=5300$ (e), and $z=z_{f_3}=5350$ (f).} 
\label{fig_add2}
\end{figure*}

We first describe numerical simulations for transmission stabilization in waveguides with 
broadband delayed Raman response and a narrowband GL gain-loss profile 
[$L(|\psi_{j}|^{2})=L_{1}(|\psi_{j}|^{2})$] for $N=2$, $N=3$, and $N=4$ sequences. 
We choose $\eta=1$ so that the desired steady state of the system is $(1, \dots, 1)$. 
The Raman coefficient is $\epsilon_{R}=0.0006$, while 
the quintic loss coefficient is $\epsilon_{5}=0.1$ for $N=2$, $\epsilon_{5}=0.15$ for and $N=3$, 
and $\epsilon_{5}=0.25$ for $N=4$. In addition, we choose $\kappa=1.2$  
and initial amplitudes satisfying $\eta_{j}(0)>(5\kappa/4-\eta^{2})^{1/2}$ 
for $1 \le j \le N$, so that the initial amplitudes belong to the basin of 
attraction of $(1, \dots, 1)$. The numerical simulations with Eqs. (\ref{global1}) and (\ref{global2}) 
are carried out up to the final distances $z_{f_1}=36110$, $z_{f_2}=21320$, 
and $z_{f_3}=5350$, for $N=2$, $N=3$, and $N=4$, respectively. 
At these distances, the onset of transmission destabilization 
due to radiation emission and pulse distortion is observed.  
The $z$ dependence of soliton amplitudes obtained by the simulations 
is shown in Figs. \ref{fig1}(a), \ref{fig1}(c), and \ref{fig1}(e)  
together with the prediction of the predator-prey model (\ref{global5}). 
Figures \ref{fig1}(b), \ref{fig1}(d), and \ref{fig1}(f)    
show the amplitude dynamics at short distances. 
Figures \ref{fig_add2}(a), \ref{fig_add2}(c), and \ref{fig_add2}(e) show the pulse patterns $|\psi_{j}(t,z)|$ 
at a distance $z=z_{r}$ before the onset of transmission instability, where $z_{r_1}=36000$ for N=2, 
$z_{r_2}=21270$ for N=3, and $z_{r_3}=5300$ for N=4.   
Figures \ref{fig_add2}(b), \ref{fig_add2}(d), and \ref{fig_add2}(f)
show the pulse patterns $|\psi_{j}(t,z)|$ at $z=z_{f}$, i.e., at the onset of transmission instability. 
As seen in Fig. \ref{fig1}, the soliton amplitudes tend to the equilibrium 
value $\eta=1$ with increasing distance for $N=2$, 3, and 4, i.e., 
the transmission is stable up to the distance $z=z_{r}$ in all three cases. 
The approach to the equilibrium state takes place along distances that are much shorter 
compared with the distances along which stable transmission is observed. 
Furthermore, the agreement between the predictions of the predator-prey model 
and the coupled-NLS simulations is excellent for $0 \le z \le z_{r}$. 
Additionally, as seen in Figs. \ref{fig_add2}(a), \ref{fig_add2}(c), and \ref{fig_add2}(e), the solitons 
retain their shape at $z=z_{r}$ despite the large number of intersequence collisions. 
The distances $z_{r}$, along which stable propagation is observed, 
are significantly larger compared with those 
observed in other multisequence nonlinear waveguide systems. 
For example, the value $z_{r_1}=36000$ for $N=2$ is larger by 
a factor of $200$ compared with the value obtained in waveguides 
with linear gain and broadband cubic loss \cite{PNC2010}. Moreover, the stable propagation distances  
observed in the current work for $N=2$, $N=3$, and $N=4$ are larger 
by factors of 37.9, 34.3, and 10.6 compared with the distances obtained 
in single-waveguide transmission in the presence of delayed Raman response and in the absence 
of nonlinear gain-loss \cite{PNT2015}. The latter increase in the stable transmission 
distances is quite remarkable, considering the fact that in Ref. \cite{PNT2015}, 
intrasequence frequency-dependent linear gain-loss was employed to further stabilize 
the transmission, whereas in the current work, the gain-loss experienced 
by each sequence is uniform. We also point out that the results of our numerical simulations 
provide the first example for stable long-distance propagation 
of $N$ soliton sequences with $N>2$ in systems described by coupled GL models.

%======

% ==== 
We note that at the onset of transmission instability, the pulse patterns become distorted, 
where the distortion appears as fast oscillations of $|\psi_{j}(t,z)|$ that are most 
pronounced at the solitons' tails [see Figs. \ref{fig_add2}(b), \ref{fig_add2}(d), and \ref{fig_add2}(f)]. 
The degree of pulse distortion is different for different pulse sequences. 
Indeed, for $N=2$, the $j=1$ sequence is significantly distorted at $z=z_{f_{1}}$, 
while no significant distortion is observed for the $j=2$ sequence. 
For $N=3$, the $j=1$ sequence is significantly distorted, 
the $j=3$ sequence is slightly distorted, while the $j=2$ sequence 
is still not distorted at $z=z_{f_{2}}$. 
For $N=4$, the $j=1$ and $j=4$ sequences are both significantly distorted 
at $z=z_{f_{3}}$, while no significant distortion is observed for the 
$j=2$ and $j=3$ sequences at this distance. 

%===
 
\begin{figure*}[ptb]
\begin{tabular}{cc}
\epsfxsize=8.2cm  \epsffile{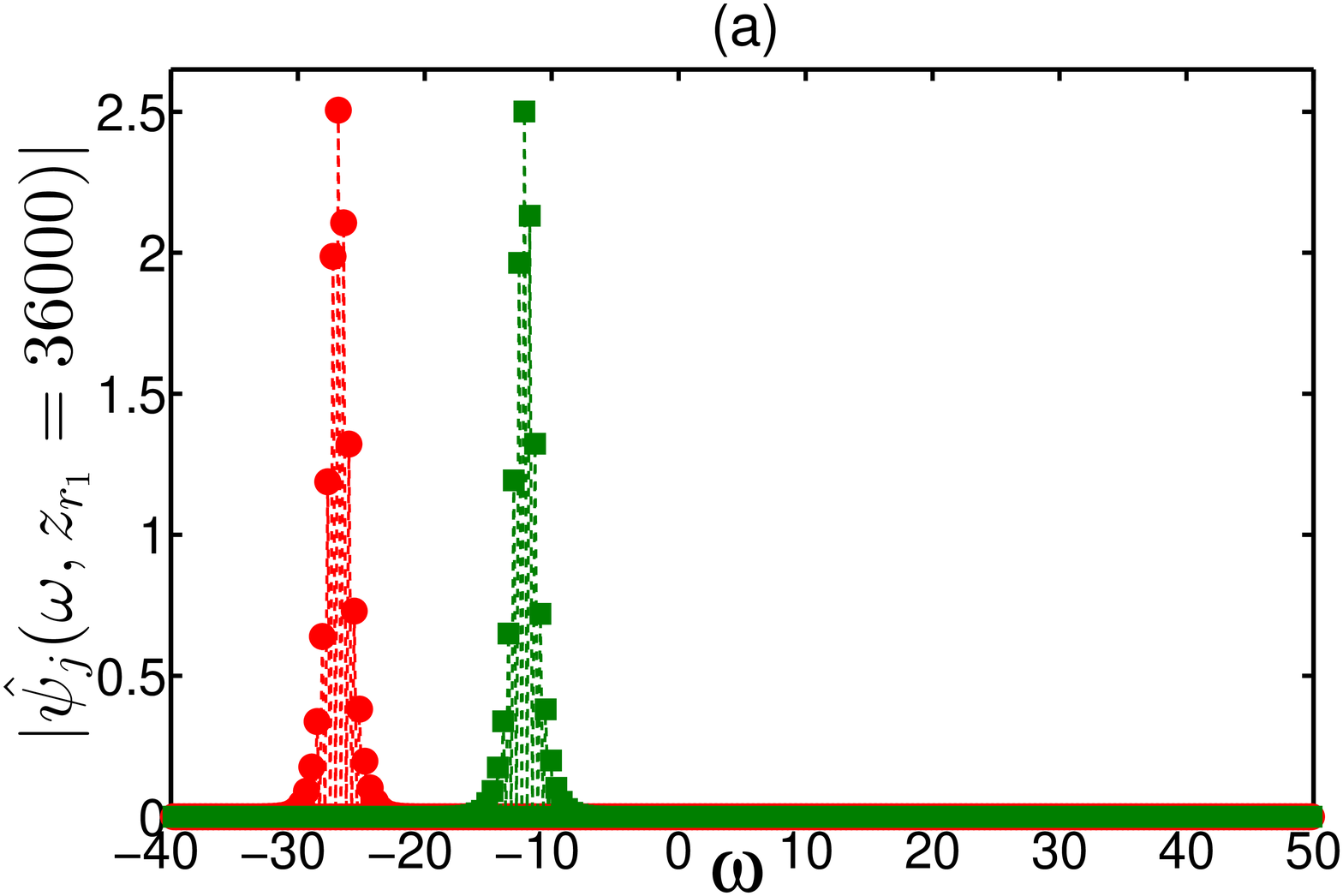} &
\epsfxsize=8.2cm  \epsffile{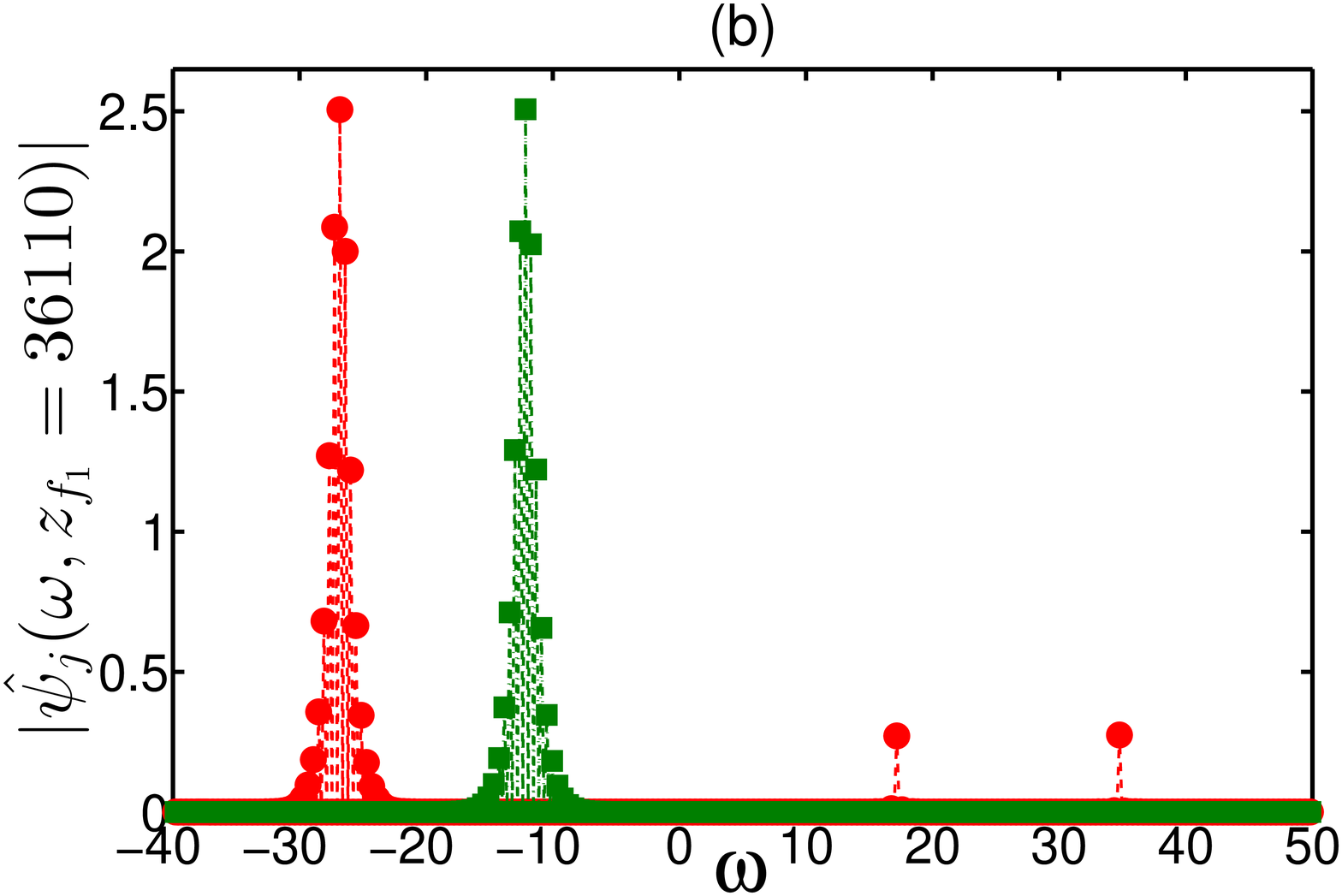} \\
\epsfxsize=8.2cm  \epsffile{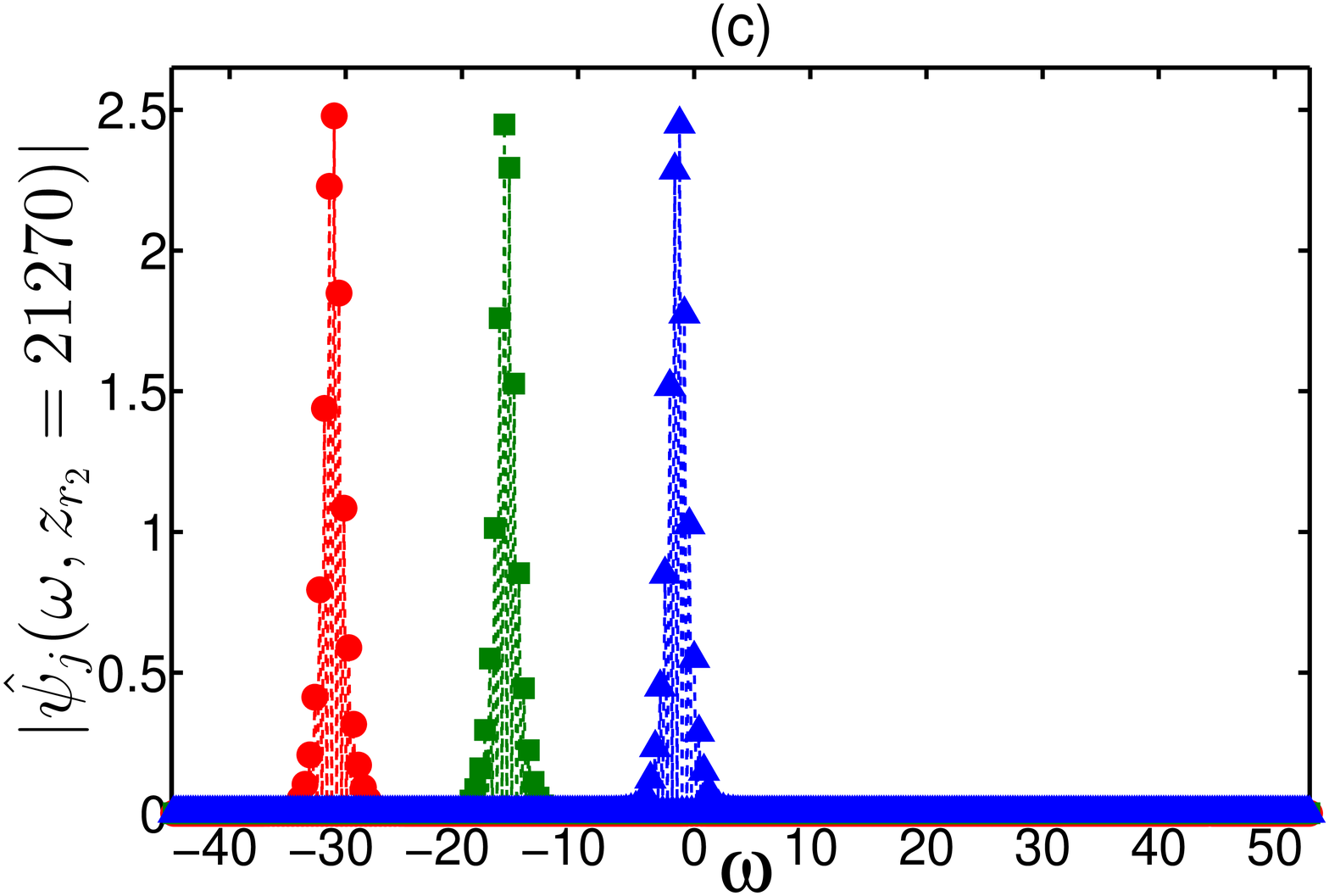} &
\epsfxsize=8.2cm  \epsffile{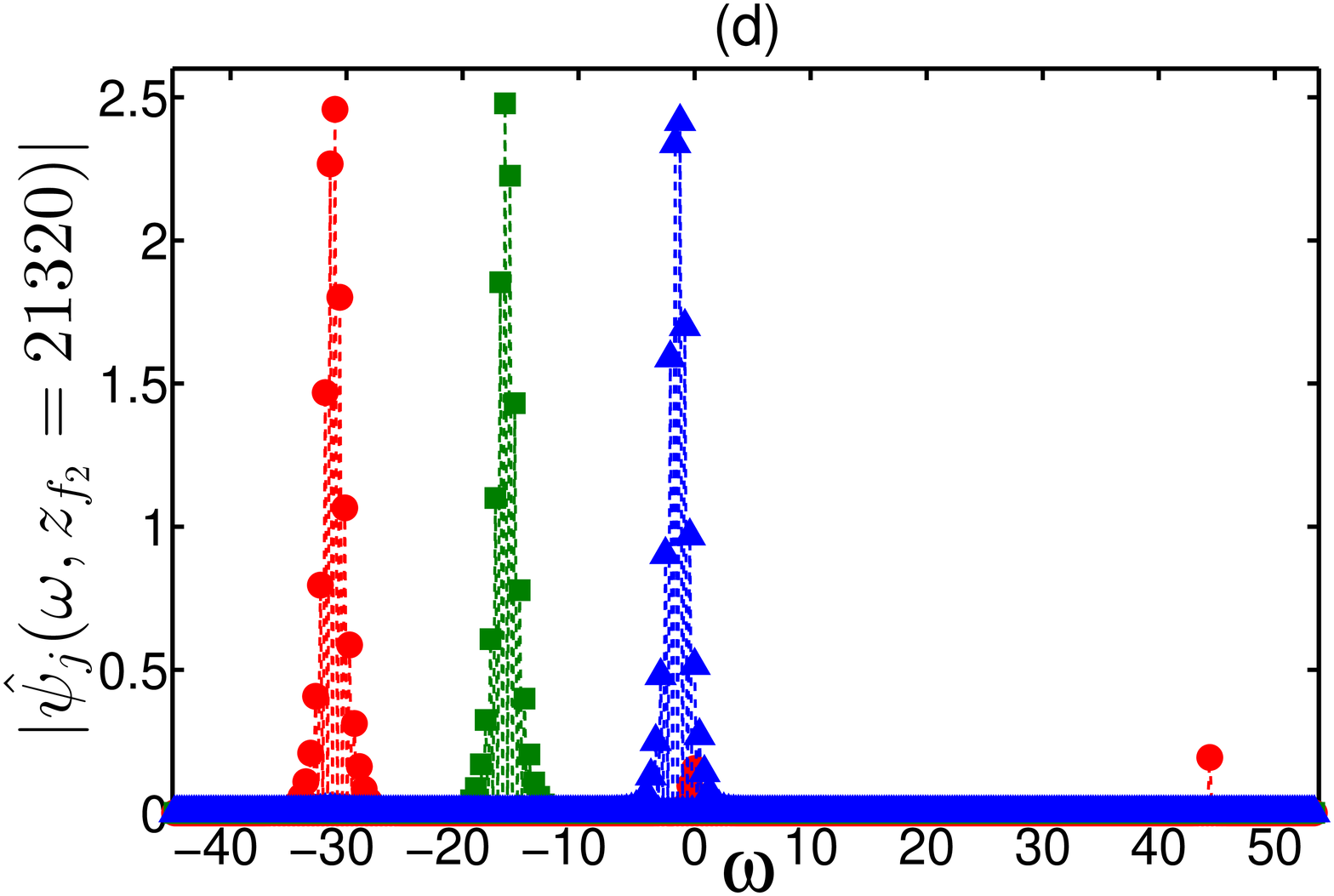} \\
\epsfxsize=8.2cm  \epsffile{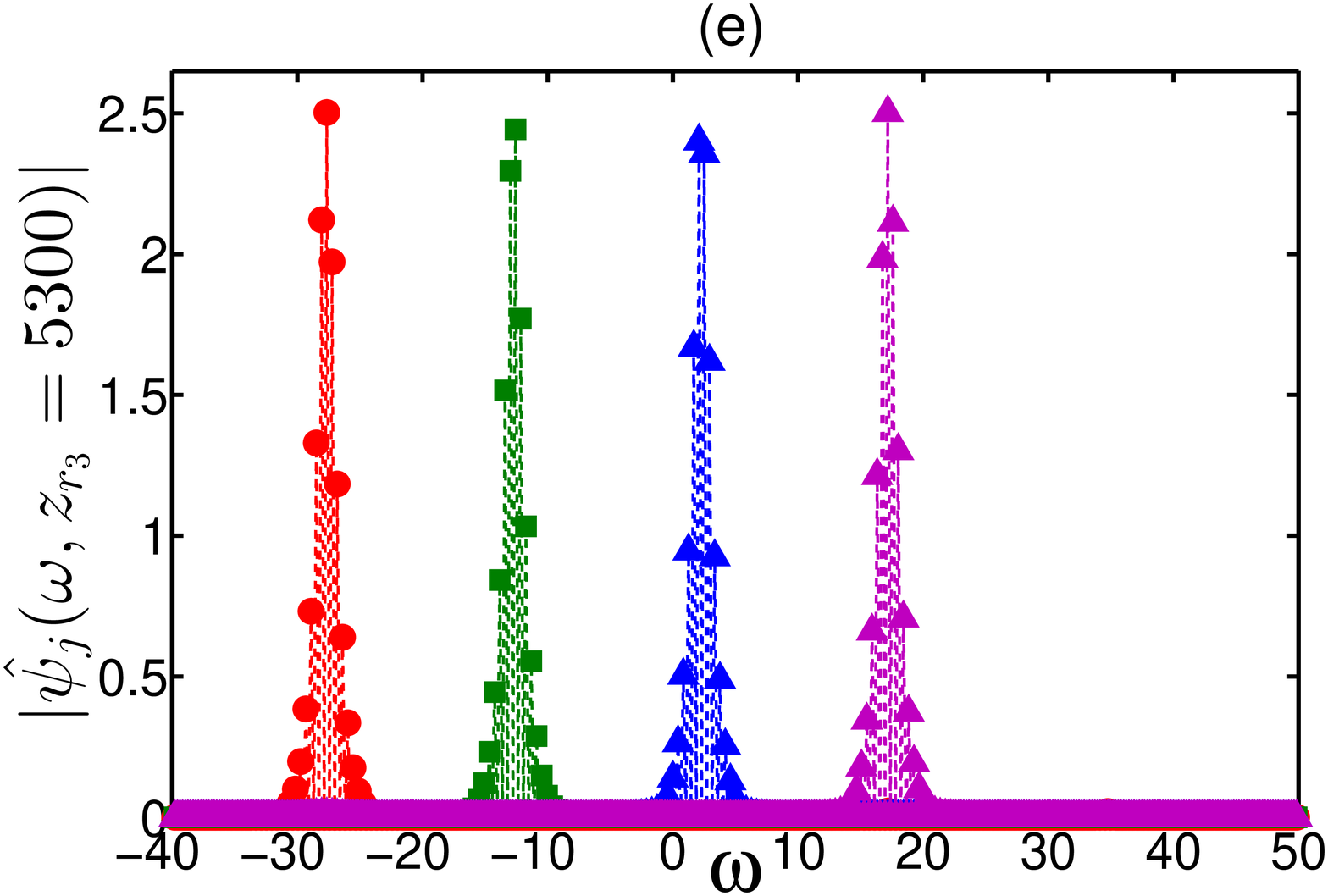} &
\epsfxsize=8.2cm  \epsffile{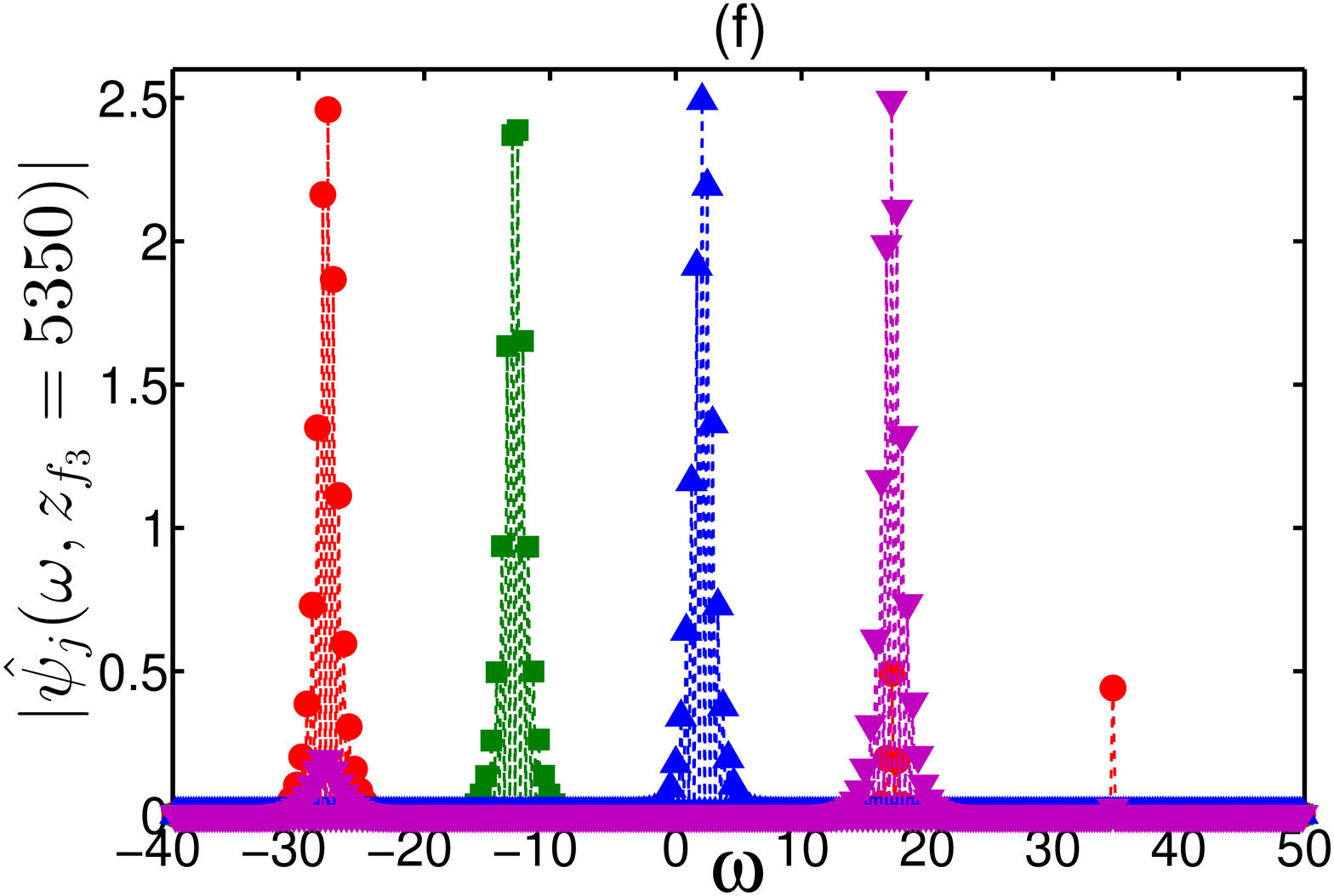}
\end{tabular}
\caption{The Fourier transforms of the pulse patterns  $|\hat\psi_{j}(\omega,z)|$ 
near the onset of transmission instability for the two-sequence [(a) and (b)], 
three-sequence [(c) and (d)], and four-sequence [(e) and (f)] transmission 
setups considered in Figs. \ref{fig1} and \ref{fig_add2}. 
Graphs (a), (c), and (e) show $|\hat\psi_{j}(\omega,z)|$ before the onset of instability, 
while graphs (b), (d), and (f) show $|\hat\psi_{j}(\omega,z)|$ at the onset of instability. 
The red circles, green squares, blue up-pointing triangles, 
and magenta down-pointing triangles represent $|\hat\psi_{j}(\omega,z)|$
with $j=1,2,3,4$, obtained by numerical solution of 
Eqs. (\ref{global1}) and (\ref{global2}). 
The propagation distances are $z=z_{r_1}=36000$ (a), $z=z_{f_1}=36110$ (b), 
$z=z_{r_2}=21270$ (c), $z=z_{f_2}=21320$ (d),        
$z=z_{r_3}=5300$ (e), and $z=z_{f_3}=5350$ (f).} 
\label{fig_add3}
\end{figure*}

\begin{figure*}[ptb]
\begin{tabular}{cc}
\epsfxsize=8.2cm  \epsffile{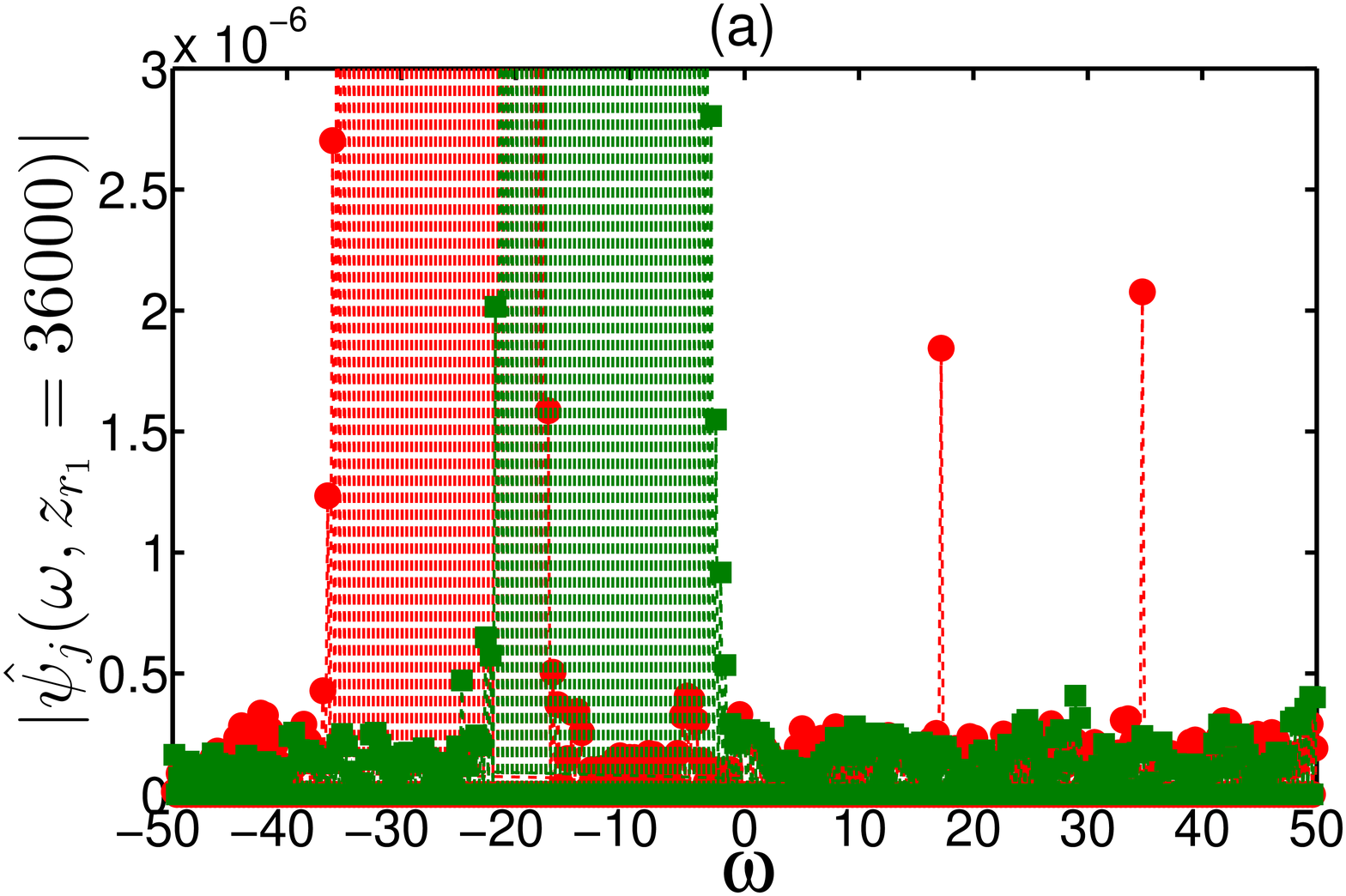} &
\epsfxsize=8.2cm  \epsffile{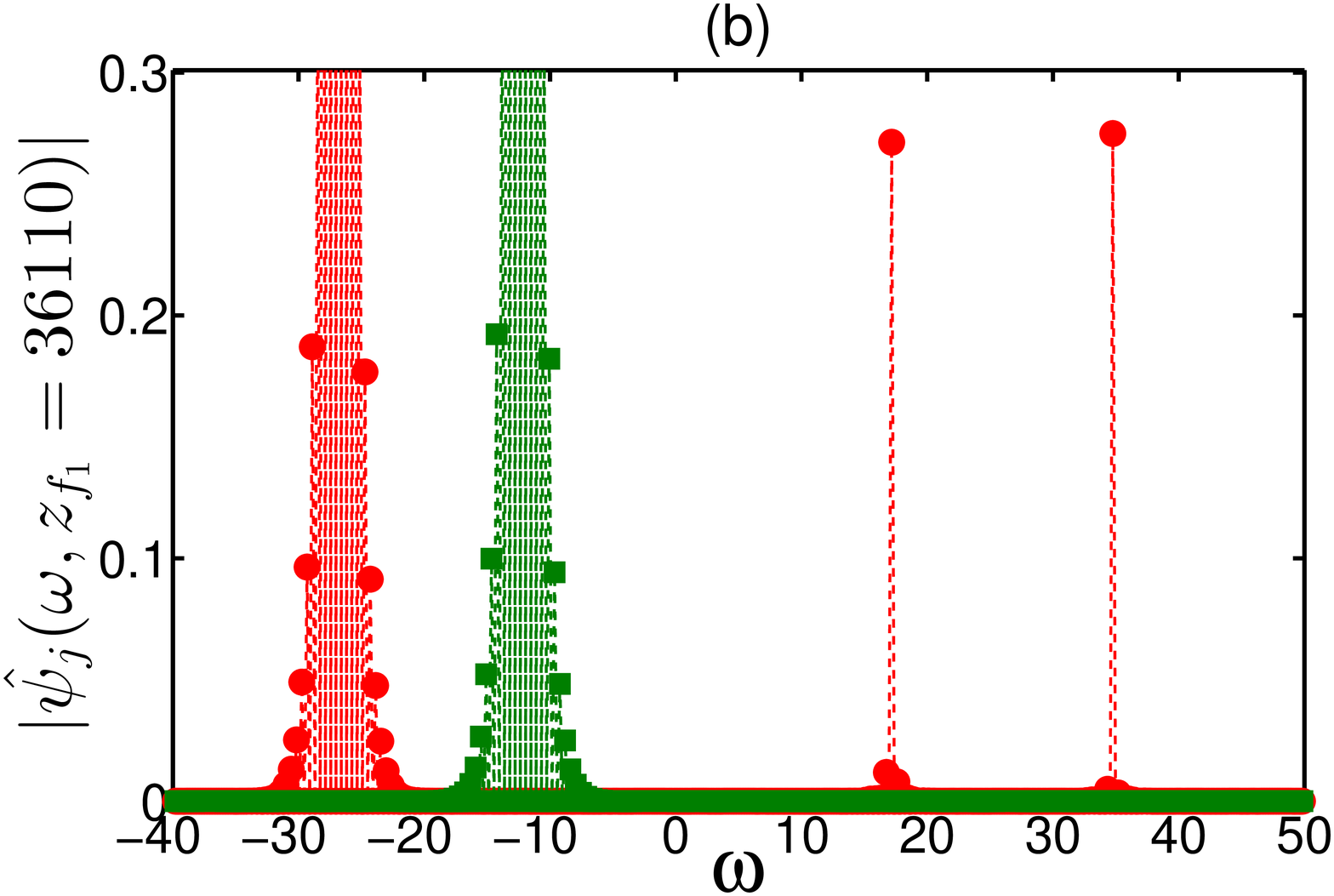} \\
\epsfxsize=8.2cm  \epsffile{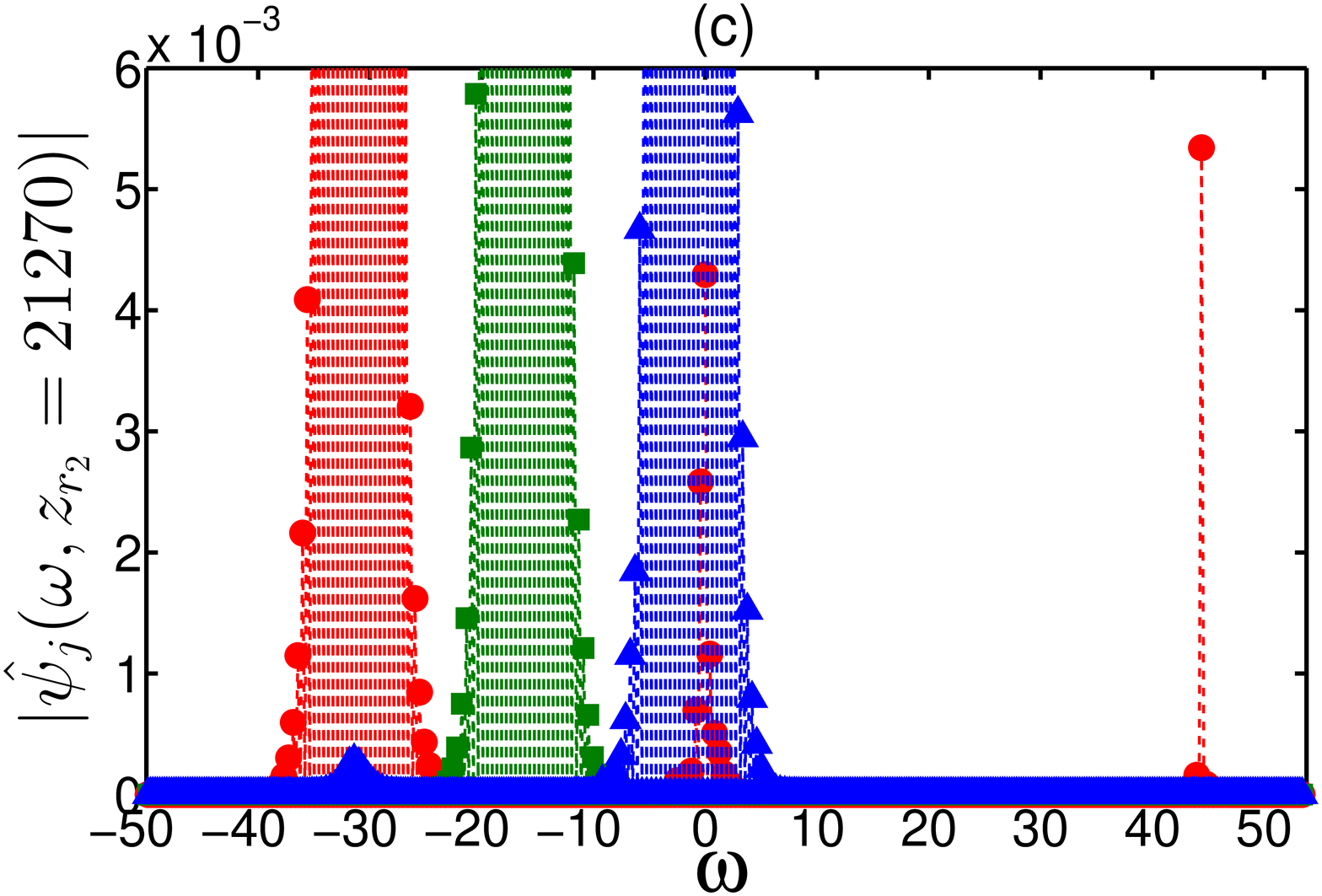} &
\epsfxsize=8.2cm  \epsffile{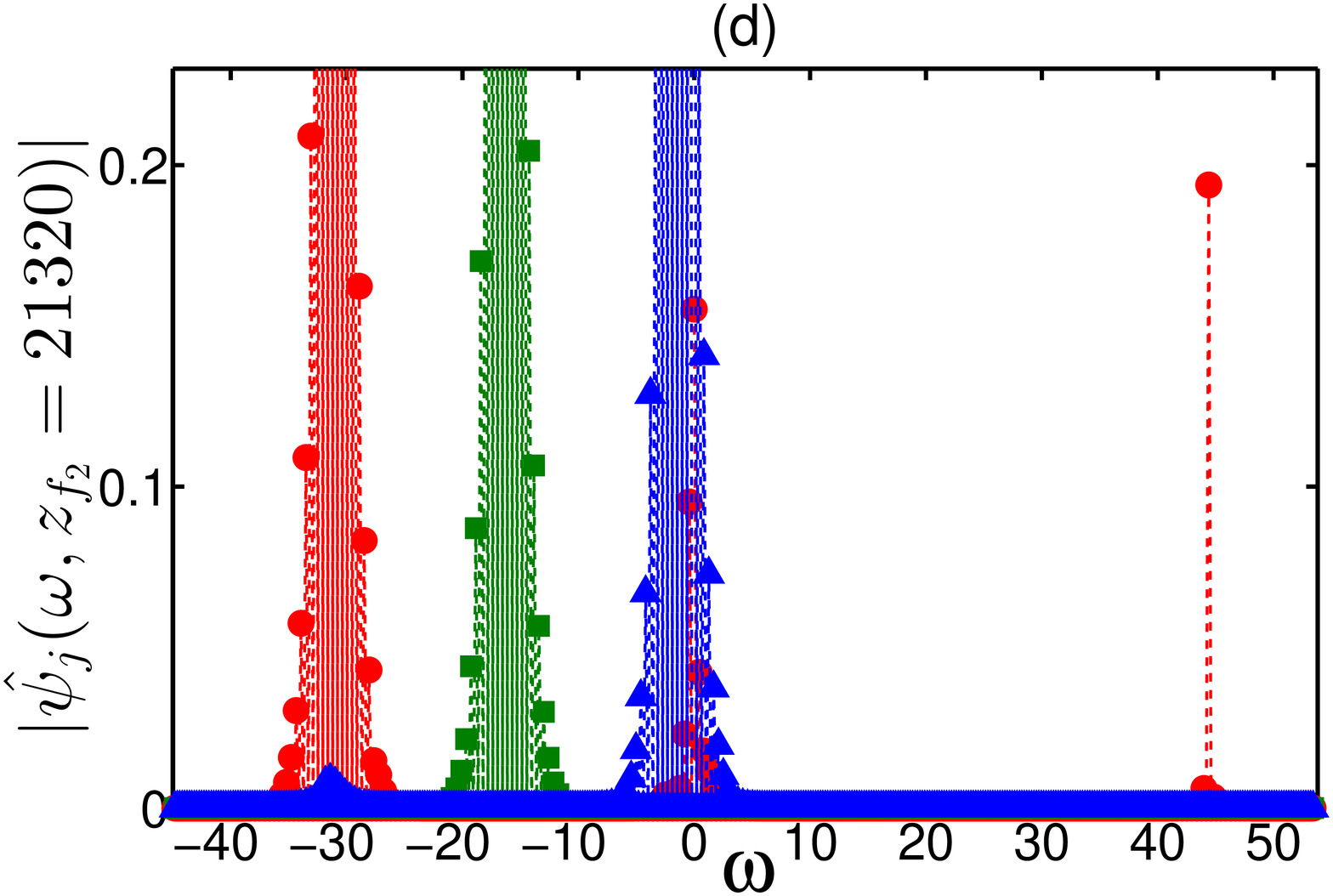} \\
\epsfxsize=8.2cm  \epsffile{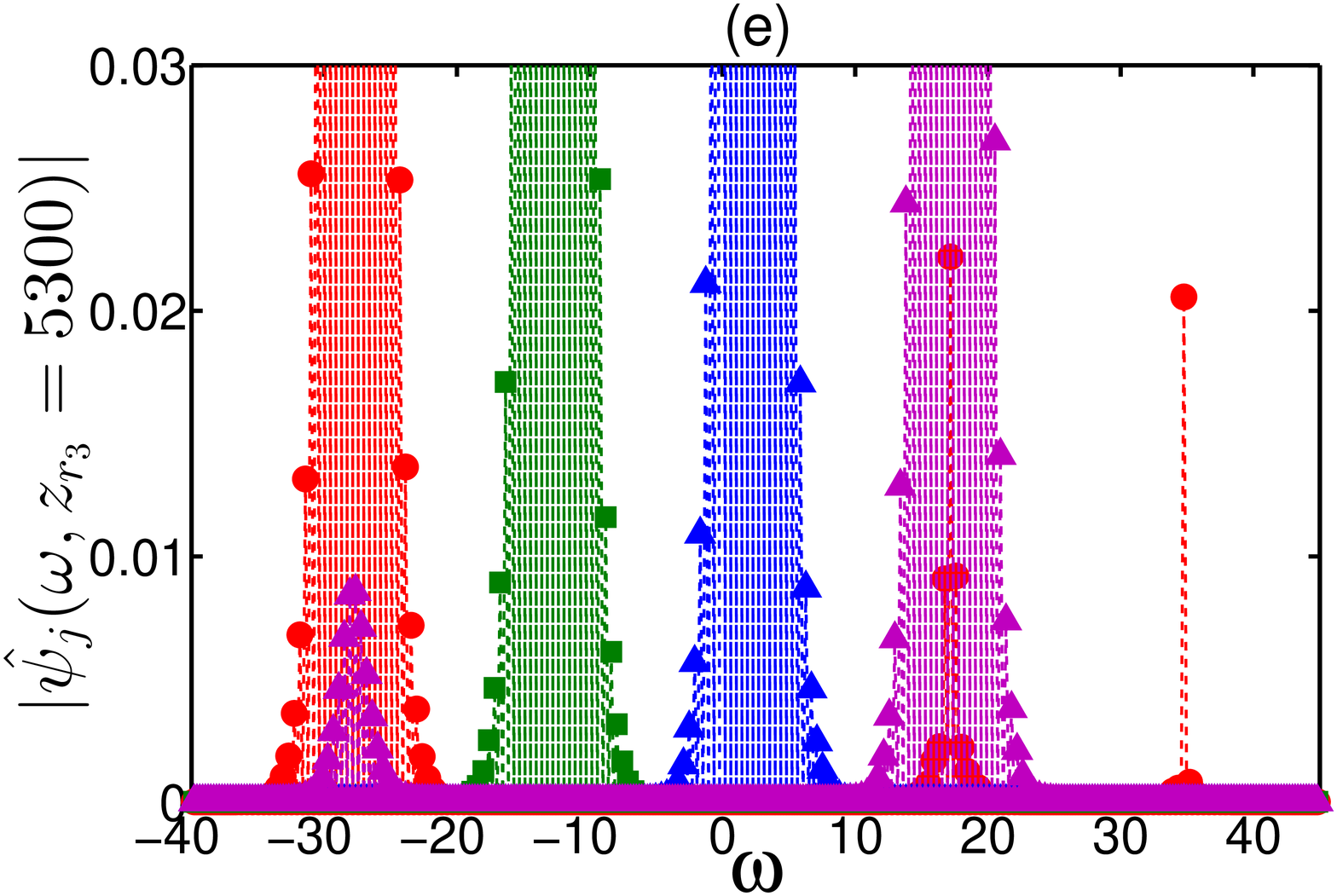} &
\epsfxsize=8.2cm  \epsffile{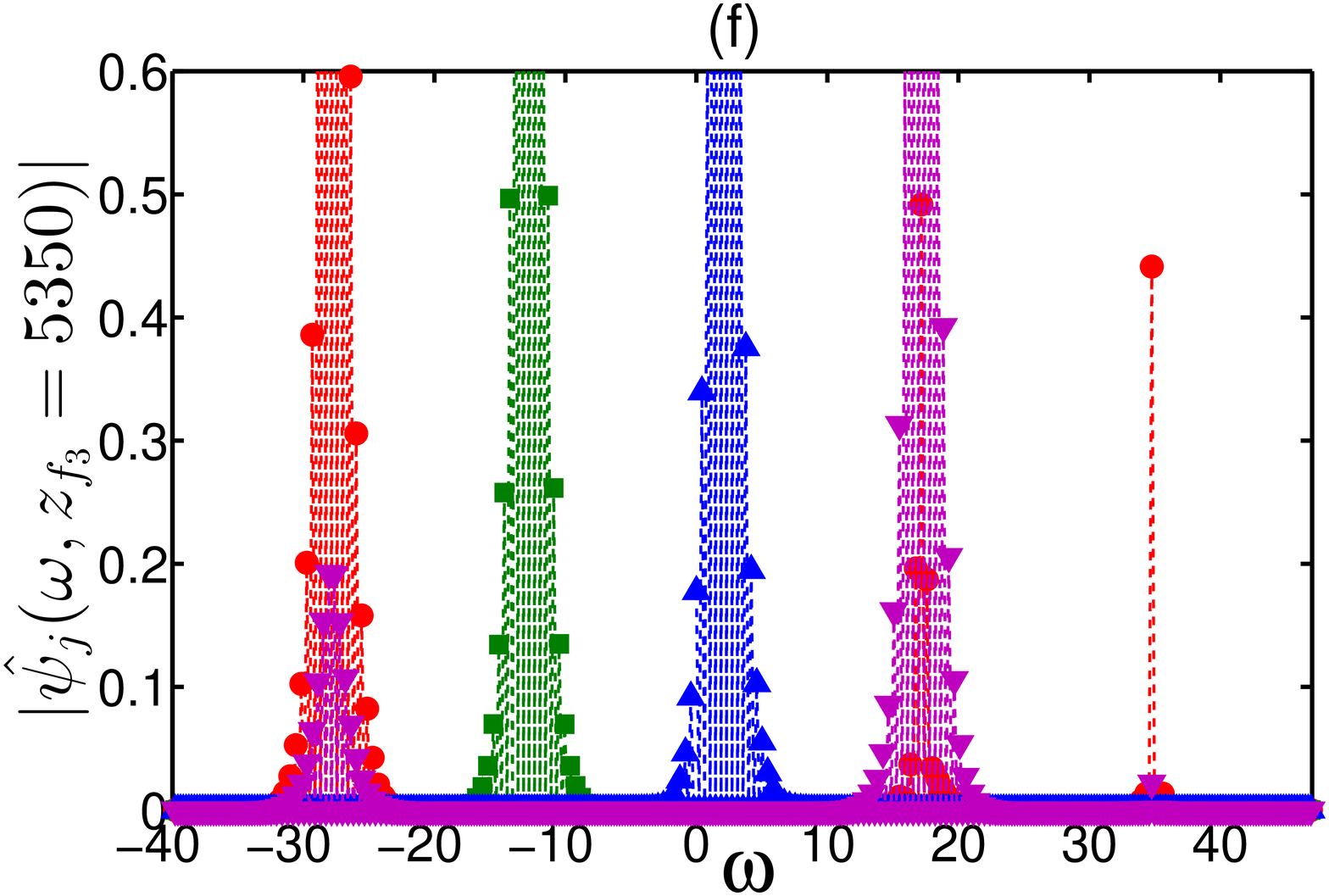}
\end{tabular}
\caption{Magnified versions of the graphs in Fig. \ref{fig_add3} for 
small $|\hat{\psi_{j}}(\omega,z)|$ values. 
The symbols and distances are the same as in Fig. \ref{fig_add3}.} 
\label{fig_add4}
\end{figure*}

The distortion of the pulse patterns and the associated transmission 
destabilization can be explained by examination of the Fourier transforms 
of the pulse patterns $|\hat\psi_{j}(\omega,z)|$. 
Figure \ref{fig_add3} shows the Fourier transforms $|\hat\psi_{j}(\omega,z)|$ 
at $z=z_{r}$ (before the onset of transmission instability) 
and at $z=z_{f}$ (at the onset of transmission instability).               
Figure \ref{fig_add4} shows magnified versions of the graphs 
in Fig. \ref{fig_add3} for small $|\hat{\psi_{j}}(\omega,z)|$ values. 
It is seen that the Fourier transforms of some of the pulse sequences 
develop pronounced radiative sidebands at $z=z_{f}$. Furthermore, 
the frequencies at which the radiative sidebands attain their maxima are 
related to the central frequencies $\beta_{j}(z)$ of the soliton sequences 
or to the frequency spacing $\Delta\beta$. The latter observation  
indicates that the processes leading to radiative sideband generation 
are resonant in nature (see also Refs. \cite{PNT2015,CPN2016}).

Consider first the Fourier transforms of the pulse patterns for $N=2$. 
As seen in Figs. \ref{fig_add3}(b) and \ref{fig_add4}(b), 
in this case the $j=1$ sequence develops radiative 
sidebands at frequencies $\omega_{s}^{(11)}=17.18$ and $\omega_{s}^{(12)}=34.76$ 
at $z=z_{f_{1}}$. In contrast, no significant sidebands 
are observed for the $j=2$ sequence at this distance. 
These findings explain the significant pulse pattern distortion of  
the $j=1$ sequence and the absence of pulse pattern distortion for the $j=2$ sequence 
at $z=z_{f_{1}}$. In addition, the radiative sideband frequencies satisfy the 
simple relations: $\omega_{s}^{(11)}-\beta_{2}(z_{r_{1}})\sim 29.3 \sim 2\Delta\beta$ 
and $\omega_{s}^{(12)} \sim 2\omega_{s}^{(11)}$.   
For $N=3$, the $j=1$ sequence develops significant sidebands at frequencies 
$\omega_{s}^{(11)}=0.0$ and $\omega_{s}^{(12)}=44.4$, 
the $j=3$ sequence develops a weak sideband at frequency  
$\omega_{s}^{(31)}=-31.42$, and the $j=2$ sequence 
does not have any significant sidebands at $z=z_{f_{2}}$
[see Figs. \ref{fig_add3}(d) and \ref{fig_add4}(d)].  
These results coincide with the significant pulse pattern distortion of the $j=1$ 
sequence, the weak pulse pattern distortion of the $j=3$ sequence, 
and the absence of pulse pattern distortion for the $j=2$ sequence at $z=z_{f_{2}}$. 
Additionally, the sideband frequencies satisfy the simple relations: 
$\omega_{s}^{(11)}\sim \beta_{3}(z_{r_{2}})$,   
$\omega_{s}^{(12)}\sim 3\Delta\beta$, 
and $\omega_{s}^{(31)}\sim \beta_{1}(z_{r_{2}})$.    
For $N=4$, the $j=1$ and $j=4$ sequences develop significant sidebands, 
while no significant sidebands are observed for the $j=2$ and $j=3$ 
sequences at $z=z_{f_{3}}$ [see Figs. \ref{fig_add3}(f) and \ref{fig_add4}(f)]. 
These findings explain the significant pulse pattern 
distortion of the $j=1$ and $j=4$ sequences and the 
absence of significant pulse pattern distortion for the $j=2$ and $j=3$ sequences 
at $z=z_{f_{3}}$. The sideband frequencies of the 
$j=1$ sequence satisfy the relations: 
$\omega_{s}^{(11)}=17.17 \sim \beta_{3}(z_{r_{3}})$, 
and $\omega_{s}^{(12)}=34.77 \sim 2\omega_{s}^{(11)}$.     
Note that the values of $\omega_{s}^{(11)}$ and $\omega_{s}^{(12)}$ 
for $N=4$ are very close to the values found for $N=2$. 
Finally, the sideband frequencies of the $j=4$ sequence satisfy the relations:    
$\omega_{s}^{(41)}=-27.65 \sim \beta_{1}(z_{r_{3}})$,  
and $\omega_{s}^{(42)}=34.35 \sim 2\omega_{s}^{(11)}$.

% =====      

% ====
                                                                                                        
% =====

We now turn to describe numerical simulations for a single transmission switching event 
in waveguides with broadband delayed Raman response and a narrowband GL gain-loss profile. 
As described in Section \ref{stability}, on-off switching of $M$ out of $N$ pulse sequences at 
a distance $z=z_{s}$  is realized by changing the value of one or more of the physical parameters, 
such that the steady state $(\eta, \dots, \eta)$ turns from asymptotically stable to unstable, 
while another steady state at $(0, \dots, 0, \eta_{s(M+1)}, \dots, \eta_{sN})$ 
is asymptotically stable. We denote the on-off switching setups by A1-A2, 
where A1 and A2 denote the sets of physical parameters used at $0 \le z<z_{s}$ 
and $z\ge z_{s}$, respectively.

Off-on switching of $M$ out of $N$ soliton sequences at $z=z_{s}$ is realized by 
changing the physical parameter values such that $(\eta, \dots, \eta)$ turns from 
unstable to asymptotically stable. As explained in Section \ref{stability}, to achieve stable long-distance 
transmission after the switching, one needs to require that the origin is an asymptotically stable  
steady state as well. Under this requirement, $\kappa$ must satisfy inequality (\ref{global8}), 
and as a result, the basin of attraction of $(\eta, \dots, \eta)$ is limited 
to $((5\kappa/4-\eta^{2})^{1/2},\infty)$ for $1 \le j \le N$.  
This leads to limitations on the turning on of the $M$ sequences, 
especially for $M \ge 2$ and $N \ge 3$. To overcome this difficulty, we 
consider a hybrid waveguide consisting of a span with a GL gain-loss profile, 
a span with linear gain-loss and cubic loss, and a second span with a GL gain-loss profile.    
The introduction of the intermediate waveguide span with linear gain-loss and cubic loss 
enables the turning on of the $M$ sequences from low amplitude values due to the 
global stability of the steady state $(\eta, \dots, \eta)$ for the 
corresponding predator-prey model. However, due to the presence of linear gain and the 
instability of the origin for the same predator-prey model, 
propagation in the waveguide span with linear gain-loss and cubic loss is unstable against emission 
of small amplitude waves. For this reason, we introduce 
the frequency dependent linear gain-loss $g(\omega,z)$ of Eq. (\ref{global3a})
when simulating propagation in the second span.
More importantly, propagation in the second span with a GL gain-loss profile 
leads to mitigation of radiative instability due to the presence of linear loss 
in all channels for this waveguide span. This enables stable long-distance 
propagation of the $N$ soliton sequences after the switching.    
We denote the off-on switching setups by A2-B-A1, 
where A2, B, and A1 denote the sets of physical parameters 
used in the first, second, and third spans of the hybrid waveguide. 
The first span is located at $[0,z_{s1})$, the second at $[z_{s1},z_{s2})$, 
and the third at $[z_{s2},z_{f}]$, where $z_{f}$ is the final propagation distance. 
Thus, off-on switching of the $M$ soliton sequences occurs at $z \ge z_{s1}$, 
while final transmission stabilization takes place at $z \ge z_{s2}$.

We present here the results of numerical simulations for on-off and off-on switching 
of two and three soliton sequences in four-sequence transmission. As discussed in 
the preceding paragraphs, on-off switching setups are denoted by A1-A2 and off-on 
switching setups are denoted by A2-B-A1. The following values of the physical parameters 
are used. The Raman coefficient is $\epsilon_{R}=0.0006$, which is the same 
value used in transmission stabilization. The other parameter values used in setup A1 
in both on-off and off-on switching are $\epsilon_{5}=0.1$, $\kappa=1.2$, and $\eta=1$.      
The parameter values used in setup A2 in on-off switching are 
$\epsilon_{5}=0.04$, $\kappa=1.8$, and $\eta=1.05$ for $M=2$ 
and $\epsilon_{5}=0.04$, $\kappa=2$, and $\eta=1.1$ for $M=3$. 
The on-off switching distance is $z_{s}=250$ for both $M=2$ and $M=3$. 
The parameter values used in setup A2 in off-on switching are 
$\epsilon_{5}=0.032$, $\kappa=2.2$, and $\eta=1.1$ for $M=2$ 
and $\epsilon_{5}=0.02$, $\kappa=2.8$, and $\eta=1.3$ for $M=3$. 
The parameter values used in off-on switching in setup B 
are $\epsilon_{3}^{(2)}=0.02$ and $\eta=1$ for both $M=2$ and $M=3$. 
To suppress radiative instability during propagation in waveguide spans with linear 
gain-loss and cubic loss (setup B), the frequency dependent linear gain-loss $g(\omega,z)$ 
of Eq. (\ref{global3a}) with $W=10$ and $g_{L}=-0.5$ is employed.    
The switching and final stabilization distances in off-on transmission switching are 
$z_{s1}=30$ and $z_{s2}=80$ for $M=2$, and $z_{s1}=30$ and $z_{s2}=90$ for $M=3$. 
We point out that similar results were obtained with other choices of the physical parameter values, 
satisfying the stability conditions discussed in Section \ref{stability}.

\begin{figure*}[ptb]
\begin{tabular}{cc}
\epsfxsize=8.2cm  \epsffile{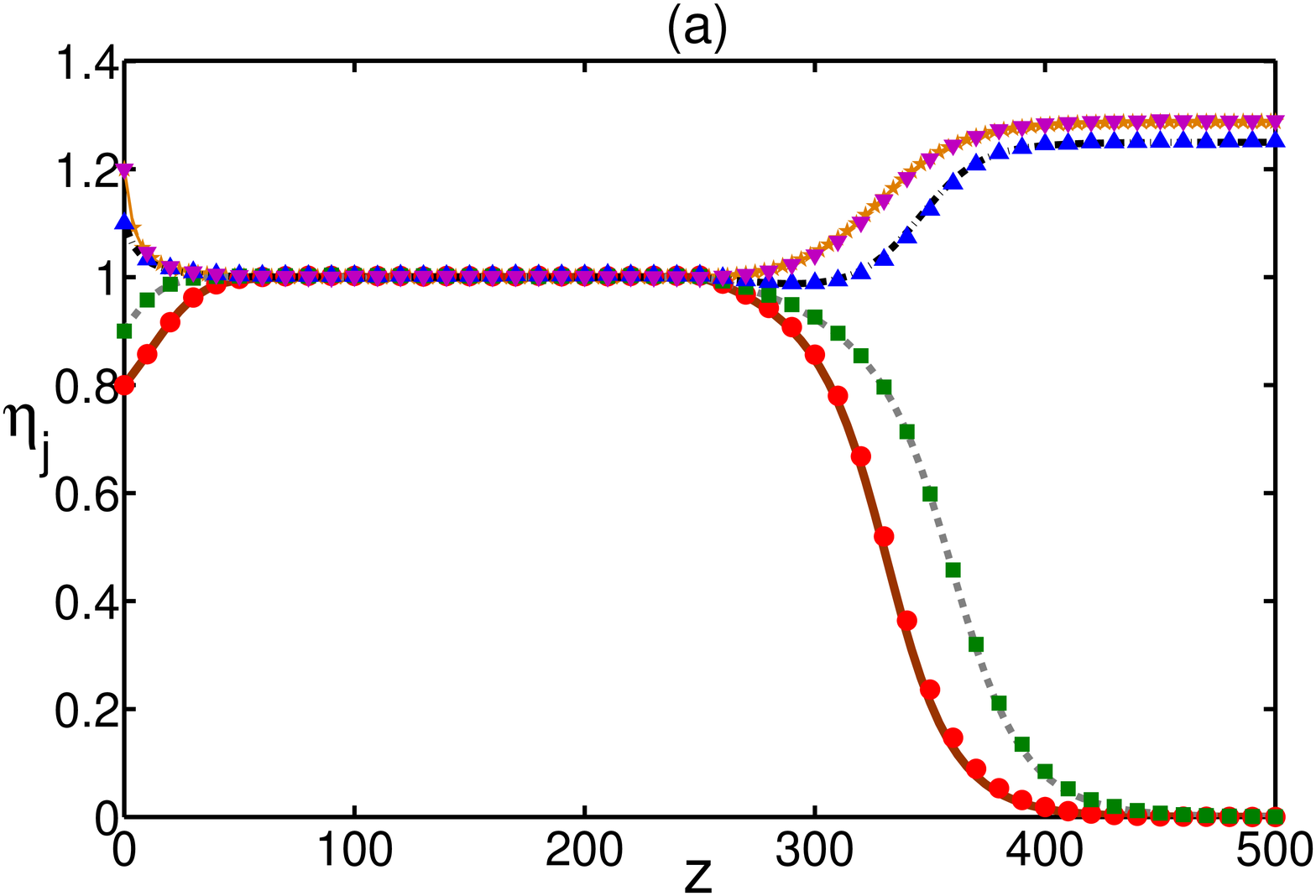} &
\epsfxsize=8.2cm  \epsffile{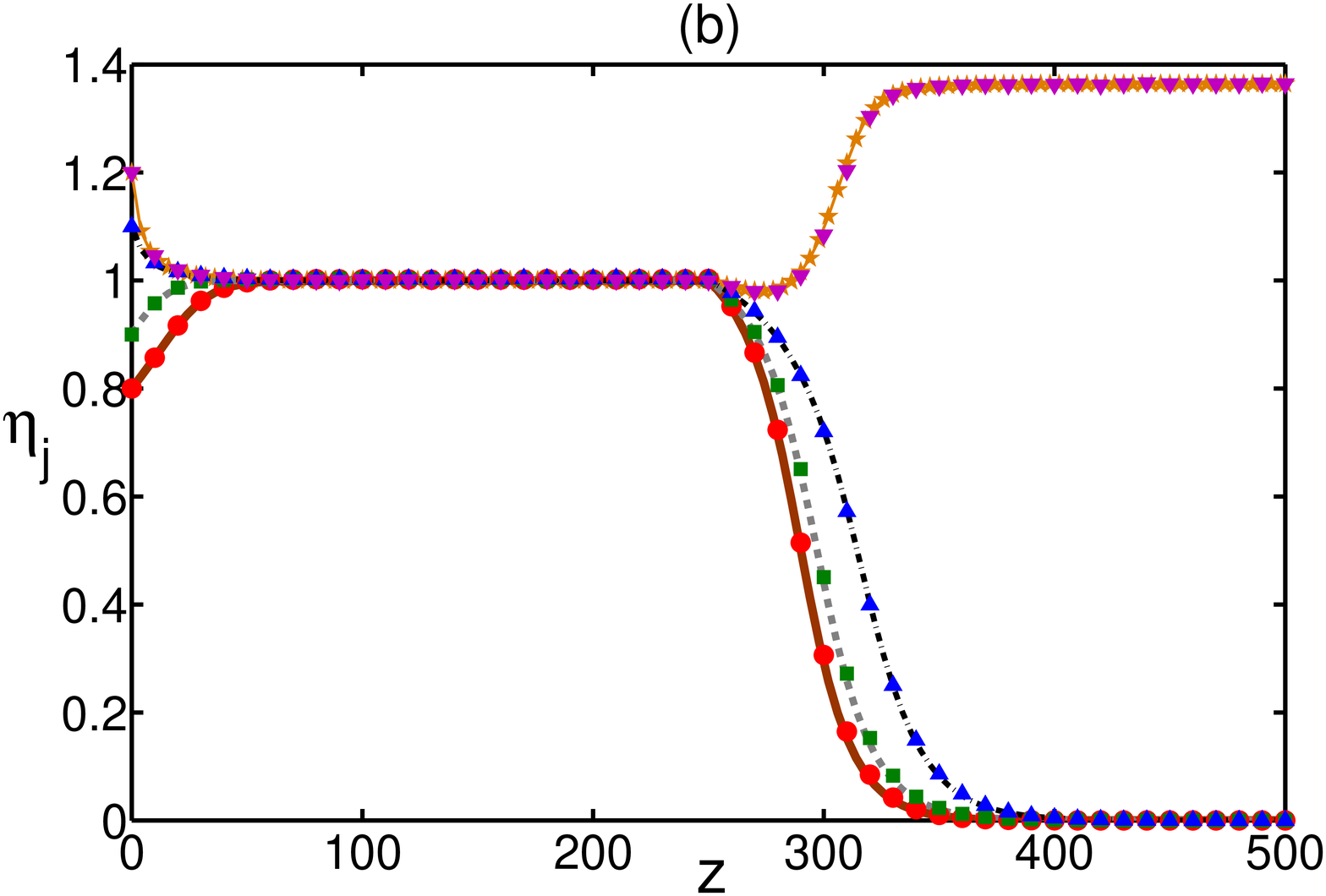} \\
\epsfxsize=8.2cm  \epsffile{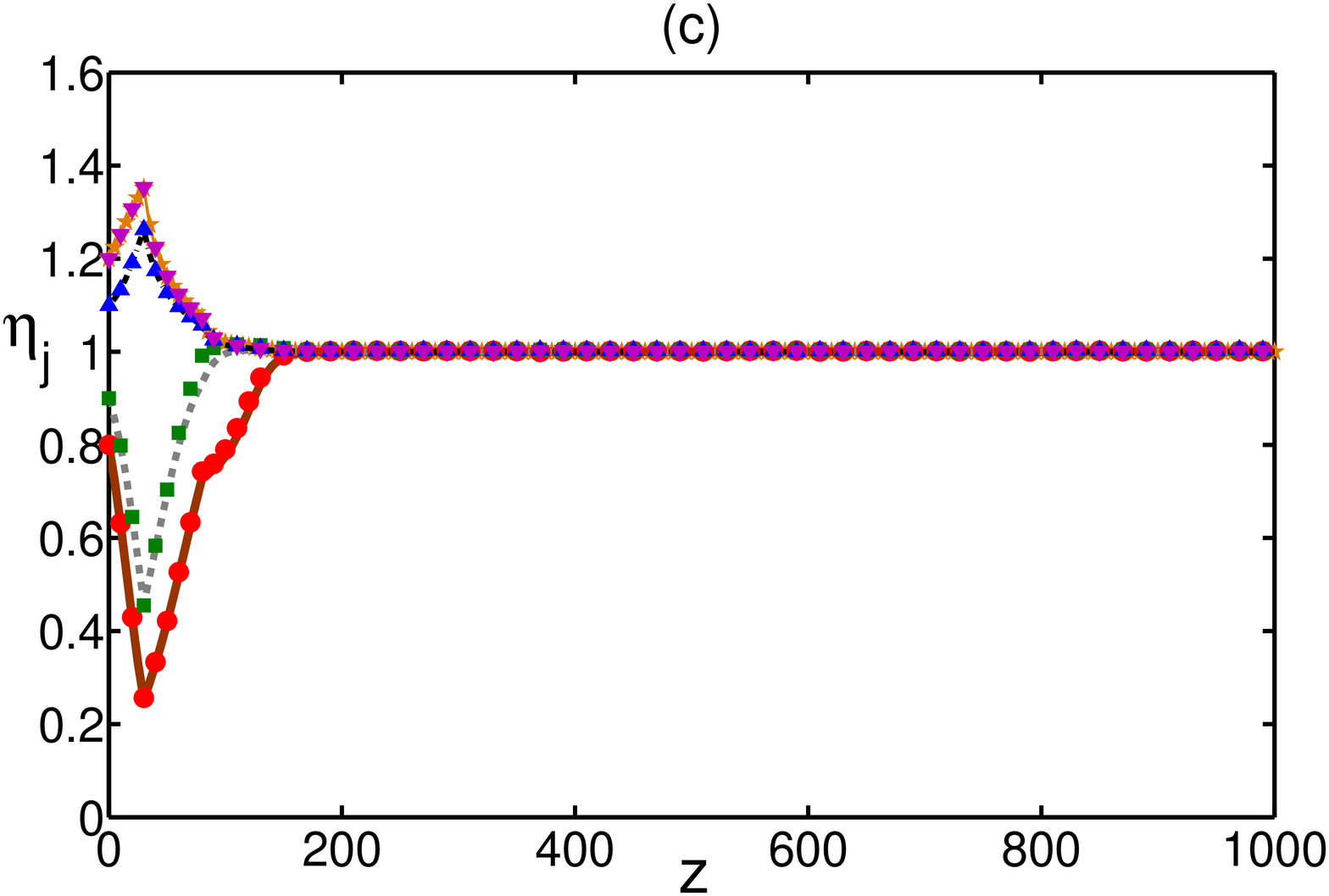} &
\epsfxsize=8.2cm  \epsffile{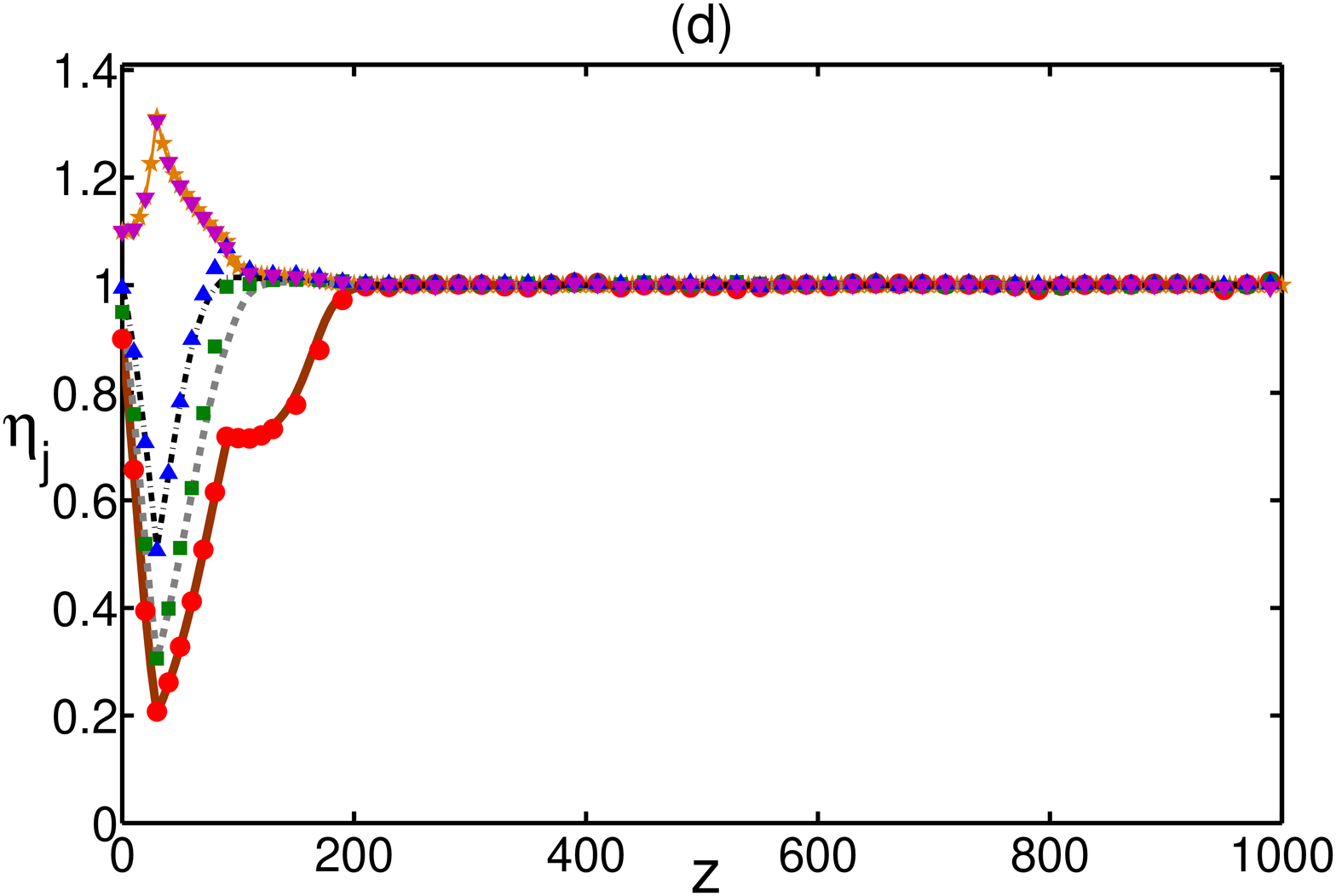} 
\end{tabular}
\caption{The $z$ dependence of soliton amplitudes $\eta_{j}$ during  
single transmission switching events in four-sequence transmission in waveguides with 
broadband delayed Raman response and narrowband nonlinear gain-loss. 
(a) and (b) show on-off switching of two and three soliton sequences in waveguides 
with a GL gain-loss profile in setup A1-A2, while (c) and (d) show off-on switching 
of two and three sequences in hybrid waveguides in setup A2-B-A1.      
The red circles, green squares, blue up-pointing triangles, 
and magenta down-pointing triangles 
represent $\eta_{1}(z)$, $\eta_{2}(z)$, $\eta_{3}(z)$, and $\eta_{4}(z)$, 
obtained by numerical simulations with Eqs. (\ref{global1}) and (\ref{global2}) in (a) and (b), 
and with Eqs. (\ref{global1})-(\ref{global3a}) in (c) and (d).    
The solid brown, dashed gray, dashed-dotted black, and solid-starred orange curves 
correspond to $\eta_{1}(z)$, $\eta_{2}(z)$, $\eta_{3}(z)$, and $\eta_{4}(z)$,  
obtained by the predator-prey model (\ref{global5}).}
 \label{fig2}
\end{figure*}

The results of numerical simulations with Eqs. (\ref{global1}) and (\ref{global2})  
for on-off switching of two and three soliton sequences in four-sequence transmission 
in setup A1-A2 are shown in Figs. \ref{fig2}(a) and \ref{fig2}(b). 
The results of simulations with Eqs. (\ref{global1})-(\ref{global3a})   
for off-on switching of two and three sequences in four-sequence transmission in setup A2-B-A1 
are shown in Figs. \ref{fig2}(c) and \ref{fig2}(d). A comparison with the predictions 
of the predator-prey model (\ref{global5}) is also presented. The agreement between 
the coupled-NLS simulations and the LV model's predictions is excellent in all four cases. 
More specifically, in on-off transmission of $M$ sequences with $M=2$ and $M=3$, 
the amplitudes of the solitons in the $M$ lowest frequency channels tend to zero, 
while the amplitudes of the solitons in the $N-M$ high frequency channels tend to 
new values $\eta_{sj}$, where $M+1 \le j \le N$. The values of the new amplitudes 
are $\eta_{s3}=1.2499$ and $\eta_{s4}=1.2878$ in on-off switching of two sequences, 
and $\eta_{s4}=1.3640$ in on-off switching of three sequences. As can be seen from 
Figs.  \ref{fig2}(a) and \ref{fig2}(b), these values are in excellent agreement with 
the predictions of the predator-prey model (\ref{global5}). 
In off-on switching of $M$ soliton sequences, the amplitudes of the solitons in 
the $M$ low frequency channels tend to zero for $z<z_{s1}$, while the amplitudes 
of the solitons in the $N-M$ high frequency channels increase with $z$ for $z<z_{s1}$. 
After the switching, i.e., for distances $z\ge z_{s1}$, the amplitudes of the 
solitons in the $M$ low frequency channels increase to the steady-state value of 1, 
while the amplitudes of the solitons in the $N-M$ high frequency channels decrease  
and tend to 1, in full agreement with the predator-prey model's predictions. 
Note that very good agreement between the coupled-NLS and predator-prey models  
is observed even when some of the soliton amplitudes are small, i.e., 
even outside of the perturbative regime, where the predator-prey model is expected to hold.  
The results of the simulations presented in Fig. \ref{fig2} and similar results      
obtained with other sets of the physical parameters demonstrate that it is indeed 
possible to realize stable scalable on-off and off-on transmission switching in the 
waveguide setups considered in the current study. Furthermore, the simulations 
confirm that design of the switching setups can be guided by stability and bifurcation 
analysis for the steady states of the predator-prey model (\ref{global5}). 
We point out that the off-on switching setups can also be employed in 
broadband transmission recovery, that is, in the simultaneous amplification 
of multiple soliton sequences, which experienced significant energy decay, 
to a desired steady-state amplitude value.

As discussed in Section \ref{stability}, an important application 
of the switching setups considered in our paper is for realizing 
efficient signal processing in multichannel transmission. 
In such application, the amplitude values $\eta_{j}$ are used 
to encode information about the type of signal processing to 
be carried out in the next processing station. As a result, 
the pulse sequences typically undergo multiple switching events 
and it is important to show that this can be realized in a stable manner. 
We therefore turn to discuss the results of numerical simulations 
with the coupled-NLS model (\ref{global1}) for multiple switching events.       
As a specific example, we consider multiple switching in a three-channel system 
in the hybrid waveguide setup A1-(A2-B-A1)-...-(A2-B-A1), where (A2-B-A1) 
repeats six times. Thus, in this case the soliton sequences first experience 
transmission stabilization in waveguide setup A1, and then undergo six successive 
off-on switching events in waveguide setup A2-B-A1. The parameter values are 
chosen such that during on switching, amplitude values in all sequences tend to 1, 
while during off switching, transmission of a single sequence  
(the lowest-frequency sequence $j=1$) is turned off. We emphasize, however, that similar 
results are obtained with other numbers of channels and in other switching 
scenarios. For the example presented here, during on switching stabilization in waveguide setup 
A1, $\epsilon_{R}=0.0006$, $\epsilon_{5}=0.15$, $\kappa=1.2$, and $\eta=1$ 
are used, while during off switching in waveguide setup A2, $\epsilon_{R}=0.0024$, 
$\epsilon_{5}=0.024$, $\kappa=2.2$, and $\eta=1.15$ are used.  
Note that the higher value of $\epsilon_{R}$ in waveguide setup A2 is required 
for realizing a faster on-off transmission switching, that is, for decreasing the 
distance along which the off switching takes place.    
Additionally, during on switching in waveguide setup B, $\epsilon_{R}=0.0006$,  
$\epsilon_{3}^{(2)}=0.02$, and $\eta=1$ are chosen.    
To suppress radiative instability during propagation in waveguide spans with linear 
gain-loss and cubic loss (setup B), the frequency dependent linear gain-loss 
$g(\omega,z)$ of Eq. (\ref{global3a}) with $W=10$ and $g_{L}=-0.5$ is employed.    
In the simulations, transmission of the $j=1$ sequence is turned off in setup A2 
at distances $z_{3m+1}=700(m+1)$ for $0 \le m \le 5$. Transmission of sequence $j=1$ is 
turned on in waveguide setup B at $z_{3m+2}=50+700(m+1)$ for $0 \le m \le 5$, 
and transmission stabilization in setup A1 starts at $z_{0}=0$ and at 
$z_{3m}=100+700m$ for $1 \le m \le 6$. 
The final propagation distance is $z_f=5000$.  
Thus, the waveguide spans are $[0,700)$,  $[700,750)$, $[750,800)$, 
$[800,1400)$, ..., $[4200,4250)$, $[4250,4300)$, and $[4300,5000]$.

\begin{figure}[ptb]
\begin{tabular}{cc}
\epsfxsize=12cm  \epsffile{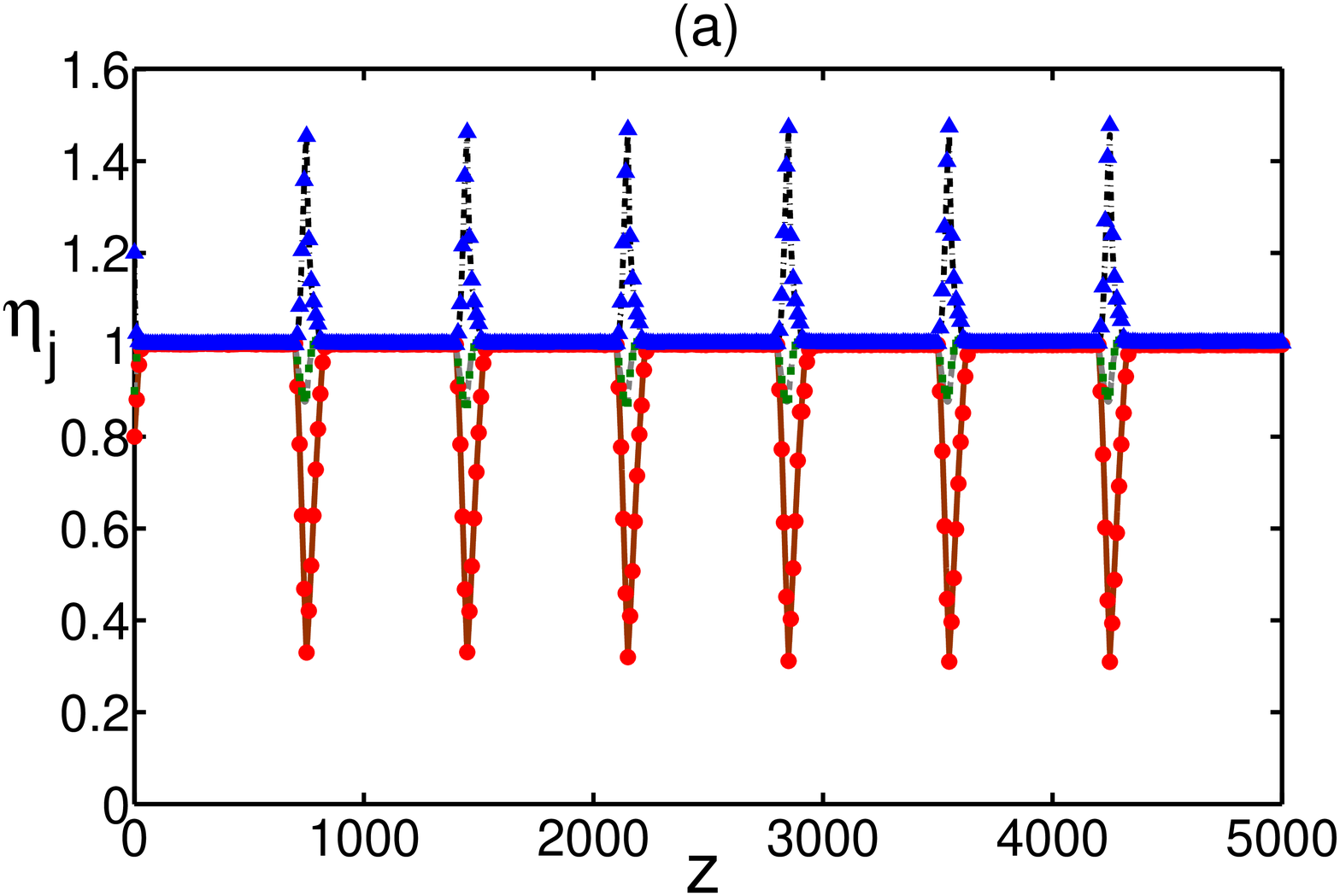}\\
\epsfxsize=12cm  \epsffile{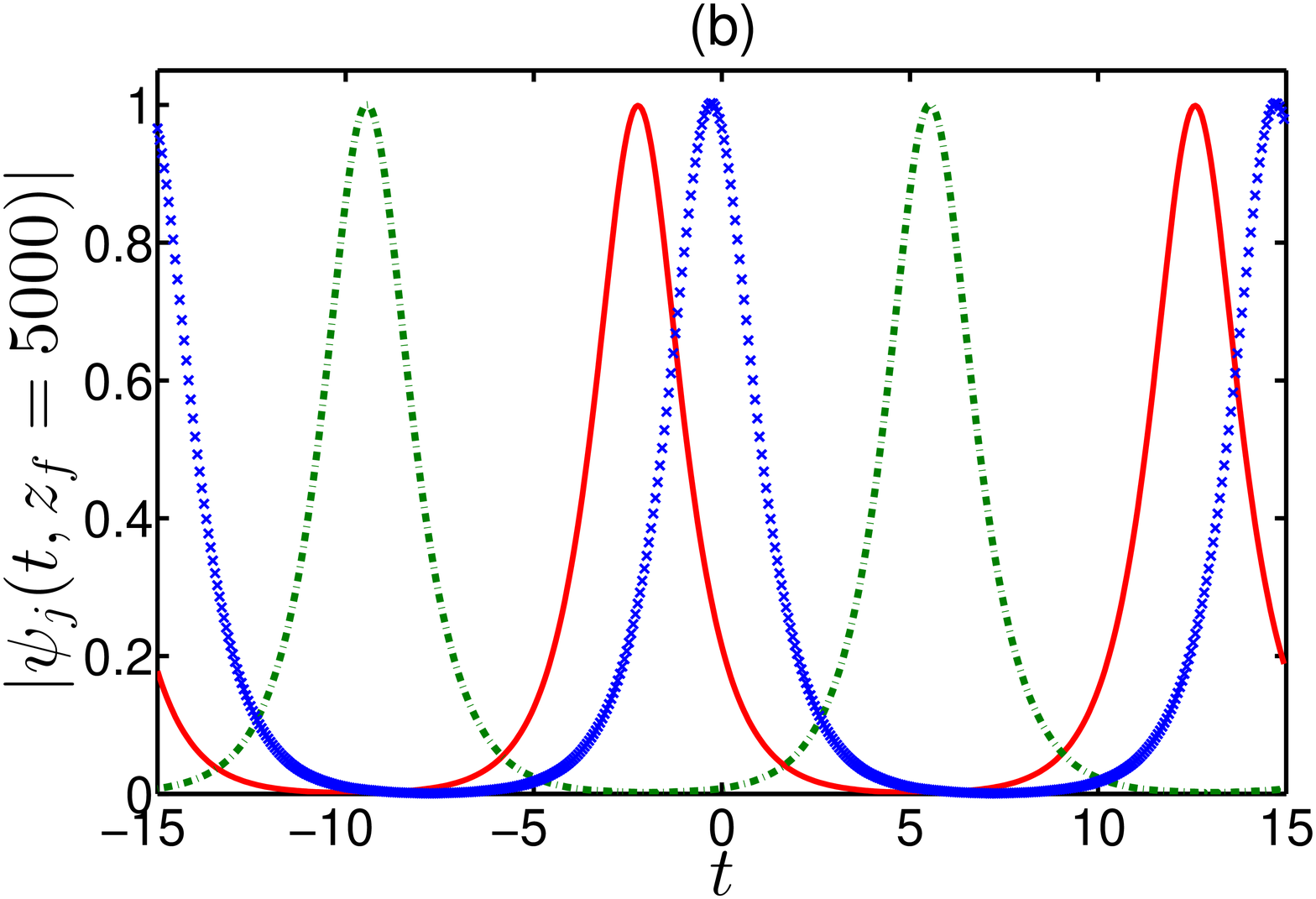}\\
\end{tabular}
\caption{(a) The $z$ dependence of soliton amplitudes $\eta_{j}$ in   
multiple transmission switching with three sequences ($N=3$) in hybrid waveguide setup 
A1-(A2-B-A1)-...-(A2-B-A1), where (A2-B-A1) repeats six times.  
In this case, the soliton sequences undergo transmission stabilization 
followed by six successive off-on switching events. 
The red circles, green squares, and blue up-pointing triangles 
represent $\eta_{1}(z)$, $\eta_{2}(z)$, and $\eta_{3}(z)$, 
obtained by numerical simulations with Eqs. (\ref{global1})-(\ref{global3a}). 
The solid brown, dashed gray, and dashed-dotted black curves  
correspond to $\eta_{1}(z)$, $\eta_{2}(z)$, and $\eta_{3}(z)$,  
obtained by the predator-prey model (\ref{global5}).
(b) The pulse patterns $|\psi_{j}(t,z)|$ at the final distance 
$z=z_{f}=5000$, as obtained by numerical solution of Eqs. (\ref{global1})-(\ref{global3a}).   
The solid red curve, dashed-dotted green curve, and blue crosses  
represent $|\psi_{j}(t,z)|$ with $j=1,2,3$.} 
 \label{fig3}
\end{figure}

The results of numerical simulations with Eqs. (\ref{global1})-(\ref{global3a}) 
for multiple transmission switching events are shown in Fig. \ref{fig3} along with 
the predictions of the predator-prey model (\ref{global5}).  
The agreement between the coupled-NLS simulations and the predator-prey model's 
predictions is excellent throughout the propagation. Furthermore, as seen in Fig. \ref{fig3}(b), 
no pulse distortion is observed at the final propagation distance $z_{f}$. 
Note that the minimal values of $\eta_{1}(z)$, 
which are attained prior to the start of the on switch, are $\eta_{1}(z_{3m+2})=0.33$. 
As a result, the value of the decision level $\eta_{th}$ for distinguishing between on and off 
transmission states can be set as low as $\eta_{th}=0.35$, which is significantly lower 
than the value $\eta_{th}=0.65$ obtained in Ref. \cite{CPJ2013} 
for transmission in a two-channel waveguide system with a broadband GL gain-loss profile.   
The results presented in Fig. \ref{fig3} along with results of 
numerical simulations with other sets of the physical parameter values 
demonstrate that stable multiple transmission switching events  
can indeed be realized over a wide range of amplitude values, using waveguides 
with broadband delayed Raman response and narrowband nonlinear gain-loss.

\section{Discussion}
\label{Discussion}
Let us discuss the reasons for the robustness and scalability of transmission stabilization 
and switchng in waveguides with  broadband delayed Raman response and narrowband 
nonlinear gain-loss. The scalability and robustness of transmission control 
can be attributed to the following properties of these waveguides. 
(1) The asymptotic stability of the steady state $(\eta, \dots, \eta)$ for 
waveguides with a GL gain-loss profile, which is independent of $N$ and $\epsilon_{R}$, 
is key to realizing scalable transmission stabilization and switching. 
(2) The presence of net linear loss in all frequency channels for waveguides 
with a GL gain-loss profile leads to mitigation of radiative instability. 
(3) Due to the narrow bandwidth of the nonlinear gain-loss, 
three-pulse interaction does not contribute to collision-induced 
amplitude shifts. As a result, the extension of the predator-prey model 
from $N=2$ to a general $N$ value is straightforward. This also makes 
the extension of waveguide setup design from $N=2$ to a general $N$ 
value straightforward. In contrast, in waveguides with broadband nonlinear gain-loss, 
three-pulse interaction gives an important contribution to 
collision-induced amplitude shifts \cite{PC2012,PNG2014}. 
Due to the complex nature of three-pulse interaction in generic 
three-soliton collisions in waveguides with broadband nonlinear gain or loss (see Ref. \cite{PNG2014}),  
it is very difficult to extend the LV model for amplitude dynamics from $N=2$ 
to a generic $N$ value in these waveguides. In the absence of an $N$-dimensional LV model, 
it is unclear how to design setups for stable transmission stabilization and switching in $N$-sequence 
waveguide systems with broadband nonlinear gain-loss.     
As a result, transmission stabilization and switching in waveguides with broadband nonlinear gain-loss 
have been so far limited to two-sequence systems \cite{PC2012,CPJ2013,NPT2015}.       
(4) The Raman-induced energy transfer in soliton collisions from high-frequency 
solitons to low-frequency solitons is an important ingerdient in the realization of 
scalable on-off switching. Indeed, to compensate for the Raman-induced energy loss or gain 
in the collisions, high-frequency sequences should be overamplified while low-frequency 
sequences should be underamplified compared to mid-frequency sequences. 
As a result, the magnitude of the net linear loss is largest for the low-frequency 
sequences, and therefore, on-off switching is easiest to realize for these 
sequences. Thus, the presence of broadband delayed Raman response introduces a preference 
for turning off the transmission of the low-frequency sequences, and by this enables 
systematic scalable on-off switching.
(5) The global asymptotic stability of the steady state $(\eta, \dots, \eta)$ for 
waveguide spans with linear gain-loss and cubic loss, which is independent of all physical parameters, 
is important for realizing robust scalable off-on switching in hybrid waveguides. 
These five waveguide properties are explained by stability and bifurcation analysis 
for the steady states of the generalized $N$-dimensional predator-prey model 
for amplitude dynamics. Thus, the analysis of the predator-prey model  is essential to 
the design of waveguide setups leading to stable scalable control of soliton-based 
multichannel transmission.

Note that waveguide setups with narrowband cubic gain and quinitc loss or with 
narrowband cubic loss can be realized by employing fast saturable absorbers with 
a bandwidth $\Delta\nu_{GL}$ satisfying: $\nu_{0} \ll \Delta\nu_{GL} \ll \Delta\nu$. 
That is, in these waveguide systems,  the bandwidth 
of the saturable absorber is larger than the spectral 
width of the optical pulses but smaller than the frequency spacing between 
adjacent frequency channels. 
Waveguide systems containing a fast saturable absorber with a finite spectral width,  
which is much larger than the spectral width of the optical pulses, have been 
studied extensively in the context of mode-locked lasers; see, for example, 
Refs. \cite{Haus75,Moores93,Haus2000} and references therein.

\section{Conclusions}
\label{conclusions}
We developed a method for achieving stable scalable control of propagation of 
multiple soliton sequences in broadband optical waveguide systems. 
The method is based on employing nonlinear waveguides with broadband 
delayed Raman response, linear gain-loss, and narrowband nonlinear gain-loss. 
We showed that the combination of Raman-induced amplitude shifts 
in interchannel collisions and single-pulse 
amplitude shifts due to linear and nonlinear gain-loss with properly chosen 
physical parameter values can be used to realize robust scalable transmission stabilization and switching. 
For this purpose, we first showed that the dynamics of soliton amplitudes 
in an $N$-sequence transmission system can be described by 
a generalized $N$-dimensional predator-prey model. 
We then carried out stability and bifurcation analysis for the steady states of the predator-prey model 
for two central cases of the gain-loss: (1) a GL gain-loss profile, (2) linear gain-loss and cubic loss. 
The stability and bifurcation analysis was then used to develop waveguide setups
that lead to robust transmission stabilization as well as on-off and off-on switching 
of $M$ out of $N$ soliton sequences.

For waveguides with a GL gain-loss profile, we obtained the Lyapunov function $V_{L}(\pmb{\eta})$ 
for the predator-prey model and used it to derive simple conditions for 
asymptotic stability and instability of the steady state with equal amplitudes 
for all sequences  $(\eta, \dots, \eta)$. 
These conditions are independent of the number of channels $N$ and the value of 
the Raman coefficient $\epsilon_{R}$, which is essential for the realization 
of scalable transmission stabilization and switching. 
We also found that the steady state at the origin is asymptotically stable, 
provided all the linear gain-loss coefficients are negative.  
Combining the requirements for asymptotic stability of both $(\eta, \dots, \eta)$ 
and the origin, we showed that the smallest value of the quintic loss coefficient 
$\epsilon_{5}$ required for robust transmission stabilization and off-on 
switching for $N \gg 1$ scales as $\epsilon_{5} \sim N^{2}\epsilon_{R}$. 
The realization of on-off switching requires stability analysis of steady states, 
for which $M$ components are equal to zero. We first gave a simple argument, 
showing that switching off of $M$ sequences is most conveniently 
realized by turning off the transmission of the low-frequency sequences $1 \le j \le M$. 
We therefore focused attention on the steady state 
$(0, \dots, 0, \eta_{s(M+1)}, \dots, \eta_{sN})$ 
and showed that for $N \gg 1$, 
stability of this steady state can be established 
by calculating only $N-M+1$ diagonal elements 
of the corresponding Jacobian matrix.

Stability analysis for the predator-prey model, describing amplitude dynamics in 
waveguides with linear gain-loss and cubic loss, was carried out in a similar manner. 
More specifically, we found that the same $V_{L}(\pmb{\eta})$ that was used for 
the predator-prey model with a GL gain-loss profile is a Lyapunov function 
for the predator-prey model with linear gain-loss and cubic loss \cite{Lyapunov}. 
Moreover, we used this Lyapunov function to show that $(\eta, \dots, \eta)$ 
is globally asymptotically stable, regardless of the values of all physical parameters. 
However, linear stability analysis showed that the origin is unstable in this case. 
The latter instability eventually leads to growth of small amplitude waves, 
and thus makes waveguides with linear gain-loss and cubic loss 
unsuitable for long-distance transmission stabilization.
On the other hand, the global asymptotic stability of $(\eta, \dots, \eta)$ means that 
waveguide spans with linear gain-loss and cubic loss can be used in hybrid waveguides 
for realizing robust off-on transmission switching.

The predictions of the generalized predator-prey model for 
scalable transmission stabilization and switching were tested by numerical 
simulations with a perturbed coupled-NLS model, 
which takes into account broadband delayed Raman response 
and a narrowband GL gain-loss profile. The coupled-NLS simulations 
for transmission stabilization were carried out with $2 \le N \le 4$ soliton sequences. 
The simulations showed stable propagation 
and excellent agreement with the predictions of the predator-prey model over 
significantly larger distances compared with those obtained in earlier works 
with other waveguide setups. More specifically, the stable propagation distances 
obtained for two-, three-, and four-sequence transmission were larger by 
factors of 37.9, 34.3, and 10.6, respectively, compared with the distances obtained 
in single-waveguide transmission in the 
presence of delayed Raman response and in the absence of 
nonlinear gain-loss \cite{PNT2015}. Furthermore, the distance along which 
stable transmission was observed in a two-channel system was larger by 
a factor of 200 compared with the distance achieved in waveguides 
with linear gain and broadband cubic loss \cite{PNC2010}. 
The enhanced stability of $N$-channel transmission through waveguides 
with broadband delayed Raman response and narrowband GL gain-loss profile
was explained in the Discussion. 
%The enhanced stability of waveguides with broadband delayed Raman 
%response and narrowband GL gain-loss profile can be attributed to the stability 
%of the predator-prey model's steady state at the origin. Physically, this 
%means that in the waveguides considered in the current study, 
%all pulse sequences propagate in the presence of net linear loss, 
%and as a result, radiative instability is mitigated.  
%In contrast, in the studies reported in Refs. \cite{PNT2015} and \cite{PNC2010}, 
%the LV model's steady state at the origin is unstable,  
%and some or all of the soliton sequences propagate under net linear gain. 
%This leads to growth of small amplitude waves 
%and to transmission destabilization at significantly shorter distances 
%compared with the distances observed in the current study.  

We demonstrated single and multiple transmission switching events 
of $M$ out of $N$ pulse sequences by carrying out numerical simulations 
with the coupled-NLS model that was described in the preceding paragraph.  
As examples, we presented the results of the simulations for the following 
setups: (a) single on-off and off-on switching events of two and three 
soliton sequences in four-sequence transmission; 
(b) six switching events in a three-sequence system, in which transmission of one soliton 
sequence was switched off and then on six consecutive times.       
The results of the coupled-NLS simulations were in excellent agreement 
with the predictions of the predator-prey model for both single and multiple 
switching events.  Furthermore, the agreement was observed even when amplitude values 
were small for some soliton sequences, i.e., even outside of the regime 
where the predator-prey model's description was expected to hold.   
Based on these results and results of simulations 
with other sets of the physical parameter values, we concluded that stable scalable transmission  
switching can indeed be realized in waveguides with broadband 
delayed Raman response and narrowband nonlinear gain-loss.

\section*{Acknowledgments}
We are grateful to T.P. Tran for help with the numerical code 
in the initial stages of this work. 
Q.M.N. and T.T.H are supported by the Vietnam National Foundation 
for Science and Technology Development (NAFOSTED) 
under Grant No. 101.99-2015.29.

\section*{Author contribution statement}
A.P. initiated the project, participated in the derivation of 
the analytic results, and took part in the analysis of the 
results of numerical simulations. 
Q.M.N. and T.T.H carried out the numerical simulations, 
participated in the derivation of the analytic results, 
and took part in the analysis of the results of numerical simulations.


\begin{thebibliography}{}
\bibitem{Agrawal2001} G.P. Agrawal, {\it Nonlinear Fiber Optics} 
(Academic, San Diego, CA, 2001).

\bibitem{Tkach97} F. Forghieri, R.W. Tkach, and A.R. Chraplyvy,
 in {\it Optical Fiber Telecommunications III}, I.P. 
Kaminow and T.L. Koch, eds., (Academic, San Diego, CA, 1997), Chapter 8.

\bibitem{Mollenauer2006} L.F. Mollenauer and J.P. Gordon,  
{\it Solitons in Optical Fibers: Fundamentals and Applications} 
(Academic, San Diego, CA, 2006).  

\bibitem{Gnauck2008} A.H. Gnauck, R.W. Tkach, A.R. Chraplyvy, 
and T. Li, J. Lightwave Technol. {\bf 26}, 1032 (2008).  


\bibitem{Essiambre2010} R.-J. Essiambre, G. Kramer, P.J. Winzer, 
G.J. Foschini, and B. Goebel, J. Lightwave Technol. {\bf 28}, 
662 (2010).  

\bibitem{multisequence} For this reason, we use the equivalent terms 
multichannel transmission, multisequence transmission, and WDM transmission 
to describe the simultaneous propagation of multiple pulse sequences with 
different central frequencies through the same optical waveguide.      

\bibitem{Agrawal2007a} Q. Lin, O.J. Painter, and G.P. Agrawal, Opt. Express {\bf 15}, 16604 (2007).

\bibitem{Dekker2007} R. Dekker, N. Usechak, M. F\"orst, and A. Driessen, J. Phys. D {\bf 40}, R249 (2007). 

\bibitem{Gaeta2008} M.A. Foster, A.C. Turner, M. Lipson, 
and A.L. Gaeta, Opt. Express {\bf 16}, 1300 (2008).

\bibitem{Chow96} J. Chow, G. Town, B. Eggleton, M. Ibsen, K. Sugden, 
and I. Bennion, IEEE Photon. Technol. Lett. {\bf 8}, 60 (1996).

\bibitem{Shi97} H. Shi, J. Finlay, G.A. Alphonse, J.C. Connolly, 
and P.J. Delfyett, IEEE Photon. Technol. Lett. {\bf 9}, 1439 (1997).

\bibitem{Zhang2009} H. Zhang, D.Y. Tang, X. Wu, and L.M. Zhao, Opt. Express {\bf 17}, 12692 (2009). 

\bibitem{Liu2013} X.M. Liu, D.D. Han, Z.P. Sun, C. Zeng, H. Lu, D. Mao, 
Y.D. Cui, and F.Q. Wang, Sci. Rep. {\bf 3}, 2718 (2013). 

\bibitem{Iannone98} E. Iannone, F. Matera, A. Mecozzi, and M. Settembre, 
{\it Nonlinear Optical Communication Networks} (Wiley, New York, 1998).

\bibitem{MM98} L.F. Mollenauer and P.V. Mamyshev,
IEEE J. Quantum Electron. {\bf 34}, 2089 (1998).

\bibitem{NP2010} Q.M. Nguyen and A. Peleg, Opt. Commun. {\bf 283}, 
3500  (2010).   

\bibitem{PNC2010} A. Peleg, Q.M. Nguyen, and Y. Chung, 
Phys. Rev. A {\bf 82}, 053830 (2010).

\bibitem{PC2012} A. Peleg and Y. Chung, 
Phys. Rev. A {\bf 85}, 063828 (2012). 

\bibitem{CPJ2013} D. Chakraborty, A. Peleg, and J.-H. Jung, 
Phys. Rev. A {\bf 88}, 023845 (2013).

\bibitem{NPT2015} Q.M. Nguyen, A. Peleg, and T.P. Tran, 
Phys. Rev. A {\bf 91}, 013839 (2015).   

\bibitem{PNT2015} A. Peleg, Q.M. Nguyen, and T.P. Tran, 
Opt. Commun. {\bf 380}, 41  (2016).   

\bibitem{Chraplyvy84}  A.R. Chraplyvy, Electron. Lett. {\bf 20}, 58 (1984).

\bibitem{Tkach95} F. Forghieri, R.W. Tkach, and A.R. Chraplyvy,
IEEE Photon. Technol. Lett. {\bf 7}, 101 (1995).

\bibitem{Ho2000} K.-P. Ho, J. Lightwave Technol. {\bf 18}, 915 (2000).

\bibitem{Yamamoto2003} T. Yamamoto and S. Norimatsu, 
J. Lightwave Technol. {\bf 21}, 2229 (2003).

\bibitem{P2004} A. Peleg, Opt. Lett. {\bf 29}, 1980 (2004).

\bibitem{P2007} A. Peleg, Phys. Lett. A {\bf 360}, 533 (2007).
   
\bibitem{CP2008} Y. Chung and A. Peleg, Phys. Rev. A  {\bf 77}, 
063835 (2008). 

\bibitem{Golovchenko2009} B. Bakhshi, L. Richardson, and E.A. Golovchenko, 
in {\it Proceedings of the Optical Fiber Communication Conference, 
San Diego, CA, 2009}, paper OThC4.    

\bibitem{PC2012b} A. Peleg and Y. Chung, Opt. Commun. {\bf 285}, 
1429 (2012).  

\bibitem{Islam2004} {\it Raman Amplifiers for Telecommunications 1: 
Physical Principles}, edited by M.N. Islam (Springer, New York, 2004).

\bibitem{Agrawal2005} {\it Raman Amplification in Fiber Optical 
Communication Systems}, edited by C. Headley and G.P. Agrawal 
(Elsevier, San Diego, CA, 2005).
 
\bibitem{Gaeta2010} Y. Okawachi, O. Kuzucu, M.A. Foster, R. Salem, 
A.C. Turner-Foster, A. Biberman, N. Ophir, K. Bergman, M. Lipson, 
and A.L. Gaeta, IEEE Photon. Technol. Lett. {\bf 24}, 185 (2012).  

\bibitem{PNG2014} A. Peleg, Q.M. Nguyen, and P. Glenn, 
Phys. Rev. E {\bf 89}, 043201 (2014). 


\bibitem{g_omega_z} Note that a similar approach for mitigation 
of radiative instability was employed in Ref. \cite{PNT2015} for 
soliton propagation in the presence of delayed Raman response 
in the absence of nonlinear gain-loss. 


\bibitem{Lotka25} A.J. Lotka, {\it Elements of Physical Biology} 
(Williams and Wilkins, Baltimore, 1925). 

\bibitem{Volterra28} V. Volterra, J. Cons. Int. Explor. Mer 
{\bf 3}, 1 (1928). 


\bibitem{AK_transmission}
Channel switching can also be implemented in 
amplitude-keyed transmission. In this case, one should define 
a second threshold level $\eta_{th2}$, satisfying $0<\eta_{th2}<\eta_{th}$. 
The larger decision level $\eta_{th}$ is then used to determine 
the transmission state of each channel for channel switching, 
while the smaller decision level $\eta_{th2}$ is used to determine 
the state of each time-slot within a given channel.         
Thus, in this case, the on and off states 
for the $j$th channel are determined by the conditions 
$\eta_{j}>\eta_{th}$ and $\eta_{th2}<\eta_{j}<\eta_{th}$, respectively, 
where $\eta_{j}$ is the common amplitude value 
for pulses in occupied time slots in the $j$th channel.  



\bibitem{Lyapunov} It is possible to show that $V_{L}(\pmb{\eta})$ of Eq. (\ref{global6}) 
is a Lyapunov function for the predator-prey model (\ref{global5})
even for an $m$th-order polynomial $L$ with a negative 
coefficient for the $m$th-order term and properly chosen values   
for the other polynomial coefficients.   


\bibitem{linear_stab} Linear stability analysis shows that 
$(\eta, \dots, \eta)$ is a stable focus when $0< \kappa < 8\eta^{2}/5$ 
and an unstable focus when $\kappa > 8\eta^{2}/5$.   

\bibitem{Negative_g_j} Here we use the fact that the origin  
is a stable node of Eq. (\ref{global5}), so that $g_j < 0$ for $1 \le j \le N$. 
  
\bibitem{Eta_s_N} These conditions should be augmented by the condition 
for asymptotic stability of $(0, \dots, 0, \eta_{sN})$.  

\bibitem{Chi89} S. Chi and S. Wen, Opt. Lett. {\bf 14}, 1216 (1989).

\bibitem{Malomed91} B.A. Malomed, Phys. Rev. A {\bf 44}, 1412 (1991).

\bibitem{Kumar98} S. Kumar, Opt. Lett. {\bf 23}, 1450 (1998).  

\bibitem{Kaup99} T.I. Lakoba and D.J. Kaup, Opt. Lett. {\bf 24}, 808 (1999).  

\bibitem{CP2005} Y. Chung and A. Peleg, Nonlinearity {\bf 18}, 1555 (2005). 

\bibitem{NP2010b} Q.M. Nguyen and A. Peleg,  J. Opt. Soc. Am. B {\bf 27}, 1985 (2010).   

\bibitem{CPN2016} D. Chakraborty, A. Peleg, and Q.M. Nguyen, 
Opt. Commun. {\bf 371}, 252 (2016).    

\bibitem{Haus75} H.A. Haus, J. Appl. Phys. {\bf 46}, 3049 (1975). 

\bibitem{Moores93} J.D. Moores, Opt. Commun. {\bf 96}, 65 (1993).

\bibitem{Haus2000} H.A. Haus, 
IEEE J. Sel. Top. Quantum Electron. {\bf 6}, 1173 (2000). 


\end{thebibliography}
\end{document}